\documentclass[ALICE,manyauthors]{cernphprep}
\usepackage{amsmath,amssymb,amsfonts}
\usepackage[comma,square,numbers,sort&compress]{natbib}
\usepackage{hyperref}
\usepackage{lineno}

\usepackage{graphicx}
\graphicspath{{figures/}}
\usepackage{csquotes}

\usepackage{tabu,booktabs}
\usepackage[table]{xcolor}
\usepackage{multi row}
\usepackage{siunitx}
\usepackage{dcolumn}
\usepackage{tablefootnote}
\usepackage{paralist}
\usepackage{enumitem}
\setlist[itemize]{leftmargin=*}
\usepackage{bm}
\usepackage[permil]{overpic}
\usepackage{txfonts}
\usepackage[capitalise]{cleveref}

\usepackage{letltxmacro}
\LetLtxMacro{\oldsqrt}{\sqrt}
\renewcommand{\sqrt}[2][\mkern8mu]{\mkern-4mu\mathop{\oldsqrt[#1]{#2}}}
\usepackage{mciteplus}
\mciteSetBstSublistMode{n}
\mciteSetSublistMode{n}

\newcommand{\nch}{N_{\mathrm{ch}}}
\newcommand{\pnch}{\mathrm{P}\left(\nch\right)}

\newcommand{\slfrac}[2]{\left.#1\right/#2}

\newcommand{\drvt}[2]{\frac{\mathrm{d}#1}{\mathrm{d}#2}}
\newcommand{\drvtinline}[2]{\slfrac{\mathrm{d} #1}{\mathrm{d}#2}}

\newcommand{\ddrvtinline}[3]{\slfrac{\mathrm{d}^2#1}{\mathrm{d}#2\,\mathrm{d}#3}}

\newcommand{\dndeta}{\drvt{\nch}{\eta}}
\newcommand{\dndetainline}{\drvtinline{\nch}{\eta}}
\newcommand{\dndetainlinez}{\ddrvtinline{\nch}{\eta}{z}}

\newcommand{\mbtrig}[1]{\text{MB$_{\text{#1}}$}\xspace}
\newcommand{\mbor}{\mbtrig{OR}}
\newcommand{\mband}{\mbtrig{AND}}

\newcommand{\inel}{\text{INEL}\xspace}
\newcommand{\nsd}{\text{NSD}\xspace}
\newcommand{\inelgt}{\text{INEL$>$0}\xspace}

\newcommand{\itstpc}{\text{ITSTPC$+$}\xspace}
\newcommand{\itssa}{\text{ITS$+$}\xspace}
\newcommand{\spd}{\text{Tracklet}\xspace}

\newcommand{\pythia}[1]{PYTHIA#1\xspace}
\newcommand{\phojet}{PHOJET\xspace}

\newcommand{\pt}{\ensuremath{p_{\mathrm{T}}}\xspace}

\makeatletter
\renewcommand{\p@subtable}{\thetable}
\renewcommand{\thesubtable}{\alph{subtable}}
\renewcommand{\@thesubtable}{\thesubtable ) \hskip\subfiglabelskip}
\makeatother

\begin{document}%

\begin{titlepage}
\PHyear{2015}
\PHnumber{259}      
\PHdate{16 September}  
%

\title{Charged--particle multiplicities\\ in proton--proton collisions at $\sqrt{s}$ = 0.9 to 8 TeV}
\ShortTitle{Charged--particle multiplicities in proton--proton collisions}   

\Collaboration{ALICE Collaboration\thanks{See Appendix~\ref{app:collab} for the list of collaboration members}}
\ShortAuthor{ALICE Collaboration} 
\vskip 3cm
\begin{abstract}
A detailed study of pseudorapidity densities and multiplicity distributions of primary charged particles produced in proton-proton collisions, at $\sqrt{s} =$ 0.9, 2.36, 2.76, 7 and 8 TeV, in the pseudorapidity range $|\eta|<2$, was carried out using the ALICE detector. Measurements were obtained for three event classes: inelastic, non-­single diffractive and events with at least one charged particle in the pseudorapidity interval  $|\eta|<1$. The use of an improved track-counting algorithm combined with ALICE{\textquoteright}s measurements of diffractive processes allows a higher precision compared to our previous publications. A KNO scaling study was performed in the pseudorapidity intervals $|\eta|<$ 0.5, 1.0 and 1.5. The data are compared to other experimental results and to models as implemented in Monte Carlo event generators PHOJET and recent tunes of PYTHIA6, PYTHIA8 and EPOS.
\end{abstract}

\end{titlepage}
\setcounter{page}{2}

\section{Introduction}
The multiplicity of emitted charged particles is one of the most basic characteristics of high-energy hadron collisions and has been the subject of longstanding experimental and theoretical studies, which have shaped the understanding of the strong interaction. Following on from earlier ALICE studies of global properties of proton-proton (pp) collisions \cite{aamodt2010first,aamodt2010charged09,aamodt2010charged7,aamodt2011strange,tagkey2014196,abelev2012multi,aliceK0Phi,Adam:2015qaa}, this publication presents a comprehensive set of measurements of the pseudorapidity density ($\dndetainline$) of primary\footnote{Primary particles are defined as prompt particles produced in the collision including all decay products, except those from weak decays of light flavour hadrons and muons.} charged particles and of their multiplicity distributions over the energy range covered by the LHC, from 0.9 to 8~TeV. The pseudorapidity density of primary charged particles was studied over the pseudorapidity range $|\eta|<2$, and their multiplicity distributions in three intervals: $|\eta|<0.5$, 1.0 and 1.5. Results are given for three conventional event classes: \begin{inparaenum}[(a)] \item inelastic (\inel) events, \item non-single diffractive (\nsd) events and \item events with at least one charged particle in $|\eta|<1$ (\inelgt). \end{inparaenum}

At LHC energies, particle production is still dominated by soft processes but receives significant contributions from hard scattering, thus multiplicity and other global event properties measurements allow to explore both components. As these properties are used as input in Glauber inspired models \cite{glauber1955cross,glauber1959lectures,glauber2006quantum,abelev2013centrality}, such studies are also contributing to a better modelling of Pb--Pb collisions. Already at 8 TeV, high multiplicity proton-proton collisions provide energy densities comparable, for instance, to energy densities in Au-Au central collisions at RHIC, allowing a comparison of nuclear matter properties in strongly interacting systems with similar energy densities but with volumes orders of magnitude smaller.

It is worth noting that, already at $\sqrt{s} = 2.36$~TeV, hadron collision models tuned to pre-LHC data failed to reproduce basic characteristics of proton-proton collisions at the LHC, such as pseudorapidity density of charged particles, multiplicity distributions, particle composition, strangeness content, transverse momentum distributions and sphericity (see for instance \cite{aamodt2010charged09}, \cite{aamodt2010charged7}, \cite{aamodt2011strange} and \cite{abelev2012transverse}). Therefore, a more precise measurement of charged-particle multiplicity distributions and a study of their energy dependence contribute to a better understanding of particle production mechanisms and serve to improve models. In turn, a better simulation of collision properties improves the determination of the detector response and background estimates of underlying event properties relevant to the study of high-\pt~phenomena.

In the Regge theory \cite{collinsintroduction,*gribov2003theory,*gribov2009strong}, one of the most successful models for describing soft hadronic interactions, the asymptotic behaviour of cross-sections for elastic scattering and multiple production of hadrons is determined by the properties of the Pomeron, the $t$-channel right-most pole, in the elastic scattering amplitude. In QCD, the Pomeron, which has vacuum quantum numbers, is usually related to gluonic exchanges in the $t$-channel. The experimentally observed increase of the total cross-section with increasing collision energy made it necessary to consider a Pomeron as a Regge trajectory with $t = 0$ intercept: $\alpha_{P}(0) = 1 + \Delta > 1$~\cite{collinsintroduction}. The energy dependence of the particle \mbox{(pseudo-)rapidity} density provides information about the Pomeron trajectory intercept parameter, $\Delta$. If interactions between Pomerons are neglected, the inclusive particle production cross-section, $\sigma_{\mathrm{Incl.}}$, is determined only by the contribution of the single (cut-)Pomeron exchange diagram. In this approximation, $\drvtinline{\sigma_{\mathrm{Incl.}}}{y}\ (\sim \drvtinline{\sigma_{\mathrm{Incl.}}}{\eta})$ at mid-rapidity is proportional to $s^{\Delta}$ \cite{abramowsky1974agk}. Thus, the energy dependence of the inclusive cross-section gives more reliable information about the value of $\Delta$ than the energy dependence of the total interaction cross-section, for which contributions from multi-Pomeron exchanges strongly modify the energy dependence of the single Pomeron exchange diagram. In the same approximation, the energy dependence of the particle (pseudo-)rapidity density in the central rapidity region is given by $\drvtinline{N}{y}\propto \slfrac{s^{\Delta}}{\sigma_{\mathrm{Int.}}}$, where $\sigma_{\mathrm{Int.}}$ is the interaction cross-section (see for instance \cite{kaidalov1982quark,kaidalov-Predictions,Kaidalov:1986tu,capellappAAmult}). Up to LHC energies, $\sigma_{\mathrm{Int.}}$ is well represented by a power law of $s$. However, for reasons of unitarity \cite{froissart1961asymptotic}, it is expected that this power law should be broken at sufficiently high energy, although well above LHC energies.
Therefore, the energy dependence of the particle (pseudo-) rapidity density in the central region at LHC, $\drvtinline{N}{y} \approx \drvtinline{N}{\eta}$, should follow the same power law trend. In this publication, this relationship is explored further for three event classes and using 5 ALICE data points.

It was more than 40 years ago that A. M. Polyakov \cite{polyakov1971similarity} and then Z. Koba, H.B. Nielsen and P. Olesen \cite{koba1972scaling} proposed that the probability distribution of producing $n$ particles in a collision, $P\left( n \right)$, when expressed as a function of the average multiplicity, $\left<n\right>$, should reach an asymptotic shape at sufficiently high energy
\begin{equation}
 P\left( n \right) = \frac{1}{\left<n\right>} \Psi \left( \frac{n}{\left<n\right>} \right)
\end{equation}
where $\Psi$ is a function supposed to describe the energy-invariant shape of the multiplicity distribution. Such scaling behaviour is a property of particle multiplicity distributions known today as “Koba-Nielsen-Olesen (KNO) scaling”.

One well identified mechanism for KNO scaling violation is the increasing probability of multi-parton scattering with increasing $\sqrt{s}$. Moreover, since the topologies and multiplicities of diffractive and non-diffractive (ND) events are different, their KNO behavior may be different. Even if KNO scaling were to be valid for each, it might not be valid for their sum. Nevertheless, KNO scaling is expected to be violated for both diffractive and non-diffractive processes \cite{abramovsky1972pis,kaidalov1982pomeron} at sufficiently high collision energies and the LHC provides the best opportunity to study the extent of these scaling violations.

Indeed, deviation from KNO scaling was already observed long ago at ISR energies (proton-proton collisions at $\sqrt{s}$ from 30.4 to 62.2 GeV), in the full phase space, for inelastic events \cite{breakstone1984charged}. On the other hand, for NSD collisions, scaling was still found to be present \cite{breakstone1984charged}, suggesting that diffractive processes might also play a role in KNO scaling violations. In e$^+$e$^-$ collisions, at $\sqrt{s}$ from 5 to 34 GeV, KNO scaling was found to hold within $\pm 20$\% \cite{Althoff:1983ew}. In proton-antiproton collisions at the CERN collider ($\sqrt{s} = $ 200, 546 and 900 GeV), KNO scaling was found to be violated for NSD collisions in full phase space \cite{Ansorge:1988kn}, \cite{Chou:1982dn}, \cite{Alpgard:1981jn}. Nevertheless, for NSD collisions, in limited central pseudorapidity intervals, KNO scaling was still found to hold up to 900 GeV, and at $\sqrt{s}$ = 546 GeV, KNO scaling was found to hold in the pseudorapidity interval $|\eta|< 3.5$ \cite{Arnison:1982rm,Kam:1985ir}. In NSD proton-proton collisions at the LHC, at $\sqrt{s} = 2.36$ and $7$~TeV and in $|\eta|<0.5$, ALICE \cite{aamodt2010charged09} and CMS \cite{Rougny:2010zz} observed no significant deviation from KNO scaling.

This publication presents a study of KNO scaling, at $\sqrt{s}$ from 0.9 to 8~TeV, in three pseudorapidity intervals ($|\eta|<$ 0.5, 1.0 and 1.5) and for a higher multiplicity reach compared to previous ALICE publications, quantified with KNO variables (moments) \cite{koba1972scaling}  as well as with the parameters of Negative Binomial Distributions (NBD) used to fit measured multiplicity distributions.

With respect to previous ALICE publications, the analysis reported here makes use of improved tracking and track-counting algorithms; better knowledge and improved simulation of diffraction processes; an expanded pseudorapidity range for $\dndetainline$ studies and better statistical precision at $\sqrt{s} = $ 0.9 and 7~TeV, extending by a factor of 2 the previously published multiplicity distribution reach. Results at $\sqrt{s} = $ 2.76 and 8~TeV are presented for the first time in this publication.

Previous measurements of both $\dndetainline$ and multiplicity distributions from CMS \cite{bCMS-eta,Khachatryan:2010nk} and UA5 \cite{Ansorge:1988kn} allow a direct comparison to our data. Others by ATLAS \cite{Aad:2010ac} and LHCb \cite{LHCb-mult} use different definitions ($\eta$ and \pt ranges) making direct comparison impossible.

This publication is organized as follows: \cref{section:subdetectors} describes the ALICE sub-detectors relevant to this study; \cref{section:experimental-conditions} provides the details of the experimental conditions and of the collection of data; \cref{section:event-selection} explains the event selection; \cref{section:track-selection-and-algorithms} describes the track selection criteria and the three track counting algorithms; \cref{section:dndeta:analysis,section:pnch:analysis} report the analyses for the measurement of the pseudorapidity density and of multiplicity distributions, respectively; \cref{section:systematics} discusses systematic uncertainties; \cref{section:experimental-results} presents the multiplicity measurements, NBD fits of the multiplicity distributions, KNO scaling and $q$-moment studies. Finally, in \cref{section:summary-and-conclusion}, the results are summarized and conclusions are given.

\section{ALICE subdetectors}
\label{section:subdetectors}
The ALICE detector is fully described in \cite{Aamodt:2008zz}. Only the main properties of subdetectors used in this analysis are summarized here. Charged-particle tracking and momentum measurement are based on data recorded with the Inner Tracking System (ITS) combined with the Time Projection Chamber (TPC) \cite{Alme2010316}, all located in the central barrel of the ALICE detector and operated inside a large solenoid magnet providing a uniform 0.5 T magnetic field parallel to the beam line.

The V0 detector \cite{1748-0221-8-10-P10016} consists of two scintillator hodoscopes, each one placed at either side of the interaction region, at $z = 3.3$~m (V0A) and at $z = -0.9$~m (V0C) ($z$ is the coordinate along the beam line, with its origin at the centre of the ALICE barrel detectors), covering the pseudorapidity ranges $2.8 < \eta < 5.1$ and $-3.7 < \eta < -1.7$, respectively. The time resolution of each hodoscope is better than 0.5~ns.

The ITS is composed of high resolution silicon tracking detectors, arranged in six cylindrical layers at radial distances to the beam line from 3.9 to 43~cm. Three different technologies are employed. For the two innermost layers, silicon pixels (SPD \cite{santoro2009alice}) are used, covering pseudorapidity ranges $|\eta| < 2$ and $|\eta| < 1.4$, respectively. The SPD is followed by two Silicon Drift Detector layers (SDD, \cite{alessandro2010operation}). The Silicon Strip Detector (SSD, \cite{nooren2009experience}) constitutes the two outmost layers consisting of double-sided silicon micro-strip sensors. The intrinsic spatial resolution ($\sigma_{r\varphi} \times \sigma_z$) of the ITS subdetectors is: 12$\times$100~$\mathrm{\mu m}^2$ for SPD, 35$\times$25~$\mathrm{\mu m}^2$ for SDD, and 20$\times$830~$\mathrm{\mu m}^2$ for SSD, where $\varphi$ is the azimuthal angle and $r$ the distance to the beam line. The ITS sensors were aligned using survey measurements, cosmic muons and collision data \cite{Rossi:2011gv}. The estimated alignment accuracy is 8~$\mathrm{\mu}$m for SPD and 15~$\mathrm{\mu}$m for SSD in the most precise coordinate ($r \varphi$). For the SDD, the intrinsic space point resolution is $\sigma_z = 30\ \mathrm{\mu m}$ in the $z$ direction and $\sigma_{r\varphi} = $ 40 to 60~$\mathrm{\mu}$m, depending on the sensor, along $r \varphi$ (drift). Because of some anomalous drift field distributions, in the reconstruction, a systematic uncertainty up to 50~$\mathrm{\mu}$m in $z$ and 500~$\mathrm{\mu}$m in $r \varphi$ was added to account for differences between data and simulation. The ITS resolution in the determination of the transverse impact parameter measured with respect to the primary vertex is typically 70~$\mathrm{\mu}$m for tracks with $\pt = 1$~GeV/$c$, including the contribution from the primary vertex position resolution.

The SPD and the V0 scintillator hodoscopes provided triggers for collecting data.

The TPC \cite{Alme2010316} is a large cylindrical drift detector with a central high voltage membrane at $z = 0$, maintained at +100~kV and two readout planes at the end-caps. The material budget between the interaction point and the active volume of the TPC corresponds to 11\% of a radiation length, when averaged over $|\eta| < 0.8$.

The TPC and the ITS were aligned relative to each other within a few hundred micrometers using cosmic-ray and proton collision data \cite{Rossi:2011gv}.

The momentum measurement is not explicitly used in this study, however, the simulation of the detector response is sensitive to the particle momentum spectrum. Since event generators used in Monte Carlo simulations do not reproduce the observed momentum distributions, the difference between data and Monte Carlo simulation is taken into account when evaluating systematic errors. For momenta lower than 2 GeV/$c$, representing the bulk of the data, the \pt resolution for tracks measured in the TPC and in the ITS, is about 0.80\% at $\pt = 1$~GeV/$c$, it increases to 0.85\% at $\pt = 2$~GeV/$c$ and to 3\% at $\pt = 0.1$~GeV/$c$.

Charged-particle multiplicities were measured using information from the TPC in $|\eta| < 0.9$ and from the ITS in $|\eta| < 1.3$. At larger pseudorapidities, the SPD alone was used to expand the range of $\dndetainline$ measurements to $|\eta| < 2.0$.

\section{Experimental conditions and data collection}
\label{section:experimental-conditions}
\subsection{Proton beam characteristics}
Data were selected during LHC collision periods at a luminosity low enough to allow the minimum bias trigger rate not to exceed 1 kHz. At $\sqrt{s} = 0.9$~TeV, the number of protons per colliding bunch varied from 9$\times$10$^9$ to 3.4$\times$10$^{11}$, while the number of colliding bunches was either 1 or 8. At $\sqrt{s} = 2.76$~TeV, the number of protons per colliding bunch varied from 5$\times$10$^{12}$ to 7$\times$10$^{12}$, while the number of colliding bunches was either 48 or 64. At $\sqrt{s} = 7$~TeV, the number of protons per colliding bunch varied from 8.6$\times$10$^9$ to 1.4$\times$10$^{12}$, resulting in a luminosity between 10$^{27}$ and 10$^{30}$~cm$^{-2}$s$^{-1}$. There were up to 36 bunches per beam colliding at the ALICE interaction point. When needed, the luminosity was kept below 10$^{30}$~cm$^{-2}$s$^{-1}$ by a transverse displacement of the beams with respect to one another. At $\sqrt{s} = 8$~TeV, there were 3 proton bunches colliding at the ALICE interaction point each containing about 1.6$\times$10$^{11}$ protons.

Data used for this study were collected at low beam currents, so that beam-induced backgrounds (beam–gas or beam-halo events) were low and could be removed offline using V0 and SPD detector information, as discussed in \cref{section:event-selection:subsection:event-background}.
\subsection{Triggers}
The ALICE trigger system is described in \cite{krivda2012alice}. Data were collected with a minimum bias trigger, \mbor, requiring a hit in the SPD or in either one of the V0 hodoscopes; i.e. essentially at least one charged particle anywhere in the 8 units of pseudorapidity covered by these detectors. Triggers were required to be in time coincidence with a bunch crossing the ALICE interaction point. Control triggers, taken for various combinations of beam and empty-beam buckets, were used to measure beam-induced and accidental backgrounds.
\subsection{Characteristics of data samples used in this study}
General characteristics of the data samples used are given in \cref{tab:data-sample-summary}.

The data at $\sqrt{s} = 0.9$~TeV were collected in May 2010, with one polarity of the ALICE solenoid magnet (solenoid magnet field pointing in the positive $z$ direction).

The first LHC data above Tevatron energy were collected in 2009, at $\sqrt{s} = 2.36$~TeV, in a run with unstable LHC beams, during which only the SPD was turned on. Therefore, in this case, the charged-particle multiplicity was measured using exclusively the SPD information. In this publication, the previously published results at $\sqrt{s} = 2.36$~TeV \cite{aamodt2010charged09} are used for comparison.

Proton-proton data were collected at $\sqrt{s} = 2.76$~TeV, an energy that matches the nucleon-nucleon centre-of-mass energy in the first Pb--Pb collisions provided by the LHC, in 2011.

Data at $\sqrt{s} = 7$~TeV were collected in 2010. About 20\% of the data were taken with a magnet polarity opposite (solenoid field pointing in the negative $z$ direction) to that of $\sqrt{s} = 0.9$~TeV data. A sample of 12.3$\times$10$^6$ events, collected without magnetic field, was used to check some of the systematic biases in track reconstruction.

\begin{table}[ht]%

\centering
 \begin{tabu} to \textwidth {>{\bfseries}X[-1,l]X[-1.5,l]X[-1.5,l]X[-1.5,l]X[-2,l]X[-1.8,l]}%
 \toprule
  \rowfont[c]{\bfseries} {}	& \multicolumn{3}{c}{MB events ($\times$10$^6$)} & \multicolumn{2}{c}{} \\
  \cmidrule{2-4}
  \rowfont[l]{\bfseries} $\sqrt{\bm{s}}$ (TeV)	& Triggered	& Reconstructed	& Selected	& $\bm{\left<\mu\right>}$	& Luminosity (nb$^{-1}$) \\
  \midrule
  0.9	& 7.4	& 6.3	& 5.6	& $0.04\pm0.01$		& $0.128\pm0.006$ \\
  2.36	& 0.097	& 0.097	& 0.04	& $<$ 0.001		&  - \\
  2.76	& 33.9	& 32.6	& 28.3	& $0.025\pm0.01$	& $0.583\pm0.013$ \\
  7	& 404.4	& 384.2	& 343.7	& $0.04\pm0.01$		& $6.05\pm0.25$ \\
  8	& 31.5	& 26.6	& 24.1	& $0.02\pm0.01$		& $0.41\pm0.02$ \\
  \bottomrule
 \end{tabu}%
 \caption{For each centre-of-mass energy: total number of minimum bias (MB) events collected; number of those events that were reconstructed; number of reconstructed events passing the selection described in the text, except for $z$ vertex quality and position; average number of interactions per bunch crossing, $\left<\mu\right>$; integrated luminosity corresponding to the number of events reconstructed.\label{tab:data-sample-summary}}%
\end{table}%
At $\sqrt{s} = 8$~TeV only a subset of runs was collected with the \mbor as a minimum bias trigger in 2012, 10 were selected for this analysis.

At 0.9 and 7~TeV, data samples are substantially larger than those available in previous ALICE publications on charged-particle multiplicities \cite{aamodt2010first,aamodt2010charged09,aamodt2010charged7}. For the charged-particle multiplicity analysis, the event sample at $\sqrt{s} = 0.9$ and 7~TeV increased by a factor of 50 and 2000, respectively, giving significant extension of the multiplicity reach and better statistical precision. The precision of $\dndetainline$ is not substantially limited by event sample size. However, the large number of runs available made it possible to study run-to-run fluctuations of the $\dndetainline$ measurements over long periods of time, thus providing a monitoring of the uniformity of the data quality.

\section{Event selection}
\label{section:event-selection}
\subsection{Background rejection}
\label{section:event-selection:subsection:event-background}
\subsubsection{Beam background}
The main sources of event background are beam gas and beam halo collisions. Such events were removed by requiring that the timing signals from the V0 hodoscopes, if present, be compatible with the arrival time of particles produced in collision events. In addition, because of the different topology of beam background events, the ratio between the number of SPD clusters and the number of SPD tracklets\footnote{A tracklet is a short track segments in the SPD, compatible with the event vertex.} is much higher in beam background events, therefore a cut on this ratio was applied. The remaining fraction of beam background events  in the data, estimated by analysing special triggers taken with non-colliding bunches or empty beam buckets, does not exceed 10$^{-4}$ for all centre-of-mass energies. The track beam background is mostly significant in the last $\eta$ bins ($|\eta| \approx 2$) where it reaches 4$\times$10$^{-3}$ in the worst case.
\subsubsection{Event pileup}
The other type of potential event background comes from multiple collision overlap. For the data used in this publication, the proton bunch spacing was 50~ns or longer, the luminosity did not exceed 10$^{30}$~cm$^{-2}$s$^{-1}$, and the probability to have collisions from different bunch crossings in the 300~ns integration time of the SPD was negligible. However, multiple collisions in the same bunch crossing, also referred to as event pileup or overlap, have to be considered in case their vertices are not distinguishable. In order to avoid or minimize corrections for event pileup, runs with a low number of interactions per bunch crossing, $\mu \leq 0.061$, were selected resulting in an average $\mu$, $\left<\mu\right> \leq 0.04$, for all data samples (\cref{tab:data-sample-summary}). This corresponds to at most 2\% probability of more than one interaction per event.

The identification of pileup events relies on multiple vertex reconstruction in the SPD, with algorithms using three basic parameters: \begin{inparaenum}[(a)]\item The distance of closest approach (DCA) to the main vertex for a SPD tracklet to be included in the search for an additional interaction: DCA $>$ 1~mm; \item The distance between an additional vertex and the main vertex, $\Delta z >$ 8~mm; \item The number of SPD tracklets ($N_{\rm trk}$) used to determine an additional vertex (number of contributors to the vertex): $N_{\rm trk} \geq 3$. \end{inparaenum}

With this choice of parameters, and with the relatively broad $z$ vertex distribution at the LHC ($\mathrm{FWHM}\geq$ 12~cm), typically only 10\% to 15\% of multiple collisions are missed, and the fraction of fake multiple collisions due to SPD vertex splitting from a single interaction is low (typically a few times 10$^{-5}$).

The pileup detection efficiency was studied both by overlapping two Monte Carlo proton-proton collisions and by measuring pileup in the data. The pileup fraction, estimated from identified pileup events in the data, is found to be consistent with what is expected from the $\mu$ values derived from trigger information (\cref{tab:data-sample-summary}).

In multiplicity measurements, pileup affects the data mainly when two vertices are not distinguishable. When they are distinguishable, the multiplicity is taken from the vertex with the highest number of tracks. The small bias induced by choosing systematically the highest multiplicity vertex is negligible in our low pileup data samples.

Comparing $\dndetainline$ measurements, for different runs, no correlation is found between $\dndetainline$ values at $\eta = 0$ and $\mu$ values. Comparing data with and without identified pileup rejection, the change in $\dndetainline$ values is smaller than 0.5\%, which is smaller than systematic uncertainties. Note that the requirements for track association to the main vertex reject a further fraction of the tracks coming from the 10\% to 15\% of unidentified pileup collisions. The conclusion is that event pileup corrections to $\dndetainline$ are negligible in these low pileup data samples.

For multiplicity distributions, even though data were selected with a low pileup probability, it is important to verify that the pileup does not distort the distributions, as the relative pileup fraction increases with multiplicity. The fraction of pileup events, which the ALICE pileup detection algorithm identifies after the event selection, is about 10$^{-2}$, with no significant differences between the four centre-of-mass energies. Moreover, tight DCA cuts allow tracks originating from the main vertex to be distinguished from those coming from a pileup vertex even when the vertices are closer than 0.8~cm in $z$. This was confirmed by simulating events, where two Monte Carlo pp collisions were superimposed, demonstrating that only 5\% of the events passing the selection had extra tracks from the secondary vertex. In 90\% of such cases, the distance along the beam line between the two vertices was $\Delta z < 0.5$~cm. In the data samples with a pileup fraction of order $\slfrac{\mu}{2} \leq 0.02$, the residual average fractions of events with pileup is at most 0.4\%. Furthermore, the simulation shows that the pileup that does affect the multiplicity of an event is rather broadly distributed across events with different multiplicity, but  becomes significant only outside the multiplicity range studied here. The multiplicity at which the pileup contribution reaches 10\% of the measured multiplicity at $\sqrt{s} = 7$~TeV is $\nch = $ 105, 170 and 310, for $|\eta| <$ 0.5, 1.0, and 1.5, respectively, which is beyond multiplicity ranges covered in this publication.

Therefore, no pileup corrections were applied. Other background contributions from cosmic muons or electronics noise are also negligible.
\subsection{Offline trigger requirement}
Both for the \inel and \inelgt normalizations, the online \mbor trigger was used. However, for the \nsd analysis, a subset of the total sample was selected offline by requiring a coincidence (\mband) between the two V0 hodoscope arrays. This corresponds to the detection of at least one charged particle in both hemispheres, in the V0 hodoscope arrays separated by 4.5 units of pseudorapidity, a topology that tends to suppress single-diffraction (SD) events; therefore, model dependent corrections and associated systematic errors are minimized.
\subsection{Vertex requirement}
The position of the interaction vertex is obtained either by correlating hits in the two silicon-pixel layers (SPD vertex), or from the distribution of the impact parameters of reconstructed global tracks\footnote{Tracks reconstructed in the TPC and matched to ITS clusters (see \cref{section:track-selection-and-algorithm:subsection:quality-requirement} and references therein)} (global track vertex) \cite{Aamodt:2008zz,Alme2010316,Rossi:2011gv,alice2006alice,doi:10.1142/S0217751X14300440}. The next step in the event selection consists of requiring the existence of a reconstructed vertex.

Two SPD vertex algorithms were used: a three-dimensional vertexer (3D-vertexer) that reconstructs the $x$, $y$ and $z$ positions of the vertex, or a one-dimensional vertexer (1D-vertexer) that reconstructs the $z$ position of the vertex. The vertex position resolution achieved depends on the track multiplicity. For the 3D-vertexer it is typically 0.3~mm both in the longitudinal ($z$) direction and in the plane perpendicular to the beam direction. The 1D-vertexer resolution in the $z$ direction is on average 30 $\mu$m. If the 3D-vertexer algorithm does not find a vertex (typically 47\% of the cases at $\sqrt{s} = 7$~TeV), then the simpler 1D-vertexer is used to determine the $z$ position of the vertex, and the $x$ and $y$ coordinates are taken from the average $x$ and $y$ vertex positions of the run. The 3D-vertexer efficiency is strongly multiplicity dependent. As the bulk of the events have a low multiplicity, this explains the relatively low average vertex finding efficiency. For the $z$ coordinate, if no reliable vertex is found (typically 14\% of the cases), either because the 1D-vertexer did not find a vertex or the 1D-vertex quality was not sufficient (the dispersion of the difference of azimuthal angles between the two hits, one in each SPD layer, of tracklets contributing to the vertex is required to be smaller than 0.02 rad), the event is rejected. For the global track vertex, the resolution is typically 0.1~mm in the longitudinal ($z$) direction and 0.05~mm in the direction transverse to the beam line.
\begin{figure}[t]%
 \centering
 \includegraphics[width=0.65\textwidth]{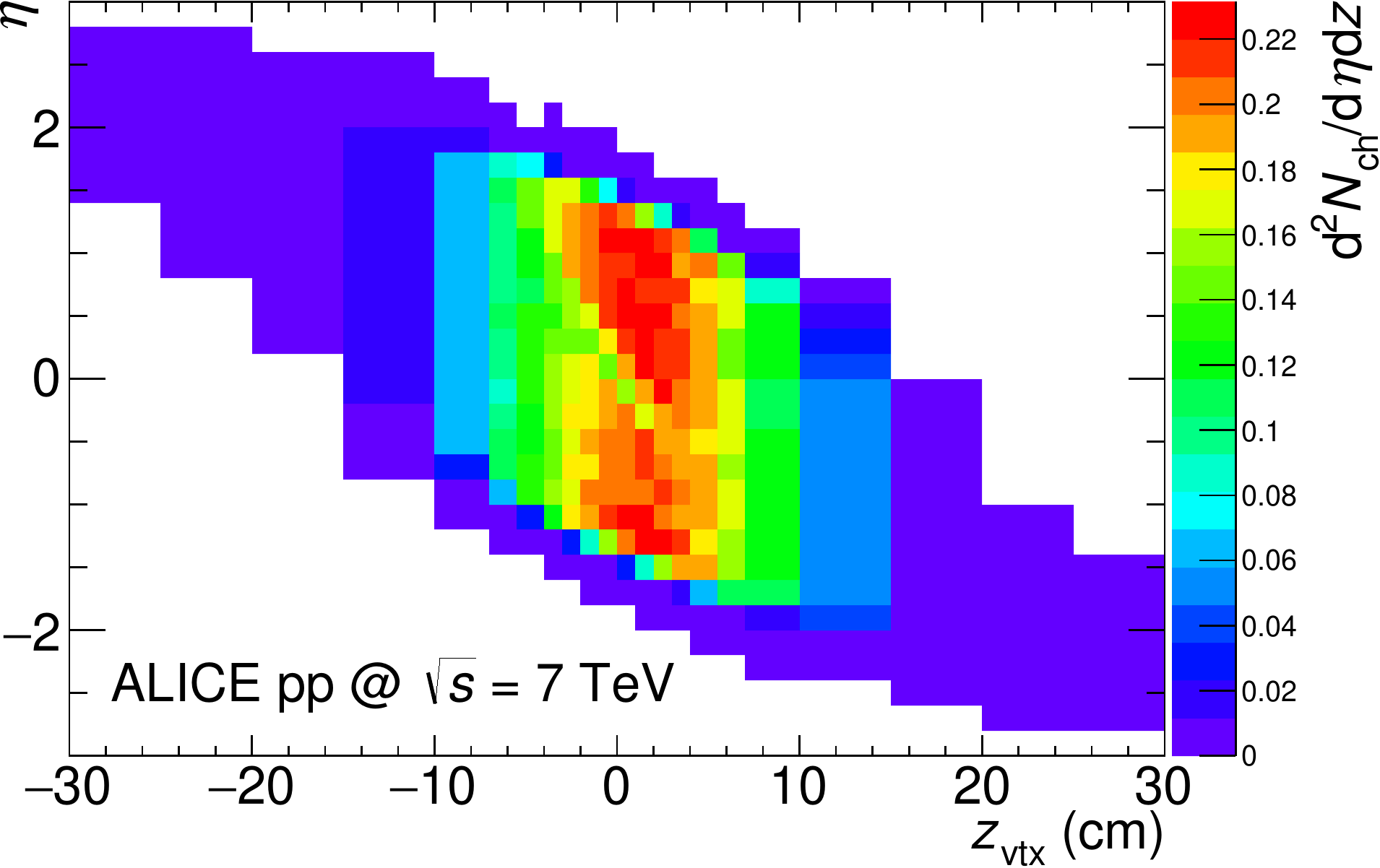}\\[-0.3cm]%
 \caption{For a data sample of events fulfilling the \mbor trigger selection, at $\sqrt{s} = 7$~TeV, the distribution of the quantity $\dndetainlinez$ is plotted for tracklets, in the plane pseudorapidity ($\eta$) vs. $z$ position of the SPD vertex ($z_{\rm vtx}$), showing the dependence of the $\eta$ acceptance on $z_{\rm vtx}$.\label{fig:eta-vs-z}}%
\end{figure}%

Both SPD and global track vertices have to be present and consistent by requiring that the difference between the two $z$ positions be smaller than 0.5~cm. If not, in 3 to 4\% of the cases, the event is rejected. The cut was chosen to be compatible with DCA$_z$ cut applied to tracks to ensure that we combine tracklets and tracks from the same collision (see \cref{section:track-selection-and-algorithms:subsection:algorithms}).  This condition removes mainly non-Gaussian tails in the columns of the detector response matrix\footnote{The response matrix is a 2-dimensional matrix obtained from the simulation giving the correspondence between generated and observed multiplicities. A response matrix column consists of the digitized distribution of the probability to measure a given multiplicity for a given generated multiplicity. } at low multiplicity, coming from the fact that SPD and track vertices, when separated, tend to have different multiplicities associated to them. In the data, this requirement also removes 80\% of pileup events with well-separated vertices.

The $\eta$ acceptance is correlated with the vertex $z$ position ($z_{\rm vtx}$) (\cref{fig:eta-vs-z}). For multiplicity distribution measurements, in order for tracks to remain within the acceptance of the SPD in the $\eta$ versus $z_{\rm vtx}$ plane, the following requirements were imposed on the vertex position along the $z$ axis: $|z_{\rm vtx}| <$ 10, 5.5 and 1.5 cm for $|\eta| <$ 0.5, 1 and 1.5, respectively. In the measurement of $\dndetainline$, the requirement on the vertex was relaxed to $|z_{\rm vtx}| <$ 30 cm, in order to allow extending the $\eta$ range to $|\eta| < 2$.
\subsection{Event selection efficiency}
As described in \cite{alice-crosssection-diffraction}, \pythia{6} \cite{Skands:2009zm,Sjostrand:1993yb,sjostrand2006pythia} and \phojet \cite{Engel:1995sb,Roesler:2000he} event generators used by ALICE were adjusted to reproduce the measured diffraction cross-sections and the shapes of the diffracted mass ($M_X$) distributions extracted from the Kaidalov-Poghosyan model \cite{kaidalov2009description}. These modified versions of event generators are referred to as \enquote{tuned for diffraction}. Typically, $\slfrac{\sigma_{\mathrm{SD}}}{\sigma_{\mathrm{\inel}}} \approx 0.20$, where $\sigma_{\mathrm{INEL}}$ is the inelastic cross-section, $\sigma_{\mathrm{SD}}$ is the SD cross-section for $M_X <$ 200 GeV/$c^2$, and $\slfrac{\sigma_{\mathrm{DD}}}{\sigma_{\mathrm{\inel}}} \approx 0.11$, where $\sigma_{\mathrm{DD}}$ is the double diffraction cross-section for $\Delta\eta > 3$ ($\Delta\eta$ is the size of the particle gap in the pseudorapidity distribution). These fractions have insignificant energy dependence between 0.9 and 7~TeV \cite{alice-crosssection-diffraction}, and the values at 7~TeV were used for 8~TeV data.

\begin{figure}[ht]%
\includegraphics[width=\textwidth]{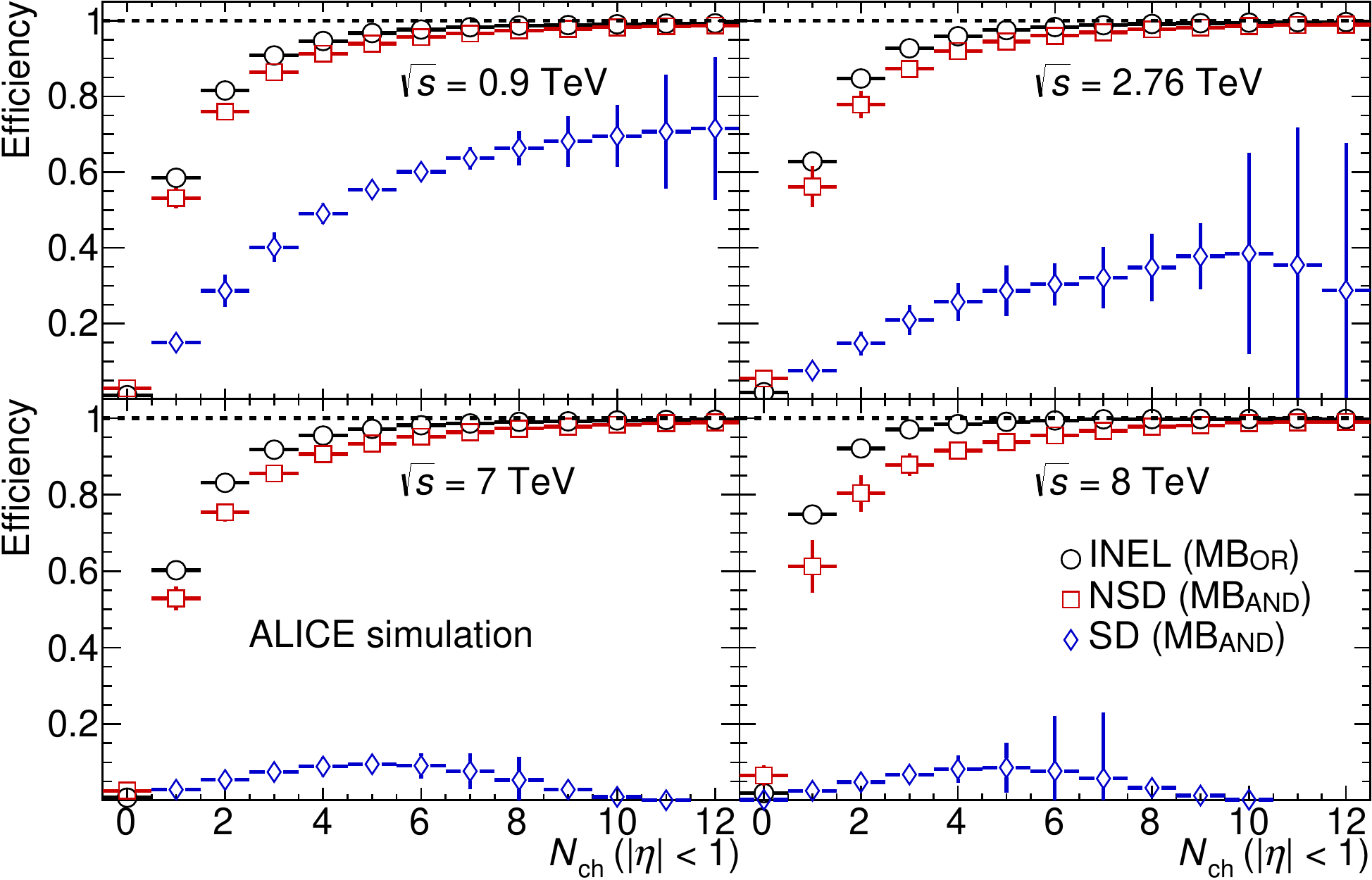}
\caption{Charged-particle multiplicity ($\nch =$ number of primary charged particles generated in $|\eta| < 1$)  dependence of the efficiency of the event selection described in \cref{section:event-selection}, obtained as the average between \pythia{6} Perugia0 and \phojet, both tuned for single diffraction defined for $M_X < 200$ GeV/$c^2$ (see \cite{alice-crosssection-diffraction}). Efficiencies are given for \inel events with \mbor trigger (open circles), \nsd events with \mband trigger (open squares), and SD events with \mband trigger (open diamonds), at $\sqrt{s} = $ 0.9~TeV (top left), 2.76 TeV (top right), 7~TeV (bottom left) and 8~TeV (bottom right). Error bars correspond to the difference between the two event generators and statistical uncertainty added in quadrature (non-negligible only for the SD events selection efficiency).\label{fig:event-selection-efficiency}}%
\end{figure}

\Cref{tab:data-sample-summary} shows the number of events selected at each centre-of-mass energy prior to the $z_{\rm vtx}$ requirement. Selection efficiencies using criteria defined above in this section, were estimated for \inel, \nsd and SD events (classified at generator level by event generator flags) as a function of the number of  generated charged particles (shown on \cref{fig:event-selection-efficiency} for the case $|\eta| < 1$ and the various centre-of-mass energies considered). The particular selection is designated by the offline trigger used to construct it, \mbor or \mband. Note that for $\dndetainline$ measurement selection efficiencies are defined in a separate way (see \cref{section:dndeta:analysis:acc-eff}). At $\sqrt{s} \geq 7$~TeV the \inel event selection efficiency based on the \mbor trigger reaches 100\% for a charged-particle multiplicity above 8.

For SD events, the efficiency of the \mband selection reduces significantly when going to higher energies (\cref{fig:event-selection-efficiency}), because the Lorentz boost of the diffracted system increases with increasing centre-of-mass energies. This implies that in the normalization to the \nsd event class, corrections for the remaining SD contribution become smaller when going to higher energies. The \mband trigger selects 84\%, 86\%, 87\% and 87\% of the \mbor triggers, and 13\%, 4\%, 1\% and 1\% of the SD events satisfy the \mband selection, at $\sqrt{s} =$ 0.9, 2.76, 7 and 8~TeV, respectively.

\section{Track selection and multiplicity algorithms}
\label{section:track-selection-and-algorithms}
\subsection{Track quality requirements}
\label{section:track-selection-and-algorithm:subsection:quality-requirement}
The following criteria were used to select reconstructed tracks associated to the main event vertex:
\begin{itemize}
 \item for tracks reconstructed from both ITS and TPC information (global tracks), the selection requires at least 70 pad hit clusters in the TPC, a good track quality ($\slfrac{\chi^2}{\mathrm{dof}} < 4$), a distance of closest approach (DCA) along the $z$ direction (DCA$_z$) $<$ 0.5~cm, and a \pt-dependent transverse DCA (DCA$_\mathrm{T}$) requirement, which corresponds to a 7 sigma selection. DCA$_\mathrm{T}$ conditions are relaxed by a factor 1.5 for tracks lacking SPD hits.
 \item for tracks reconstructed with ITS information only (ITS-only tracks) the number of ITS hit clusters associated to the track must be larger than 3, among the 6 layers of the ITS, and $\slfrac{\chi^2}{\mathrm{dof}} < 2.5$. The DCA$_z$ and DCA$_\mathrm{T}$ requirements are the same as for global tracks.
 \item for SPD tracklets, the association to the vertex is ensured through a $\chi^2$ requirement. Using the SPD vertex as the origin, differences in azimuthal ($\Delta\varphi = \varphi_2 - \varphi_1$, bending plane) and polar ($\Delta\theta = \theta_2 - \theta_1$, non-bending direction) angles are calculated between hits in the inner (layer~1) and in the outer (layer~2) SPD layers. Hit combinations, called tracklets, are selected with the following condition
 \begin{equation}
 \label{eq:tracklet-chi2}
  \chi^2 \equiv \frac{\left( \Delta\varphi \right)^2}{\sigma^2_{\varphi}} + \frac{1}{\sin^2 \left( \frac{\theta_1 + \theta_2}{2} \right)}\times \frac{\left( \Delta\theta \right)^2}{\sigma^2_{\theta}} < 1.6
 \end{equation}
 where $\sigma_{\varphi} = 0.08$~rad, $\sigma_{\theta} = 0.025$~rad and the $\sin^2$ factor takes into account the $\theta$ dependence of $\Delta\theta$. The $\chi^2$ value 1.6 was chosen to lie well within the part of the $\chi^2$ distribution of the data correctly reproduced by the simulation. The cut imposed on the difference in azimuthal angles rejects charged particles with a transverse momentum below 30 MeV/c; however, the effective transverse-momentum cut-off is determined mostly by particle absorption in the material and is approximately 50 MeV/c, in $|\eta| < 1$. If more than one hit in an SPD layer matches a hit in the other layer, only the hit combination with the smallest $\chi^2$ value is used.
\end{itemize}

Some of the SPD elements had to be turned off, resulting in lower efficiency in some regions of the $\eta$ versus azimuthal angle plane. In order to reach the best possible precision in the measurement of $\dndetainline$, fiducial cuts were applied to both tracks and tracklets, excluding azimuthal regions where the tracking efficiency corrections are relatively large. These fiducial cuts vary with data taking periods, following the evolution of the SPD acceptance. At $\sqrt{s} =$ 0.9, 2.76, 7, and 8~TeV, the fractions of the acceptance removed were 64\%, 68\%, 65\%, and 35\%, respectively. Some of the SPD elements could be recovered before collecting 8~TeV data, explaining the improvement.

For multiplicity distribution studies, fiducial cuts were not applied because they increase statistical uncertainty, hence limiting the high multiplicity reach.
\subsection{Track counting algorithms}
\label{section:track-selection-and-algorithms:subsection:algorithms}
In previous ALICE publications \cite{aamodt2010first,aamodt2010charged09,aamodt2010charged7}, the charged-particle multiplicity was measured in $|\eta| < 1.3$ using only SPD tracklets built from SPD pixel hits. In order to extend the pseudorapidity range to $|\eta| < $~2, an improved tracklet algorithm, initially used in \cite{aamodt2010charged}, was introduced to take into account the $\theta$ dependence of the uncertainty in the $\chi^2$ (\cref{eq:tracklet-chi2}). With this improvement, the efficiency for detecting SPD tracklets became uniform as a function of pseudorapidity and $z$ position of the vertex, which allowed vertices further away from the nominal interaction point along the beam direction to be used, thereby extending significantly the pseudorapidity range.

To be less sensitive to the SPD acceptance, track counting algorithms were developed, that make use of tracking information from other ALICE detectors, the SDD, the SSD and the TPC. Each track is counted as primary if it fulfills the transverse DCA requirements listed in \cref{section:track-selection-and-algorithm:subsection:quality-requirement} and it is not associated to a secondary vertex identified by a dedicated algorithm \cite{alice2006alice} tuned to tag  $\gamma$-conversions, K$^0$  and $\Lambda$ decays.

Three multiplicity estimators were developed by ALICE using three different samples of tracks:
\begin{itemize}
 \item SPD tracklets, with $|\eta|<2$ (referred to as \spd algorithm)\footnote{Potentially $|\eta|\lesssim 3$ can be reached using event vertices displaced from the detector center at distances $|z_{\rm vtx}| \gtrsim 30$ cm (see \cref{fig:eta-vs-z}), however the sample size of such events is too small.}. The \spd algorithm stores, for each tracklet, references to ITS or global track candidates using at least one of its pixel clusters.
 \item ITS-only tracks, with $|\eta| < 1.3$, obtained using all hit clusters in this detector, plus tracklets ($|\eta|<2$) built out of SPD pixel clusters not matched to any ITS track (referred to as \itssa algorithm).
 \item TPC tracks, with $|\eta| < 0.9$, matched to hits in the ITS, plus ITS-only tracks (up to $|\eta|<1.3$) built out of silicon hit clusters not matched to any TPC track, plus tracklets ($|\eta|<2$) built out of SPD pixel clusters not matched to any ITS or TPC track (referred to as \itstpc algorithm).
\end{itemize}
\begin{figure}[ht]%
 \centering
 \includegraphics[width=0.62\textwidth]{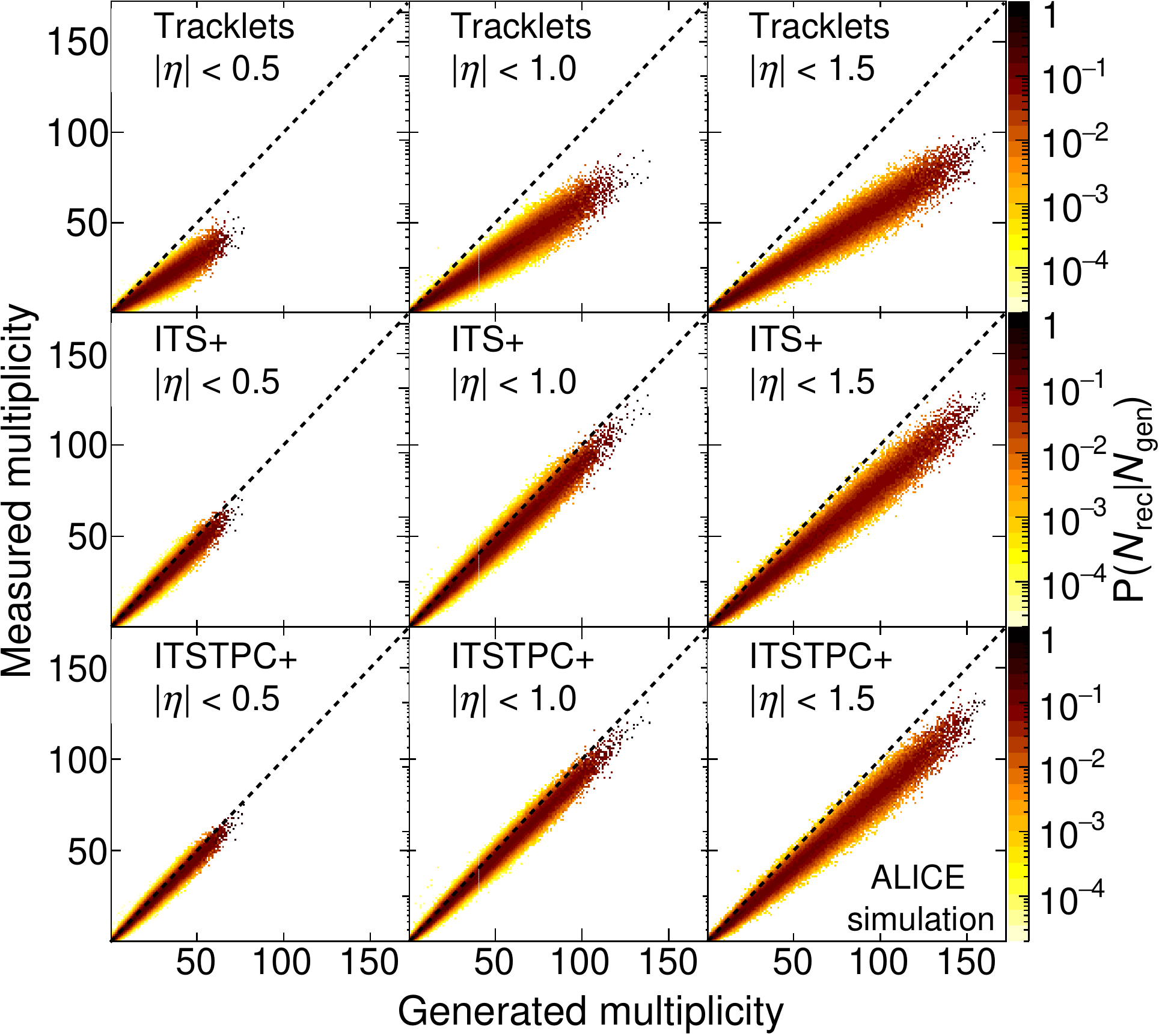}\\[-0.3cm]%
 \caption{Graphical representation of the detector response matrices obtained with \pythia{6} CSC \cite{Skands:2009zm} combined with a simulation of the ALICE detector, at $\sqrt{s} = 7$~TeV, for three pseudorapidity intervals ($|\eta| <$ 0.5, 1.0, and 1.5 from left to right, respectively), and for the three track counting algorithms, \spd, \itssa and \itstpc, from top to bottom, respectively. Horizontal axes show generated primary charged-particle multiplicities and vertical axes measured multiplicities.\label{fig:response-all-eta-all-algorithms}}%
\end{figure}%

In order to keep away from the edges of the detectors, where the acceptance is less precisely known, ITS and TPC tracks used in this study are limited to $|\eta| < 1.3$ and $|\eta| < 0.9$, respectively.

\begin{figure}[t]%
 \centering
 \includegraphics[width=0.9\textwidth]{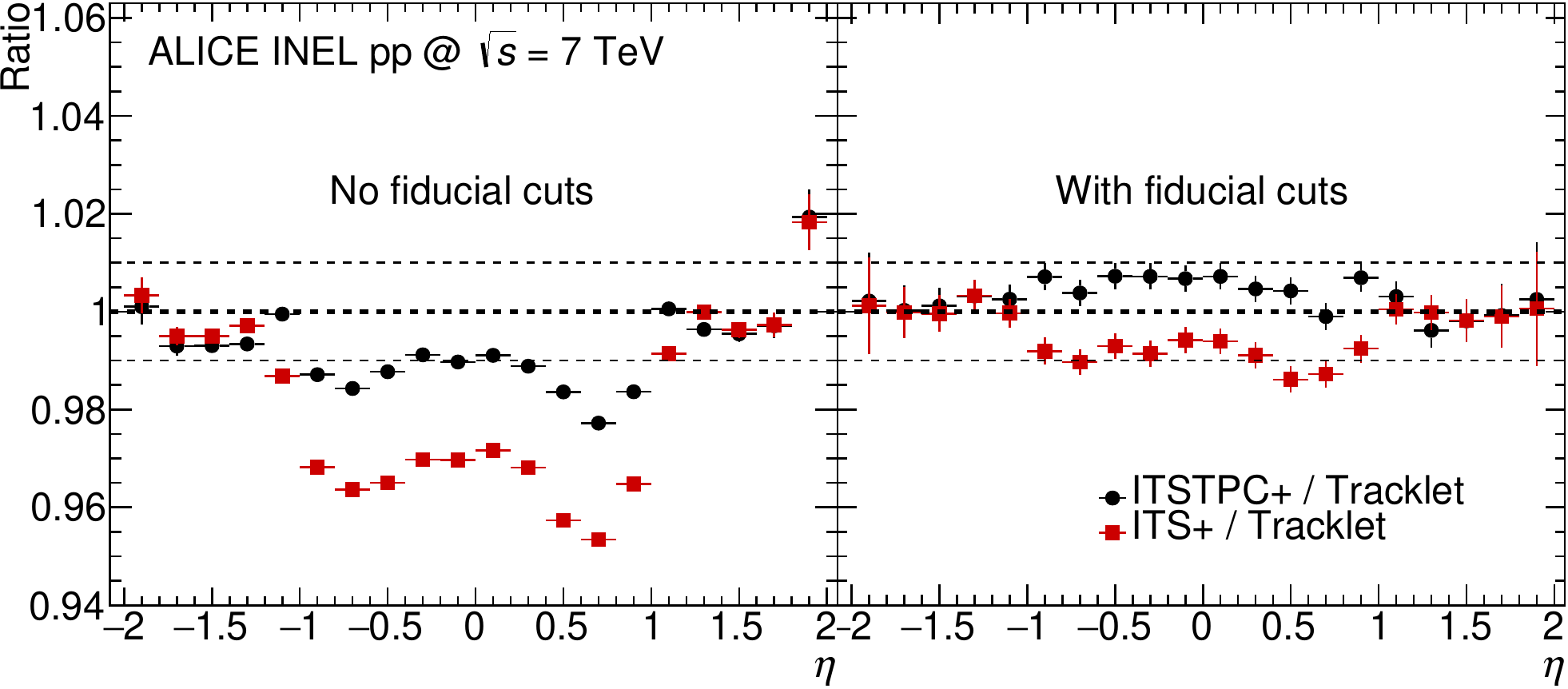}
 \caption{The three multiplicity algorithms are compared, after full correction, (left) without and (right) with fiducial cuts in azimuthal angles. Ratios of $\dndetainline$ measurements with different algorithms are shown: \itstpc over \spd (black circles) and \itssa over \spd (red squares).\label{fig:azimuthal-cut-effect-dndeta}}%
\end{figure}

Properties of the three track counting algorithms are compared in \cref{fig:response-all-eta-all-algorithms}, showing that, going from \spd to \itssa and to \itstpc algorithms, the detector response matrix becomes narrower and has a topology closer to that of a diagonal matrix. When going from $|\eta| < 0.5$ to $|\eta| < 1.5$, the response matrix becomes broader and has a less diagonal topology, as geometrical acceptance effects become more important, and dominated by the SPD with significant inefficiency due to some missing modules. Note that by restricting the azimuthal angle to good regions of the SPD, the difference between algorithms in $\dndetainline$ measurements is of order $\pm1$\% in the central region (\cref{fig:azimuthal-cut-effect-dndeta}). However, the result with the \spd algorithm is not sensitive to this cut, and, as it is needed to measure multiplicities beyond $|\eta| = 1.3$, it is used alone for $\dndetainline$ measurement. For multiplicity distribution measurements all three algorithms are used without the $\varphi$ region restrictions with a corresponding systematic uncertainty contribution.

In the pseudorapidity region $|\eta| < 0.9$, the TPC accounts for 90\% of the tracks, the ITS complement 9\% and the SPD complement 1\%. These fractions vary with the $\eta$ range. Outside $|\eta| < 1.3$, SPD tracklets are the only contribution. The small fluctuations between points in $|\eta| > 1.3$ come from the slightly different number of events used for averaging between algorithms, after efficiency corrections in each $\eta$ bin.

\section{Pseudorapidity density of primary charged particles: analysis}
\label{section:dndeta:analysis}
Raw $\dndetainline$ distributions have to be corrected for detector and trigger acceptance and efficiency, and for contamination from daughters of strange particles. Note that this section only describes the particularities of $\dndetainline$ measurement, unless specifically stated otherwise. For charged particle multiplicity distribution measurement see \cref{section:pnch:analysis}.
\subsection{Acceptance and efficiency corrections}
\label{section:dndeta:analysis:acc-eff}
Three types of corrections have to be applied to the raw data: \begin{inparaenum}[(a)]\item a track-to-particle correction to take into account the difference between measured tracks and \enquote{true} charged primary particles. This correction mainly depends on acceptance effects and on detector and reconstruction efficiencies; \item corrections for the bias coming from the vertex reconstruction requirement, at both track and event levels (vertex reconstruction correction). This bias exists on both the number of tracks and the events used, since events without a reconstructed vertex are not selected, and tracks from those events therefore do not contribute; \item corrections at both track and event levels, to take into account the bias  due to the \mbor trigger required for \inel and \inelgt event classes or the \mband offline selection for the \nsd event class. \end{inparaenum}

In practice, the number of tracks is corrected as a function of $\eta$ and $z_{\rm vtx}$ and the number of events is corrected as a function of reconstructed track multiplicity and $z_{\rm vtx}$. The number of events without trigger or without reconstructed vertex is estimated from the simulation and included in the corrected number of events. Finally, the quantity $\dndetainline$, averaged over all events, is obtained for each $\eta$ bin. The range of $z_{\rm vtx}$ contributing to the multiplicity varies with $\eta$ (\cref{fig:eta-vs-z}). For instance, at $\eta = 2$, tracks originate mostly from vertices in the range: -30~cm $< z_{\rm vtx} <$ -5~cm. Therefore, for each $\eta$ bin, a $z_{\rm vtx}$ acceptance correction is applied. See \cite{Janthesis} for details of the procedure.
\subsection{Strangeness correction}
Since ALICE{\textquoteright}s definition of primary charged particles excludes particles originating from the weak decays of strange particles, data have to be corrected for cases when daughter particles of such decays pass the track selection. Current Monte Carlo event generators have a strangeness content which differs from data by a factor approaching 2. Therefore, the strangeness content in the Monte Carlo simulation was normalized to data using ALICE{\textquoteright}s K$^0$ and $\Lambda$ measurements in $|\eta| < 0.9$ \cite{aamodt2011strange}, that were extrapolated to $|\eta|\leq2$ using the shape from simulation. The ratios of strangeness contents between data and Monte Carlo generators are slightly centre-of-mass energy dependent. For $\sqrt{s}$ varying from 0.9 to 8~TeV they increase from 1.6 to 1.85 according to \pythia{6}, and from 1.4 to 1.6 according to \phojet. The uncertainty on these ratios coming from the uncertainty in ALICE measurements of strange particle production \cite{aamodt2011strange}, is estimated to be 5\%. The strangeness contamination is slightly $\eta$ dependent, and varies from 1.7\% at $\eta = 0$ to 2.5\% at $\eta = 2$ at $\sqrt{s} = 7$~TeV. The strangeness correction is about 1\%, has no significant $\eta$ variation in $|\eta| < 2$ and no significant energy dependence between $\sqrt{s} =$ 0.9 and 8~TeV. This correction is explained in more detail in the \cref{section:systematics:common:strangeness-particle-composition}, where the corresponding systematic uncertainty is discussed.
\subsection{Event class normalization}
The final correction applied to the data is the normalization to one of the three event classes defined in this study: \nsd, \inel and \inelgt. In the normalization to \nsd, corrections have to be made for the fraction of SD events remaining in the selection and for the fraction of double-diffraction (DD) events not included in the selection. In the normalization of results to the \inel event class, corrections have to be made for the fraction of single- and double-diffractive events not included in the selection. The \inelgt class is of interest because it minimizes diffractive corrections. In addition, ALICE measurements of SD and DD cross-sections \cite{alice-crosssection-diffraction} reduced the systematic uncertainties coming from diffraction. Corrections for higher order diffractive processes associated with events with two or more pseudorapidity gaps (regions devoid of particles) are neglected in the normalization to \inel, \nsd and \inelgt classes, as their contribution to inelastic collisions is expected to be smaller than 1\% \cite{collinsintroduction,kaidalov2009description}. Furthermore, such events tend to have a high trigger efficiency, which makes corresponding corrections even smaller.
\begin{table}[b]%
 \centering
 {\tabulinesep=0.6mm
 \begin{tabu}{>{\bfseries}lll}
 \rowfont[c]{\bfseries} $\sqrt{\bm{s}}$ (TeV)	& \mband	& \mbor \\
 \midrule
 0.9	& 0.94$^{+0.02}_{-0.02}$	& 0.91$^{+0.03}_{-0.01}$ \\
 2.76	& 0.93$^{+0.03}_{-0.03}$	& 0.88$^{+0.06}_{-0.035}$ \\
 7	& 0.93$^{+0.02}_{-0.02}$	& 0.85$^{+0.06}_{-0.03}$ \\
 8	& 0.93$^{+0.02}_{-0.02}$	& 0.85$^{+0.06}_{-0.03}$ \\
 \end{tabu}}%
  \caption{\mband trigger efficiencies for \nsd events and \mbor trigger efficiencies for inelastic events at four centre-of-mass energies, obtained from diffraction-tuned versions of \pythia{6} Perugia0 \cite{Skands:2009zm} and \phojet \cite{Engel:1995sb}. Uncertainties listed are total uncertainties. Statistical errors are negligible. The asymmetry of the \mbor errors is due to the asymmetric uncertainties in the diffraction efficiencies.\label{tab:trigger-efficiencies}}
\end{table}%

To normalize measurements to a given event class, trigger biases must be corrected for, both at event and track levels. For the \inel and \inelgt classes, the correction is straightforward using the \mbor trigger efficiency (\cref{tab:trigger-efficiencies}).

For the \nsd event class, contamination of the event sample by SD events must be taken into account. The measured quantity may be re-written as:
\begin{multline}
 \label{eq:mband-selected-dndeta}
 \frac{1}{\left(N_{\rm ev}\right)_{\mband}} \drvt{\left( \sum N_{\rm trk} \right)_{\mband}}{\eta} = \\ \frac{1}{\left( N^{\mathrm{NSD}}_{\rm ev} \right)_{\mband} + \left( N^{\mathrm{SD}}_{\rm ev} \right)_{\mband}} \left( \drvt{\left(\sum N_{\rm trk}^{\mathrm{NSD}}\right)_{\mband}}{\eta} + \drvt{\left( \sum N_{\rm trk}^{\mathrm{SD}} \right)_{\mband}}{\eta}\right)
\end{multline}
where $\left( \sum N_{\rm trk}^{\mathrm{Class}} \right)_{\mathrm{Trigger}}$ is the number of tracks aggregated over all \rm events $\left( N_{\rm ev}^{\mathrm{Class}} \right)_{\mathrm{Trigger}}$ of a given class (superscript) selected with a given trigger type (subscript outside the parentheses). Given that $\left( N_{\rm ev}^{\mathrm{SD}} \right)_{\mband} \propto \varepsilon^{\mathrm{SD}}_{\mband}\sigma^{\mathrm{SD}}$ and $\left( N_{\rm ev}^{\mathrm{NSD}} \right)_{\mband} \propto \varepsilon^{\mathrm{NSD}}_{\mband}\sigma^{\mathrm{NSD}}$, where $\varepsilon$ and $\sigma$ are efficiencies and cross-sections, respectively, for SD or NSD \rm events \cite{alice-crosssection-diffraction}, one obtains:
\begin{multline}
 \label{eq:nsd-dndeta-solution}
 \frac{1}{\left( N^{\mathrm{NSD}}_{\rm ev} \right)_{\mband}} \drvt{\left(\sum N_{\rm trk}^{\mathrm{NSD}}\right)_{\mband}}{\eta} = \\
 \left( 1 + \frac{\varepsilon^{\mathrm{SD}}_{\mband}\sigma^{\mathrm{SD}}}{\varepsilon^{\mathrm{NSD}}_{\mband}\sigma^{\mathrm{NSD}}} \right) \frac{1}{\left(N_{\rm ev}\right)_{\mband}} \drvt{\left( \sum N_{\rm trk} \right)_{\mband}}{\eta} - \frac{\varepsilon^{\mathrm{SD}}_{\mband}\sigma^{\mathrm{SD}}}{\varepsilon^{\mathrm{NSD}}_{\mband}\sigma^{\mathrm{NSD}}} \frac{1}{\left( N^{\mathrm{SD}}_{\rm ev} \right)_{\mband}} \drvt{\left( \sum N_{\rm trk}^{\mathrm{SD}} \right)_{\mband}}{\eta}
\end{multline}
The coefficient in front of the single diffraction term in \cref{eq:nsd-dndeta-solution}, varies from 0.04 at $\sqrt{s} = 0.9$~TeV to 0.003 at $\sqrt{s} = 8$~TeV. As the single diffraction term is not measured, but corresponds to a relatively small correction, this term was calculated using the simulation. The corresponding uncertainty was estimated by varying the single diffraction term conservatively between extreme cases, assuming either no SD, or assuming that all events are from SD. The last step consists of correcting for the \mband trigger efficiency to obtain the desired quantity, $\slfrac{1}{\left( N^{\mathrm{NSD}}_{\rm ev} \right)_{\mband}} \drvtinline{\left(\sum N_{\rm trk}^{\mathrm{NSD}}\right)_{\mband}}{\eta}$.

The DD event content of the \mbor and \mband data samples, is small, of the order of 5.5\% and 4.5\%, respectively. These fractions do not vary significantly between 0.9 and 8~TeV. The corrections for DD efficiency are included in the general efficiency correction. For the \inel and \inelgt event classes, the \mbor trigger efficiency for DD events as a function of multiplicity is the same as for the other inelastic events. The \mband selection, which is used for the \nsd event sample, has an efficiency for DD events that is lower than that of the other inelastic events. However, we checked in the simulation that the average efficiency correction for the \nsd event class gives the same result as separate efficiency corrections implemented for DD and ND events.

\section{Multiplicity distributions of primary charged particles: analysis}
\label{section:pnch:analysis}
\subsection{Unfolding multiplicity distributions}
\label{section:pnch:analysis:unfolding}
The data samples used in these measurements are described in \cref{tab:data-sample-summary}. The next step in the analysis consists of correcting the raw distributions for detector acceptance and efficiencies, using an unfolding method.

The unfolding procedure follows the same approach as in \cite{aamodt2010charged09}, i.e. the corrected distribution is constructed by finding the vector $U$, which minimizes a $\chi^2$ given by
\begin{equation}
 \label{eq:unfolding-chi2}
 \chi^2 = \sum\limits_m \left( \frac{M_m - \sum_t R_{mt} U_t}{s_m} \right)^2 + \beta\times F(U)
\end{equation}
where $M$ represents the raw multiplicity distribution vector with uncertainty vector $s$, $U$ the unfolded multiplicity distribution vector, and $R$ the detector response matrix. Indices $m$ and $t$ run from 0 to the maximum number of multiplicity bins, in raw and corrected distributions respectively. The regularization term $\beta\times F(U)$ is used to decrease the sensitivity of the unfolding to statistical fluctuations. For $F(U)$ a usual Tikhonov-type of function \cite{cowan-stat-analysis} was used, which has a smoothing effect on the unfolded distribution
\begin{equation}
 \label{eq:regularization-term}
 F(U) = N \times \sum\limits_{t=1}^{N-1} \left( \frac{U_{t+1} - 2U_t + U_{t-1}}{U_t} \right)^2
\end{equation}
where $N$ is the number of unfolded multiplicity bins, evaluated with the help of the response matrix, from the maximum raw multiplicity.

The weight $\beta$ (\cref{tab:unfolding-regularization-weights}) was chosen to minimize the mean squared error \cite{cowan-stat-analysis}. The solution is found to be stable over a broad range of $\beta$ values ($\pm 50$\%), and the correct minimum was ensured in each case by scanning $\beta$ over few orders of magnitude. The particular values of optimal weights depend on many features of the unfolding problem, such as distribution size, a pattern of fluctuations in the input raw data, properties of the response matrix and the regularization term. The most obvious dependence was eliminated by factorizing $N$ in \cref{eq:regularization-term}.

\begin{table}[hb]%

 \centering
 \begin{tabu}{>{\bfseries}llll}%
 \rowfont[c]{\bfseries} $\sqrt{\bm{s}}$	& \multicolumn{3}{c}{$\Delta\bm{\eta}$} \\
 \cmidrule{2-4}
 \rowfont[c]{\bfseries} (TeV)	& 0.5	& 1	& 1.5 \\
 \midrule
 0.9	& $10$		& $10$		& $10^2$ \\
 2.76	& $10$		& $10^2$	& $10^2$ \\
 7	& $10$		& $10^2$	& $10^3$ \\
 8	& $50$		& $10^2$	& $10^2$ \\
 \end{tabu}%
 \caption{Values of the weight parameter $\beta$ used in the regularization term (\cref{eq:unfolding-chi2}), for each centre-of-mass energy and for each pseudorapidity range.\label{tab:unfolding-regularization-weights}}
\end{table}%

\begin{figure}[t]%
 \centering
 \includegraphics[width=0.9\textwidth]{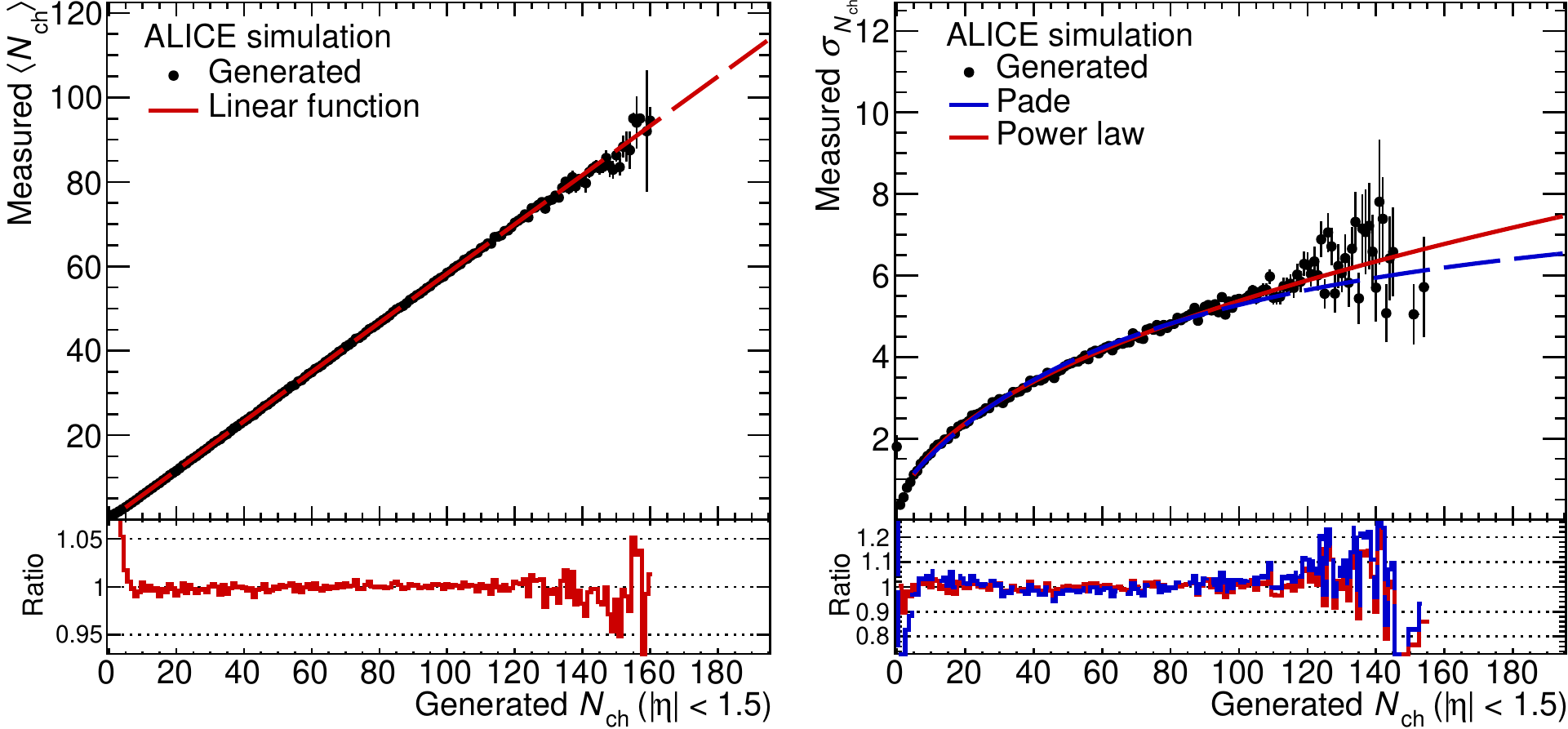}
 \caption{Example of Gaussian parameterization of the response matrix, at $\sqrt{s} =$ 7~TeV, for $|\eta| <1$: (left) parameterization of the mean values, with a linear function (red dashed line); (right) parameterization of the widths, with a Pad\'{e} function (red solid line) and a power law function (blue dashed line). The bottom parts of the figures show the ratios between data and fits.\label{fig:response-parameterization}}%
\end{figure}

For each generated multiplicity bin $N^{\rm gen} = t$, the response matrix column $R_{mt}$ consists of the distribution of the probability to measure multiplicity $\nch = m$. To extend the response matrix to the highest multiplicities encountered in this study, beyond the reach of the available simulation, probability distributions were parameterized and extrapolated towards high multiplicities (\cref{fig:response-parameterization}). In the low-$N^{\rm gen}$ region ($N^{\rm gen} <$ 10 to 20, depending on the $\eta$ range) the response matrix was taken directly from the simulation. In the large $N^{\rm gen}$ region ($N^{\rm gen} \geq$ 10 to 20), the column $R_{mt}$ is well described by a Gaussian distribution and mean values follow a linear trend (\cref{fig:response-parameterization}). Widths were parameterized using two different functions, a Pad\'{e} function and a power law
 \begin{align}
 \label{eq:response-width-parameterization}
  W(t) & = C_0 \frac{1 + C_1 t + C_2 t^2}{1 + C_3 t} 	&	& \text{Pad\'e}   &\\
  W(t) & = C_0 + C_1 t^{\gamma} 				&	& \text{Power law} &
 \end{align}
$C_0$, $C_1$, $C_2$, $C_3$ and $\gamma$ are constants to be fitted. These functions have different asymptotic behaviours (\cref{fig:response-parameterization}), however, using either function makes a difference only for multiplicities above 100 (in $|\eta|<1.5$).

The switch to parameterization occurs at $N^{\rm gen} = $ 10, 15 and 20, for $|\eta| <$ 0.5, 1 and 1.5, respectively, for all energies. These values ensure that using the parameterized response matrix introduces no distortions in the low multiplicity region.

The range of multiplicities in the final unfolded distribution was further restricted by requiring that the bias (an estimate of how far is the result from the true solution \cite{cowan-stat-analysis}) is less than 10\% in each bin. As unfolding is performed for each correction scenario (see \cref{section:systematics:subsection:unfolding-systematics} on systematic uncertainties), in the end the multiplicity range is limited by the unfolding resulting in the shortest range. The quality of the unfolding was verified by comparing the raw distribution $M$ to the product $R \otimes U$.
\subsection{Event class normalization}
After the unfolding step, distributions have to be corrected for event selection efficiency (including trigger efficiency and vertex reconstruction efficiency).

For the \inel and \inelgt event classes this is straightforward, given that
\begin{equation}
 M_m = \sum\limits_t \varepsilon_t R_{mt} U_t
\end{equation}
where $\varepsilon_t$ is the selection efficiency for true multiplicity $t$. Thus the unfolded distribution can be normalized to the inelastic event class by dividing the contents of each multiplicity bin by the corresponding efficiency.

For the NSD event class the procedure is different as the unfolded distribution, $U^{*}$, includes a contamination by SD events
\begin{equation}
 \label{eq:mband-selected-nsd-pnch}
 U^{*}_t = \varepsilon_t^{\mathrm{NSD}}U_t^{\mathrm{NSD}} + \varepsilon_t^{\mathrm{SD}} U_t^{\mathrm{SD}}
\end{equation}
where upper indices denote the event class. The overall fraction of SD in inelastic collisions ($\alpha_t$) was measured by the ALICE Collaboration \cite{alice-crosssection-diffraction}
\begin{equation}
 \label{eq:sd-fraction-alpha-t}
 U_t^{\mathrm{SD}} = \alpha_t U_t^{\mathrm{INEL}} = \alpha_t \left( U_t^{\mathrm{NSD}} + U_t^{\mathrm{SD}} \right)
\end{equation}
The desired unfolded distribution normalized to the NSD event class is obtained by combining \cref{eq:mband-selected-nsd-pnch,eq:sd-fraction-alpha-t}
\begin{equation}
 U_t^{\mathrm{NSD}} = \frac{U_t^{*}}{\varepsilon_t^{\mathrm{NSD}} + \frac{\alpha_t}{1 - \alpha_t}\varepsilon_t^{\mathrm{SD}}}
\end{equation}

\section{Study of systematic uncertainties}
\label{section:systematics}
\subsection{Common sources of systematic uncertainties}
\subsubsection{Material budget}
The material budget in the ALICE central barrel was checked in the range $|\eta| < 0.9$, by comparing measured and simulated gamma conversion maps. The conclusion is that, in this pseudorapidity range, the material budget is known with a precision of 5\% \cite{doi:10.1142/S0217751X14300440}.
The corresponding systematic uncertainty was obtained by varying the material budget in the simulation, conservatively over the whole pseudorapidity range by $\pm 10$\%, which induces a variation of $\dndetainline$ of $\pm 0.2$\% for all event classes. For multiplicity distributions, the $\eta$ range considered does not exceed $\pm 1.5$, making the effect of the higher material budget uncertainty outside $|\eta| < 0.9$ relatively small. The systematic uncertainty from material budget is negligible compared to other sources of uncertainty.
\subsubsection{Magnetic field}
The magnetic field map was measured and simulated with finite precision. To check the sensitivity of the detector response to the precision of the simulation of the ALICE solenoidal magnetic field, data samples collected at $\sqrt{s} = 7$~TeV with opposite polarities were compared. The differences in $\dndetainline$ values are consistent with observed fluctuations between runs within the data-taking period at this energy. Therefore, the contribution from systematic uncertainties associated with the magnetic field are smaller than, or of the order of, the run-to-run fluctuations, and have been neglected.
\subsubsection{Strangeness correction and particle composition}\label{section:systematics:common:strangeness-particle-composition}
The main sources of uncertainty associated to the correction for strange particles originate from: \begin{inparaenum}[(a)]\item the difference ($\leq 5$\%) in K$^0$ and $\Lambda$ detection efficiency between data and simulation \cite{aamodt2011strange}; \item the difference in \pt distributions of strange particles in data compared to simulation, which implies a difference in the fractions of daughter particles meeting the vertex association condition; and \item the uncertainty in the simulation of the strange particle content. \end{inparaenum}

For $\dndetainline$ measurements, the systematic uncertainty from strangeness correction is found to have a small $\eta$ variation, it is slightly larger at $\eta = 0$ compared to $|\eta| = 2$. These uncertainties have a small energy dependence, they increase slightly with increasing $\sqrt{s}$, from 0.14\% at $\sqrt{s} =$ 0.9~TeV to 0.16\% at $\sqrt{s} =$ 8~TeV. The uncertainties at $\eta = 0$ are listed in \cref{tab:dndeta-systematics-at-0}.

For multiplicity distributions, the strangeness contamination was studied with the Monte-Carlo simulation by evaluating the survival probability of strange particle decay products for the track selection used in the analysis. The probability that a track from strange particle decay passes the track requirements is less than 0.1\% on average, leading to a negligible contribution to the uncertainty on multiplicity distributions.

The particle composition affects the efficiency estimate, because different particle species have different efficiencies and effective \pt~cut off. The influence of the uncertainty in particle composition was estimated by varying, in the simulation, the relative fractions of charged kaons and protons with respect to charged pions by $\pm$30\%, which covers conservatively the uncertainties in the measured particle composition at the LHC \cite{Abelev:2012cn}, and found to range, for $\dndetainline$, between 0.1\% for the \inel events class and 0.2\% for the \nsd and \inelgt event classes (\cref{tab:dndeta-systematics-at-0}). The effect is negligible for multiplicity distributions.

\begin{table}[tb]%
 \centering
 {\small
 \begin{tabu} to \textwidth {>{\bfseries}X[3,c]X[0.7,l]X[1,r]X[1,r]X[1,r]X[1,r]}%
 \toprule
 \rowfont{\bfseries} {}				& \multirow{2}{\linewidth}{Event class}				& \multicolumn{4}{c}{$\sqrt{\bm{s}}$ (TeV)} \\
 \cmidrule[\heavyrulewidth]{3-6}
 \rowfont{\bfseries} {}				& {} 		& 0.9 	& 2.76 	& 7 	& 8  \\
 \midrule
 Material budget				& All				& 0.2		& 0.2		& 0.2		& 0.2	\\
 \midrule
 Strangeness corrections			& All				& 0.14		& 0.17		& 0.16		& 0.16	\\
 \midrule
 \multirow{2}{*}{$\bm{p_{\mathrm{T}}}$~uncertainties}	& \multirow{2}{*}{All}		& $+$1.0	& $+$1.0	& $+$1.0	& $+$1.0 \\
 {}						& {}				& $-$0.5	& $-$0.5	& $-$0.5	& $-$0.5 \\
 \midrule
 \multirow{3}{*}{Particle composition}		& \inel				& 0.1		& 0.1		& 0.1		& 0.1 \\
 {}						& \nsd				& 0.2		& 0.2		& 0.2		& 0.2 \\
 {}						& \inelgt			& 0.2		& 0.2		& 0.2		& 0.2 \\
 \midrule
 \multirow{6}{*}{Event generator dependence}	& \multirow{2}{*}{\inel}	& $+$3.5	& $+$6.7	& $+$7.3	& $+$7.3 \\
 {}						& {}				& $-$1.2	& $-$4.0	& $-$3.5	& $-$3.5 \\
 \cmidrule{2-6}
 {}						& \multirow{2}{*}{\nsd}		& $+$4.0	& $+$3.9	& $+$2.2	& $+$2.2 \\
 {}						& {}				& $-$2.0	& $-$3.0	& $-$2.0	& $-$2.0 \\
 \cmidrule{2-6}
 {}						& \multirow{2}{*}{\inelgt}	& $+$0.5	& $+$0.5	& $+$0.6	& $+$0.6 \\
 {}						& {}				& $-$0.5	& $-$0.5	& $-$0.6	& $-$0.6 \\
 \midrule
 SPD simulation					& All				& 0.8		& 1.1		& 0.6		& 0.7 \\
 \midrule
 \multirow{6}{*}{Total systematic uncertainty}	& \multirow{2}{*}{\inel}	& $+$3.8	& $+$6.9	& $+$7.4	& $+$7.4 \\
 {}						& {}				& $-$1.6	& $-$4.2	& $-$3.6	& $-$3.6 \\
 \cmidrule{2-6}
 {}						& \multirow{2}{*}{\nsd}		& $+$4.2	& $+$4.2	& $+$2.6	& $+$2.6 \\
 {}						& {}				& $-$2.3	& $-$3.3	& $-$2.2	& $-$2.3 \\
 \cmidrule{2-6}
 {}						& \multirow{2}{*}{\inelgt}	& $+$1.5	& $+$1.7	& $+$1.5	& $+$1.5 \\
 {}						& {}				& $-$1.2	& $-$1.4	& $-$1.2	& $-$1.2 \\
 \midrule
 \multirow{3}{*}{Run-to-run fluctuations}	& \inel				& 0.4		& 0.2		& 0.4		& 0.4 \\
 {}						& \nsd				& 0.4		& 0.3		& 0.3		& 0.3 \\
 {}						& \inelgt			& 0.3		& 0.2		& 0.4		& 0.4 \\
 \bottomrule
 \end{tabu}%
 }
  \caption{Relative contributions, in percent, to systematic uncertainties in the measurement of $\dndetainline$, at $\eta =$~0, for centre-of-mass energies considered in this study and for the three event classes defined in the text; the \pt~uncertainties combine the contributions from the difference between MC models and data with the uncertainty on the \pt~distribution below 50 MeV/c. Run-to-run fluctuation contributions indicated here for comparison are not included in total uncertainties.\label{tab:dndeta-systematics-at-0}}%
\end{table}%
\subsubsection{Detector Simulation}
The systematic uncertainty related to the limited precision with which the detector performance is simulated was evaluated by varying the threshold on parameters used to select the various types of tracks, over a range obtained from the observed difference in the distributions of these parameters between simulation and data. For $\dndetainline$ measurements based on tracklets, the $\chi^2$ cut was varied between 1.3 and 4. The spread of $\dndetainline$ values over the range of $z$-positions of the vertex covered by a given $\eta$ bin was used as a measure of the bias introduced by the $z$-dependence of the tracklet reconstruction efficiency. The corresponding uncertainty ranges from 0.6\% to 1.1\% depending on the data sample (\cref{tab:dndeta-systematics-at-0}).

Since for multiplicity distribution measurements all three track-counting algorithms are used, track parameters were also varied for global tracks, and for ITS-only tracks (details may be found in \cite{doi:10.1142/S0217751X14300440}):
\begin{itemize}
\item for global tracks, the minimum number of TPC clusters was varied between 60 and 80 and the $\chi^2$ cut between 3 and 5.
\item for ITS-only tracks, the $\chi^2$ cut was varied between 2 and 3.
\end{itemize}

The uncertainty on the DCA distribution is not considered here, as it is included implicitly in the uncertainty on the strangeness content, which affects the DCA distribution.

We find that, in the case of multiplicity distributions, uncertainties in the detector simulation are negligible compared to other sources of uncertainties.
\subsubsection{Model dependence}
The remaining SD fraction in the sample selected with the \mband trigger in view of the normalization to the NSD event class is 3\%, 1\% and negligible at $\sqrt{s} =$ 0.9, 2.76 and $\geq$ 7~TeV, respectively. Uncertainties coming from diffraction contributions are included in the trigger efficiency uncertainties (\cref{tab:trigger-efficiencies}) obtained in \cite{alice-crosssection-diffraction}, for $\sqrt{s} \leq 7$~TeV. At $\sqrt{s} = 8$~TeV, the efficiency values were taken to be the same as for $\sqrt{s} = 7$~TeV. In addition, the model uncertainties at $\sqrt{s} =$ 0.9, 2.76, 7 and 8~TeV were obtained from the difference between \pythia{6} Perugia0 and \phojet. A test of the efficiency evaluation was obtained by comparing simulated \mband to \mbor trigger efficiency ratios to the measured values. Excellent agreement was found at all energies \cite{alice-crosssection-diffraction}.

For multiplicity distributions, the systematic uncertainty from the model dependence is included in both efficiency correction and \pt dependence uncertainties, as different event generators and tunes are used to estimate independently efficiencies and response matrices (see \cref{section:systematics:subsection:unfolding-systematics}).
\subsubsection{\pt~dependence}
None of the MC generators used in the detector simulation reproduces correctly the \pt~distribution of charged particles observed in the data \cite{Aamodt:2010my,Abelev:2013ala}. This introduces an uncertainty in the determination of the detector response, as it is integrated over transverse momenta, and the probability of detecting a particle decreases with decreasing \pt. This affects in particular, together with the uncertainty on the material budget and the magnetic field, extrapolations of measurements to $\pt = 0$.

\begin{figure}[t]%
\includegraphics[width=\textwidth]{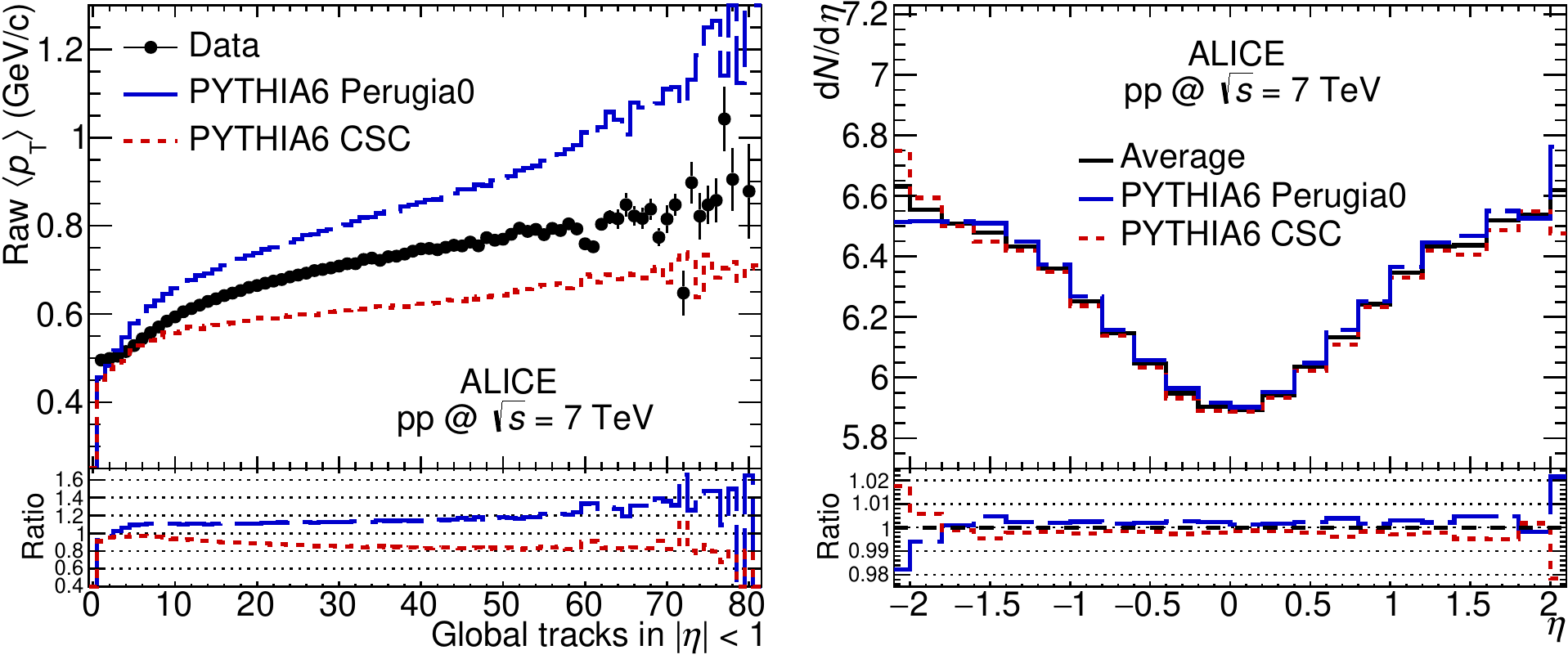}
 \caption{(left) At $\sqrt{s} = 7$~TeV, average raw \pt~vs. raw charged track multiplicity in $|\eta| <1$, for data (black circles), \pythia{6} Perugia0 (blue dashed line) and ATLAS-CSC (red dotted line). The bottom part of the figure shows the ratios of the two simulated distributions to the data; (right) Comparison of $\dndetainline$ evaluations, as a function of $\eta$, using correction maps obtained with (a) \pythia{6} Perugia0 (blue dashed line), (b) \pythia{6} ATLAS-CSC (red dotted line) and (c) the average between (a) and (b); The bottom part of the figure shows the ratios of corrected distributions from (a) and (b) to the average.\label{fig:pt-vs-nch-effect}}
\end{figure}

In order to study \pt~spectrum effects, two different tunes of the \pythia{6} event generator were used, ATLAS CSC and Perugia0, which give an average \pt~versus charged-particle multiplicity respectively below and above the data (\cref{fig:pt-vs-nch-effect} (left)), for most of the multiplicity range ($\nch > 2$). The difference between measurements obtained with response matrices corresponding to each of these Monte Carlo generators, is used as the corresponding systematic error contribution. \Cref{fig:pt-vs-nch-effect} (right) shows that this procedure introduces an uncertainty which is weakly dependent on $\eta$, and amounts to $\pm$0.3\%, when averaged over $|\eta| \leq 2$. 

For undetected particles, below a threshold of about 50 MeV/c, a value chosen to coincide with a track detection efficiency of 50\%, the corresponding systematic uncertainty is obtained by varying their fraction by a conservative amount ($-$50\% and $+$100\%). (The fraction of the \pt~spectrum below 50 MeV/c is about 1\% of the total, for both \pythia{6} and \phojet.) The resulting systematic uncertainty on $\dndetainline$ is $\eta$ dependent, and found to range from $-0.5$\% and $+1.0$\%, at $\eta = 0$, to -0.75\% and +1.5\%, at $|\eta| = 2$.

The systematic uncertainty induced on $\pnch$ by the difference in \pt~between data and simulation is slightly sensitive to the tune of \pythia{6} considered: for instance, at $\nch = 90$, varying the \pt~spectrum below 50 MeV/c by -50\% and +100\% induces a change of -5\% and +9\%, respectively for the ATLAS CSC tune and of -4\% and +8\%, respectively, for the Perugia0 tune.

\subsection{Systematic uncertainties in unfolding of multiplicity distributions}
\label{section:systematics:subsection:unfolding-systematics}
\subsubsection{Uncertainty evaluation}
The results of the unfolding procedure and unfolding uncertainty estimate were cross-checked, in standard ways:
\begin{itemize}
\item changing the regularization term by either varying $\beta$ or the function $F(U)$;
\item using two alternative unfolding procedures: Bayesian \cite{D'Agostini:1994zf} and singular value decomposition \cite{Hocker:1995kb}.
\end{itemize}
The changes in unfolded distributions due to variations of the different elements used in the unfolding procedure, each having their own systematic uncertainty, were studied by considering all possible combinations: counting algorithm (3 cases), event generator used for efficiency correction (2 cases: \pythia{6} and \phojet tuned for diffraction), event generator used for the response matrix (2 cases \pythia{6} CSC and \pythia{6} Perugia0), low \pt spectrum extrapolation below 50~MeV/c (3 cases: varying the integral below 50~MeV/c by $-50$\%, 0\% and $+100$\%), response matrix parameterization (2~cases: Power law, and Pad\'e), which correspond to 72 separate measurements that are correlated as a result of the unfolding procedure. To take these correlations into account, for a given energy and a given $\eta$ range, the systematic uncertainty from all the sources considered was estimated as the overall spread between the resulting distributions. The average between these distributions (center of the band covered by the 72 curves) was used as the measurement. The total uncertainty originating from the unfolding procedure (evaluated for each multiplicity bin as a linear sum of the unfolding bias \cite{cowan-stat-analysis} and the unfolded distribution covariance calculated from the statistical uncertainty of the raw distribution) was added linearly to the systematic uncertainty defined as half the size of the band. This takes into account the fact that the unfolded distributions come from the same raw multiplicity histogram, hence the raw data statistical fluctuations are propagated to each of the unfolded distributions in a similar way and affect the spread of distributions uniformly. The resulting systematic uncertainty is expected to be highly correlated with multiplicity.

The systematic uncertainties from material budget, tracklet and track selection, detector alignment (evaluated by changing the geometry within alignment uncertainties), particle composition, strangeness corrections are found to be negligible. Among the non-negligible contributions to multiplicity distribution systematic uncertainties, the contribution from selection efficiency (\cref{tab:pnch-efficiency-uncertainty})  can be evaluated separately as it is a multiplicative correction. All the other contributions, from the uncertainty on $\left<\pt\right>$, the extrapolation down to $\pt = 0$, the counting algorithms, and the model dependence including the contribution from diffraction, are all mixed together through the procedure described above. For \nsd and \inel event classes, the diffraction contribution is mainly significant in the zero multiplicity bin, which is absent for the \inelgt event class.
\begin{table}[t]%
 \centering
  {\small
  \begin{tabu} to 0.9\textwidth {>{\bfseries}X[0.5,c]X[0.1,r]X[0.6,l]X[1.5,c]X[1.5,c]X[1.5,c]X[1.6,c]}
   \toprule
\rowfont{\bfseries}   \multicolumn{2}{c}{\multirow{2}{2.8cm}{Source of uncertainty}}	& \multirow{2}{\linewidth}{Event class}	& \multicolumn{4}{c}{$\sqrt{\bm{s}}$ (TeV)} \\
   \cmidrule{4-7}
   {}	&	{}	& {}	& {\bfseries 0.9}	& {\bfseries 2.76}	& {\bfseries 7}	& {\bfseries 8} \\
   \midrule
   \multirow{9}{*}{\rotatebox[origin=c]{90}{\parbox{3.2cm}{\centering\textbf{Efficiency uncertainty (\%)}}}}	&	\multirow{3}{*}{$|\eta|<0.5$}	& \inel	& 0.96 -- 0.7 -- 0.5	& 3.05 -- 2.1 -- 1.2	&  2.11 -- 1.5 -- 0.9	&  2.52 -- 1.7 -- 1.0 \\
   {}	& {}	& \nsd		&  4.68 -- 3.5 -- 2.4     & 9.20 -- 6.3 -- 3.7      &  4.55 -- 3.2 -- 1.9     & 13.45 -- 9.1 -- 5.3 \\
   {}	& {}	& \inelgt	& -- -- 0.7 -- 0.5	& -- -- 2.1 -- 1.2	& -- -- 1.5 -- 0.9	& -- -- 1.7 -- 1.0 \\
   \cmidrule{2-7}
   {}	& \multirow{3}{*}{$|\eta|<1$}	& \inel	& 0.14 -- 0.2 -- 0.2     & 0.49 -- 0.5 -- 0.5      &  0.18 -- 0.2 -- 0.3     &  0.02 -- 0.2 -- 0.3 \\
   {}	& {}	& \nsd		& 0.42 -- 0.8 -- 1.2	& 1.66 -- 2.1 -- 2.3	&  0.31 -- 0.8 -- 1.1	&  2.91 -- 3.1 -- 3.1 \\
   {}	& {}	& \inelgt	& -- -- 0.2 -- 0.2	& -- -- 0.5 -- 0.5	& -- -- 0.2 -- 0.3	& -- -- 0.2 -- 0.3 \\
   \cmidrule{2-7}
   {}	& \multirow{3}{*}{$|\eta|<1.5$}	& \inel	& 0.02 -- 0.2 -- 0.3     & 0.03 -- 0.1 -- 0.2      &  0.03 -- 0.3 -- 0.4     &  0.03 -- 0.2 -- 0.4 \\
   {}	& {}	& \nsd		& 0.10 -- 0.4 -- 0.7	& 0.12 -- 1.2 -- 2.0      &  0.01 -- 0.4 -- 0.7     &  0.19 -- 1.6 -- 2.8 \\
   {}	& {}	& \inelgt	& -- -- 0.2 -- 0.3	& -- -- 0.1 -- 0.2	& -- -- 0.3 -- 0.4	& -- -- 0.2 -- 0.4 \\
   \bottomrule
  \end{tabu}
  \label{tab:pnch-efficiency-uncertainty}
   \caption{Systematic uncertainties, in percent, due to efficiency corrections including trigger and reconstruction efficiency, event generator dependence and diffraction, for \inel, \nsd and \inelgt event classes, in pseudorapidity intervals $|\eta| <$ 0.5, 1.0 and 1.5, at $\sqrt{s} =$ 0.9, 2.76, 7 and 8~TeV. Numbers are given, from left to right, for multiplicities 0, 1 and 2. For the \inelgt event class, the efficiencies are the same as for the \inel class, except that there is no bin with multiplicity 0. Note that the uncertainty listed here for bin 0 is not the total uncertainty in that bin. Bin 0 values are recalculated separately from simulation, which adds a significant spread of values that is seen in the total uncertainty estimate.}
  }
 \end{table}
 \begin{table}[t]%
   {\small
  \begin{tabu} to 0.9\textwidth {>{\bfseries}X[0.5,c]X[0.1,r]X[0.6,l]X[1.5,c]X[1.5,c]X[1.5,c]X[1.6,c]}
   \toprule
\rowfont{\bfseries}   \multicolumn{2}{c}{\multirow{2}{2.8cm}{Source of uncertainty}}	& \multirow{2}{\linewidth}{Event class}	& \multicolumn{4}{c}{$\sqrt{\bm{s}}$ (TeV)} \\
   \cmidrule{4-7}
   {}	&	{}	& {}	& {\bfseries 0.9}	& {\bfseries 2.76}	& {\bfseries 7}	& {\bfseries 8} \\
   \midrule
   \multirow{9}{*}{\rotatebox[origin=c]{90}{\parbox{3.8cm}{\centering\textbf{Total systematic uncertainty (\%)}}}}	&	\multirow{3}{*}{$|\eta|<0.5$}
		& \inel		& 4 -- 4 -- 16	& 2 -- 2 -- 6	& 1 -- 3 -- 4	& 22 -- 9 -- 13 \\
   {}	& {}	& \nsd		& 16 -- 6 -- 28	& 28 -- 9 -- 15	& 19 -- 7 -- 9	& 35 -- 8 -- 12 \\
   {}	& {}	& \inelgt	& 8 -- 3 -- 14	& 3 -- 2 -- 6	& 3 -- 4 -- 4	&  3 -- 1 --  4 \\
   \cmidrule{2-7}
   {}	& \multirow{3}{*}{$|\eta|<1$}
		& \inel		& 7 -- 7 -- 21		& 3 -- 3 -- 8	& 1 -- 3 -- 5	& 29 -- 11 -- 13 \\
   {}	& {}	& \nsd		& 38 -- 10 -- 32	& 52 -- 8 -- 15	& 38 -- 8 -- 11	& 59 --  8 -- 14 \\
   {}	& {}	& \inelgt	& 14 -- 6 -- 20		& 7 -- 3 -- 8	& 10 -- 3 -- 5	&  4 --  2 --  4 \\
   \cmidrule{2-7}
   {}	& \multirow{3}{*}{$|\eta|<1.5$}
		& \inel		& 11 -- 9 -- 27	& 3 -- 7 -- 12	& 3 -- 5 -- 8	& 34 -- 9 -- 14 \\
   {}	& {}	& \nsd		& 62 -- 8 -- 34	& 76 -- 12 --24	& 57 -- 8 -- 14	& 80 -- 9 -- 17 \\
   {}	& {}	& \inelgt	& 12 -- 7 -- 27	& 7 -- 7 -- 11	& 10 -- 6 -- 8	&  8 -- 1 --  6 \\
   \bottomrule
  \end{tabu}
  \label{tab:pnch-total-uncertainty}
  \caption{Total systematic uncertainties in charged-particle multiplicity distribution measurements, in percent, for \inel, \nsd and \inelgt event classes, in pseudorapidity intervals $|\eta| <$ 0.5, 1.0 and 1.5, at $\sqrt{s} =$ 0.9, 2.76, 7 and 8~TeV. Values for \inel and \nsd event classes are given for three characteristic multiplicities, from left to right: zero multiplicity - mean multiplicity - five times the mean multiplicity. For the \inelgt class, the first value is for $\nch  = 1$, as there is no entry for $\nch = 0$.}
  }%
\end{table}%
\subsubsection{Bin-to-bin correlations in systematic uncertainty}
\label{section:systematics:subsection:unfolding-systematics:correlation}
Due to specific nature of the correction procedure, the final multiplicity distributions contain bin-to-bin correlations coming from various sources with different properties. These sources can be categorized by their effect:
\begin{itemize}
 \item statistical correlations, resulting from the propagation of the raw distribution statistical uncertainties through unfolding process, their characteristics largely depend on the response matrix structure;
 \item fully correlated shift of the distribution as a result of the uncertainty in normalization;
 \item long-range correlations in systematic uncertainties as a result of multiplicity scale change (that is determined by the position of the response matrix bulk relative to the diagonal, see \cref{section:pnch:analysis:unfolding}) in the unfolded distribution.
\end{itemize}

We found that first category correlations (statistical) are negligible compared to the last category (scaling), while the second category correlations can be easily factorized for fitting the multiplicity distribution. The three main sources of correlated uncertainty, falling into the third category, are the change of counting algorithm, the change of event generator tune used to produce the response matrix (see \cref{fig:pt-vs-nch-effect}) and the variation of $\pt$ distribution under the detection threshold. In order to evaluate the effect on the final distributions we construct 18 intermediate distributions corresponding to all possible combinations of counting algorithm, event generator tunes and variations of $\pt$ spectra. All other sources of systematics are included (except the uncertainty corresponding to bin $\nch = 0$ renormalization, as it is not applied) with the same procedure described for the final distribution previously in this section. These 18 distributions are then treated independently in \cref{section:experimental-results:subsection:NBD-parameterization} to evaluate the effect of correlated uncertainties on NBD fits.

\subsection{Summary of systematic uncertainties}
\subsubsection{Pseudorapidity density}
The various contributions to systematic uncertainties in $\dndetainline$ are summarized in \cref{tab:dndeta-systematics-at-0} for the three event classes and the four centre-of-mass energies studied in this publication. For the \inelgt event class, a precision of ~1.5\% is achieved, as the sensitivity to diffraction is negligible and the $\eta$ range is reduced in the definition of this event class.

In the $\eta$ range covered in this study ($|\eta|<2$), we find that systematic uncertainties show essentially no $\eta$ variation (\cref{fig:dndeta-run-to-run-uncertainty}), and are therefore strongly correlated bin-to-bin.
\subsubsection{Multiplicity distributions}
The efficiency uncertainties (\cref{tab:pnch-efficiency-uncertainty}) are only relevant at low multiplicity. For multiplicities above 8 to 9, the efficiency reaches 100\% and the corresponding systematic uncertainty becomes negligible. Therefore, efficiency uncertainties are only given for a few characteristic low multiplicities. The total systematic uncertainties vary with multiplicity, therefore they are given in \cref{tab:pnch-total-uncertainty} only for a few characteristic values of the multiplicity.
\subsubsection{Consistency checks}
In the measurement of $\dndetainline$, statistical errors are negligible. Therefore, the study of run-to-run fluctuations (measured RMS of $\dndetainline$ results in different runs) provides a check that all run dependent corrections are properly handled.

\begin{figure}[tb]%
\includegraphics[width=\textwidth]{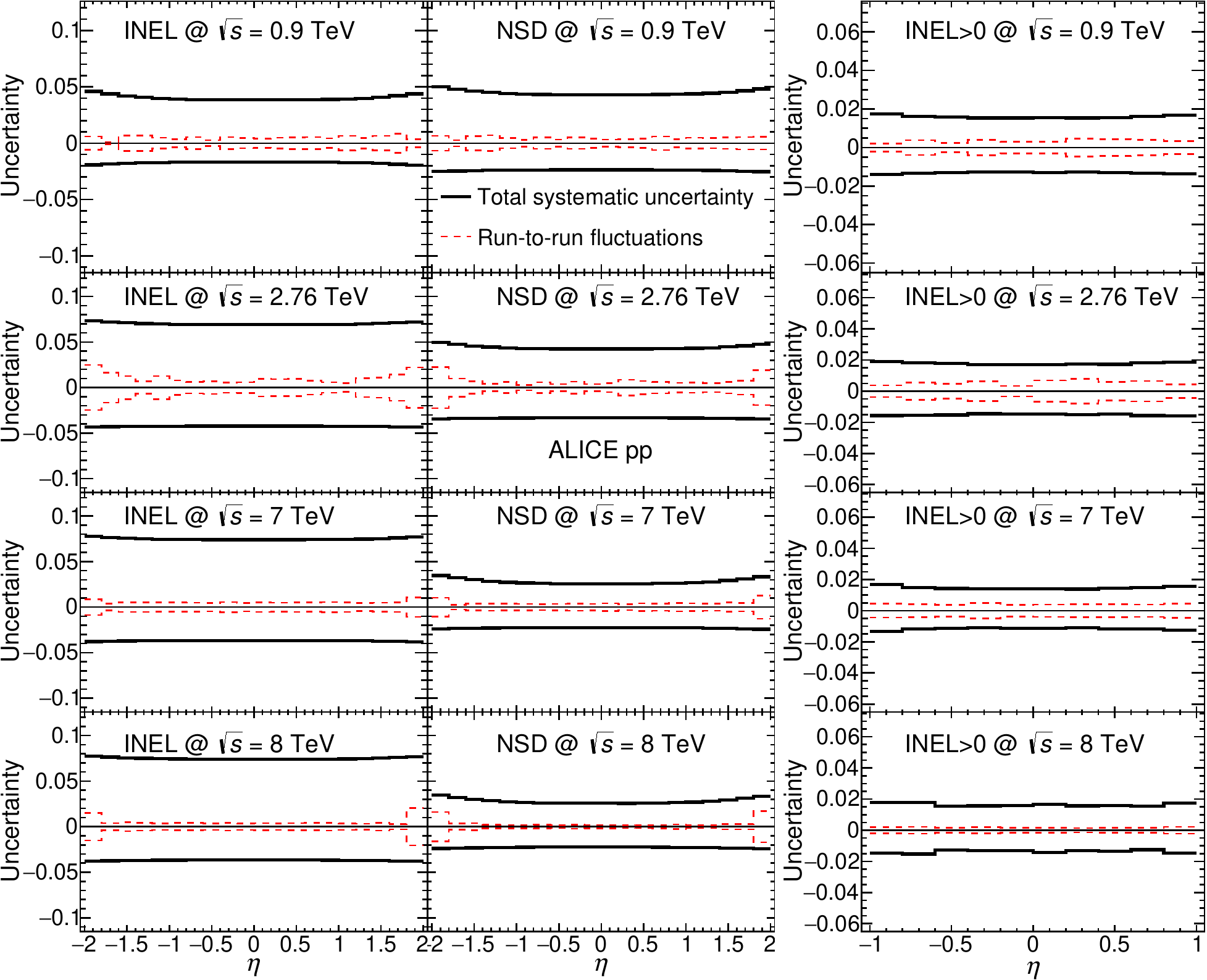}
 \caption{Total relative systematic uncertainty on $\dndetainline$ (thick black lines), as a function of pseudorapidity, compared to run-to-run fluctuations (thin red dashed lines) at $\sqrt{s} = 0.9$~TeV (top row), 2.76~TeV (second row), 7~TeV (third row) and 8~TeV (bottom row), for the \inel, \nsd and \inelgt event classes, as indicated.\label{fig:dndeta-run-to-run-uncertainty}}%
\end{figure}

For the \inel and \nsd event classes, contributions from run-to-run fluctuations are significantly smaller than the total systematic error (\cref{fig:dndeta-run-to-run-uncertainty}). For the \inelgt event class, for which the precision is highest, the relative importance of run-to-run fluctuations is larger than for the \inel and \nsd event classes (\cref{fig:dndeta-run-to-run-uncertainty}), but reaches at most 5\% of the total systematic uncertainty.

As data correction procedures are significantly different between $\dndetainline$ and multiplicity distribution measurements, a test was performed to verify the consistency of the two measurements. At the four centre-of-mass energies and for the three pseudorapidity intervals used in this study, integrals of the multiplicity distributions were found to be consistent with the direct measurements of $\dndetainline$, within errors.

\section{Experimental results}
\label{section:experimental-results}
\subsection{Pseudorapidity density of primary charged particles: measurements}
In their common $\eta$ range, $|\eta| < 0.9$, the three track counting algorithms discussed in \cref{section:track-selection-and-algorithms:subsection:algorithms} achieve a similar precision of 1\%, and were found to give consistent results. The main difference is that for ITS+ and ITSTPC+, there is a detector calibration contribution to systematics for the TPC and the SDD, not present for the SPD. To achieve the largest possible $\eta$ range, in the measurement of $\dndetainline$ versus $\eta$, the \spd algorithm is used alone.

\begin{figure}
\centering
\begin{overpic}[width=0.8\textwidth]{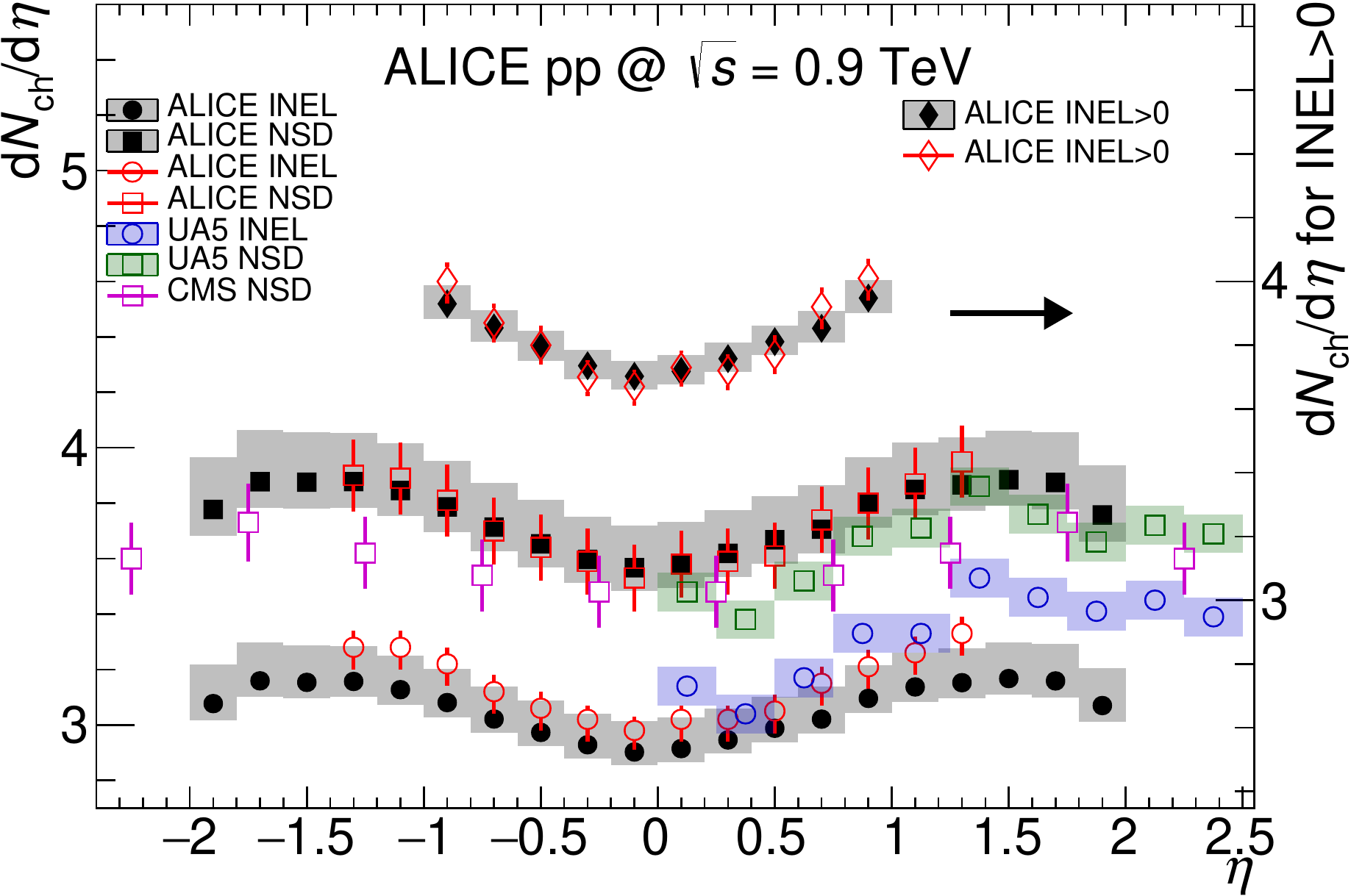}%
 \put(253,534){\scriptsize{\cite{aamodt2010charged09}}}
 \put(250,511){\scriptsize{\cite{aamodt2010charged09}}}
 \put(232,487){\scriptsize{\cite{bUA5}}}
 \put(230,466){\scriptsize{\cite{bUA5}}}
 \put(234,441){\scriptsize{\cite{bCMS-eta}}}

 \put(868,544){\scriptsize{\cite{aamodt2010charged7}}}
 \end{overpic}\\[-0.3cm]%
 \caption{$\dndetainline$ vs. $\eta$ at $\sqrt{s} = 0.9$~TeV, for the three normalizations defined in the text, and a comparison with ALICE previous measurements \cite{aamodt2010charged09,aamodt2010charged7}, UA5 \cite{bUA5} and CMS \cite{bCMS-eta}. Note that to avoid overlap of data points on the figure, the \inelgt data were displaced vertically, and for these data the scale is to be read off the right-hand side vertical axis. Systematic uncertainties on previous data are shown as error bars (except for UA5, with coloured bands), while they are shown as grey bands for the data from this publication.\label{fig:dndeta-result-sqrts-09}}%
\end{figure}

The measurement at $\sqrt{s} = 0.9$~TeV is shown in \cref{fig:dndeta-result-sqrts-09} compared with previous results. At $|\eta| > 0.9$ the measurement for the \inel event class is slightly lower than in ALICE{\textquoteright}s previous publication \cite{aamodt2010charged09}. The difference comes mainly from: \begin{inparaenum}[(a)]\item the tuning of the MC generators for diffraction, as larger pseudorapidities are more sensitive to SD; \item the subtraction of particles coming from the decay of strange particles was improved, using ALICE{\textquoteright}s measurement of strangeness \cite{aamodt2011strange}; and \item the improvement of the $\eta$ dependence of the \spd algorithm. \end{inparaenum}

\begin{figure}[!p]%
 \centering
 \subfigure{
 \begin{overpic}[width=0.65\textwidth]{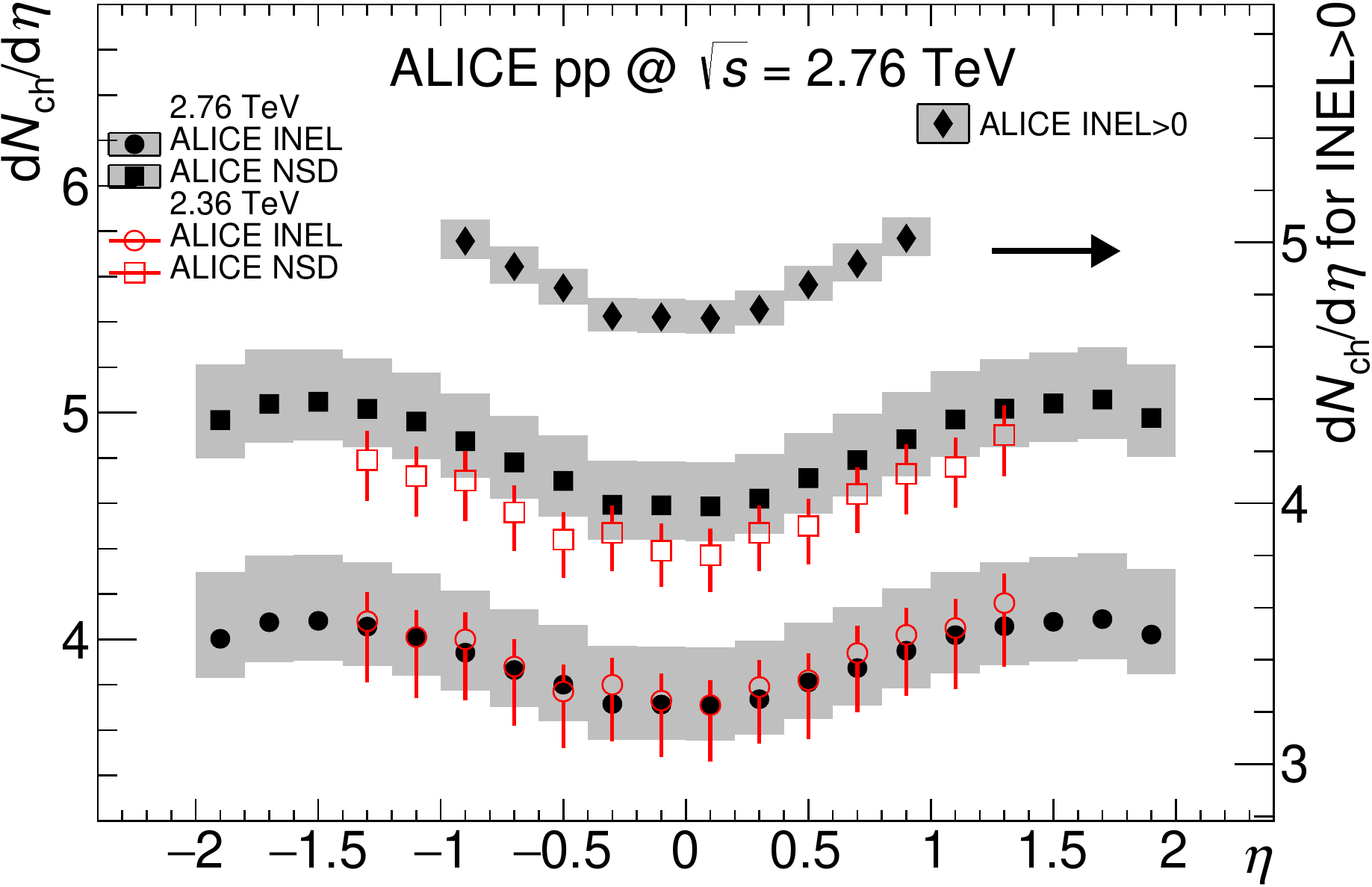}
 \put(255,468){\scriptsize{\cite{aamodt2010charged09}}}
 \put(255,444){\scriptsize{\cite{aamodt2010charged09}}}
 \end{overpic}
 }\\[-0.64cm]%
 \subfigure{
 \begin{overpic}[width=0.65\textwidth]{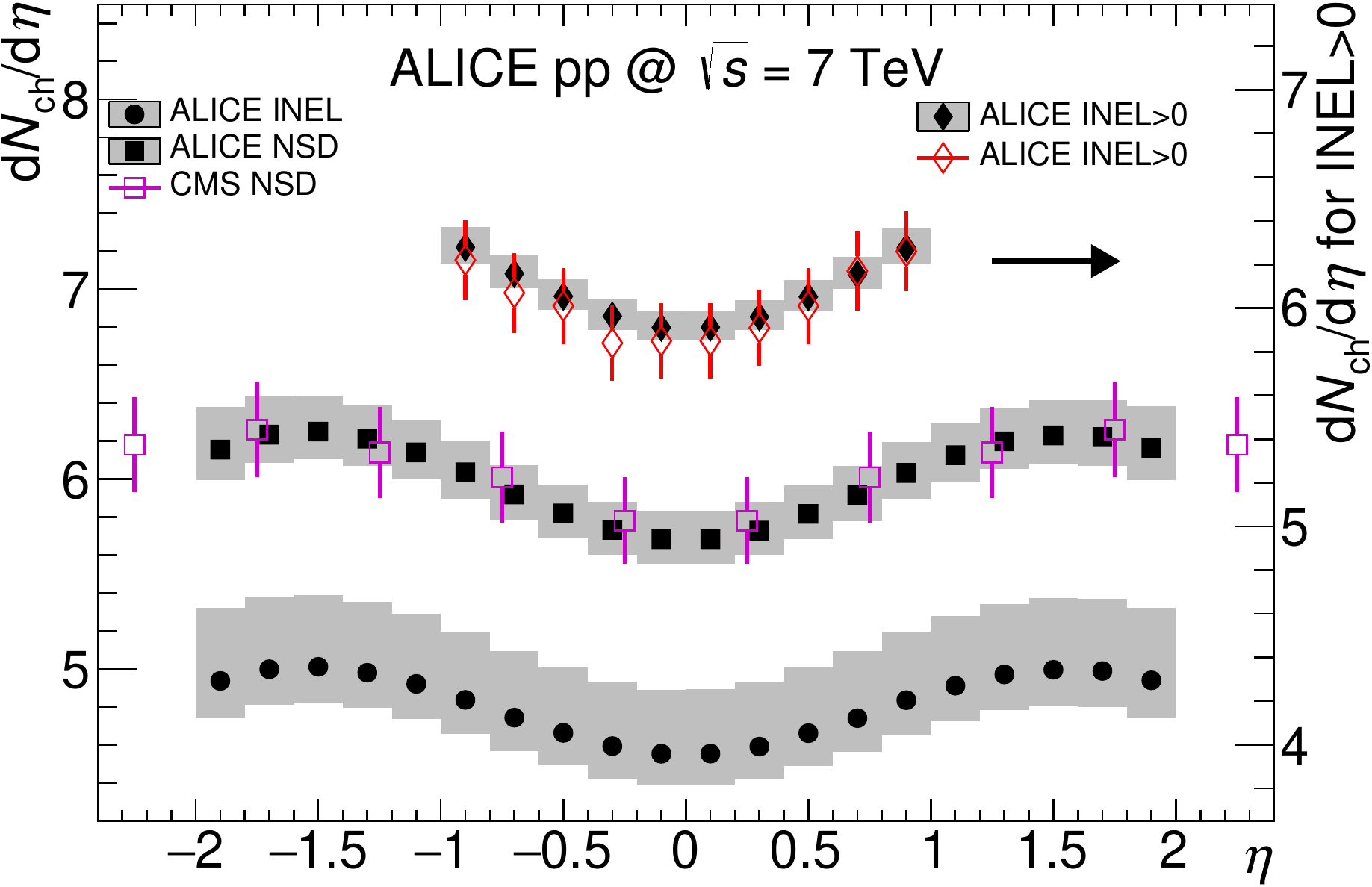}
  \put(235,506){\scriptsize{\cite{bCMS-eta}}}
  \put(867,527){\scriptsize{\cite{aamodt2010charged7}}}
 \end{overpic}
 }\\[-0.64 cm]%
 \subfigure{\includegraphics[width=0.652\textwidth]{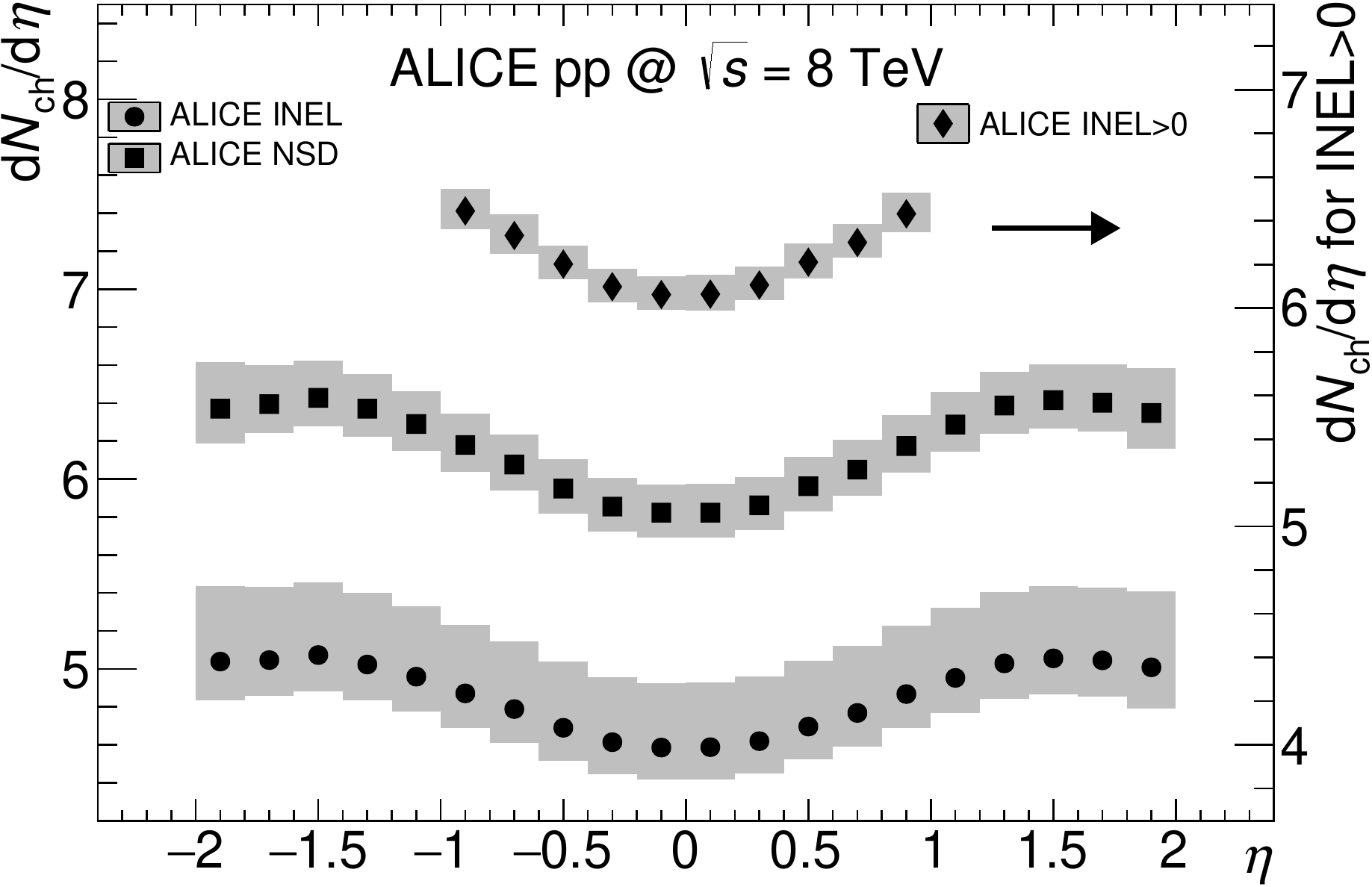}}\\[-0.3cm]%
 \caption{$\dndetainline$ vs. $\eta$ measurements: $\sqrt{s} = 2.76$~TeV compared with $\sqrt{s} = 2.36$~TeV taken from ALICE \cite{aamodt2010charged09} (top); $\sqrt{s}=$~7~TeV and comparison with CMS \cite{bCMS-eta} and ALICE \cite{aamodt2010charged7} data (middle); $\sqrt{s} = 8$~TeV (bottom). Systematic uncertainties are shown as error bars for the previous data and as grey bands for the data from this publication. The scale is to be read off the right-hand side axis for \inelgt.\label{fig:dndeta-result-sqrts-276-7-8}}%
\end{figure}

\begin{figure}[ht]%
 \centering%
 \includegraphics[width=\textwidth]{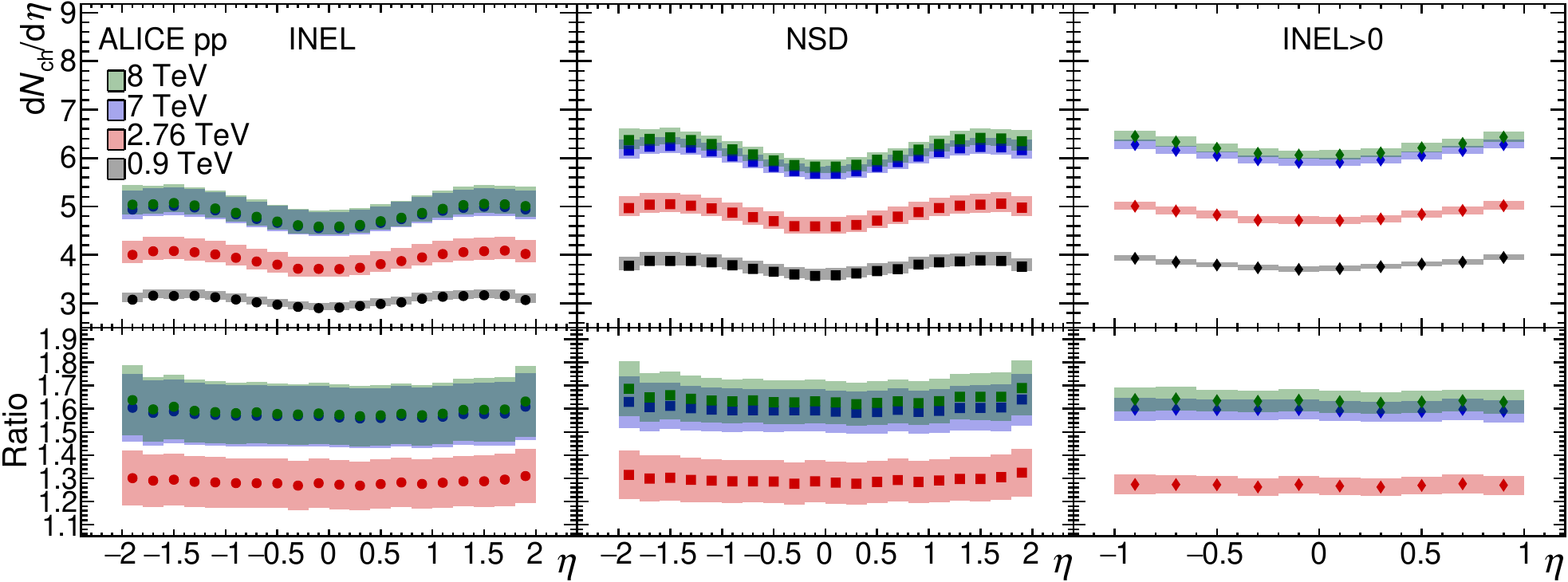}\\[-0.3cm]%
 \caption{Comparison of $\dndetainline$ vs. $\eta$ measurements between the various centre-of-mass energies considered in this study: \inel (left), \nsd (middle), and \inelgt (right). The lower parts of the figures show the ratios of data at energies indicated to the data at 0.9~TeV, with corresponding colours. Systematic uncertainties are indicated as coloured bands.\label{fig:dndeta-results-energy-evolution-full}}%
\end{figure}

The discrepancy with UA5 for the \inel event class at large $\eta$ could perhaps be related to the fact that UA5 used a $\slfrac{1}{M_X}$ variation of the single-diffractive cross section (see \cite{kaidalov2009description}). Note also that UA5 data seem to be internally inconsistent (see discussion in \cite{poghosyan1806two}). The measurement at $\sqrt{s} = 2.76$~TeV is shown in \cref{fig:dndeta-result-sqrts-276-7-8} top, and found to be consistent with the ALICE measurement at $\sqrt{s} = 2.36$~TeV \cite{aamodt2010charged09}, as expected because of the small change in center-of-mass energy. The new measurements of $\dndetainline$, at $\sqrt{s} = 7$~TeV (\cref{fig:dndeta-result-sqrts-276-7-8} middle), show agreement both with the previous ALICE results for \inelgt \cite{aamodt2010charged7}, and with CMS \nsd data \cite{bCMS-eta}. The measurements of $\dndetainline$ at $\sqrt{s} = 8$~TeV (\cref{fig:dndeta-result-sqrts-276-7-8} bottom) show the 3\% increase with respect to the 7~TeV data, which corresponds to what is obtained in the extrapolation from lower energy data. Comparisons of the $\eta$ distributions at the four centre-of-mass energies (\cref{fig:dndeta-results-energy-evolution-full}), for the three events classes, show no significant change of shape and a smooth increase of the charged-particle density with increasing energy.

\begin{figure}[t]%
 \centering
 \includegraphics[width=\textwidth]{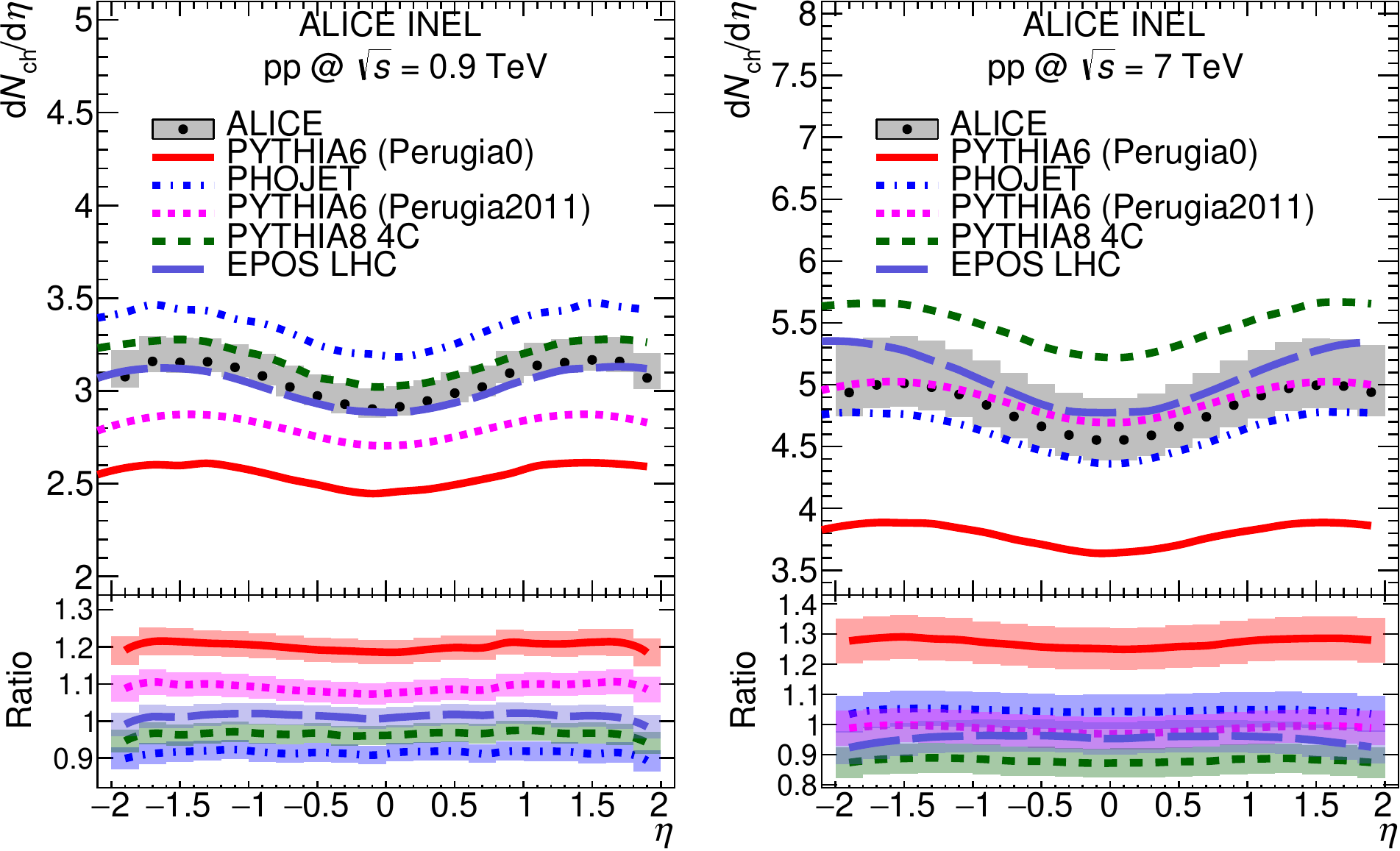}
 \caption{Comparison with models of ALICE measurements of $\dndetainline$ versus $\eta$ for the \inel event class, at $\sqrt{s} = 0.9$~(left) and 7~TeV (right): ALICE data (black circles with grey band), \pythia{6} tune Perugia0 \cite{Skands:2009zm} (red continuous line), \phojet \cite{Engel:1995sb} (blue dot-dashed line), \pythia{6} tune Perugia 2011 \cite{Skands:2009zm} (pink dashed line), \pythia{8} 4C \cite{Sjostrand:2007gs,Corke:2010yf} (green dashed line), EPOS LHC \cite{pierog2013epos} (long dashed light blue line). The lower parts of the figures show ratios of data to simulation. Systematic uncertainties on ratios are indicated by coloured bands.\label{fig:dndeta-results-model-comparison}}%
\end{figure}

The data for the \inel event class at $\sqrt{s} =$ 0.9 and 7~TeV were compared to simulations with current event generators (\cref{fig:dndeta-results-model-comparison}). At $\sqrt{s} =$ 0.9~TeV, EPOS LHC \cite{pierog2013epos} and \pythia{8} 4C \cite{Sjostrand:2007gs,Corke:2010yf} are consistent with the data. \phojet overestimates the data, while \pythia{6} Perugia0 and Perugia 2011 underestimate the data. At $\sqrt{s} = 7$~TeV, EPOS LHC, \phojet and \pythia{6} Perugia 2011 are consistent with the data. \pythia{8} 4C overestimates the data, while \pythia{6} Perugia0 underestimates the data. Note that \pythia{6} Perugia 2011, \pythia{8} 4C and EPOS LHC were tuned using LHC data.

\subsection{Energy dependence of $\dndetainline$ at $\eta = 0$}%
The traditional definition  for $\dndetainline$ at $\eta = 0$ is an integral of the data over the pseudorapidity range $|\eta| < 0.5$
\begin{equation}
 \left.\dndeta\right|_{\eta=0} \equiv \int\limits_{-0.5}^{+0.5} \dndeta \mathrm{d}\eta
\end{equation}
\begin{table}[htb]%
 \centering
 {\tabulinesep=0.6mm
 \begin{tabu}{>{\bfseries}llll}
\rowfont[c]{\bfseries}  $\sqrt{\bm{s}}$ (TeV)	& \inel				& \nsd				& \inelgt \\
  \midrule
  0.9				& $ 2.94^{+0.11}_{-0.05} $	& $3.61^{+0.17}_{-0.16}$	& $3.75^{+0.06}_{-0.05}$ \\
  2.36$^{\dagger}$	& $ 3.77^{+0.25}_{-0.12} $	& $4.43^{+0.17}_{-0.12}$	& {---} \\
  2.76				& $ 3.75^{+0.26}_{-0.16} $	& $4.63^{+0.30}_{-0.19}$	& $4.76^{+0.08}_{-0.07}$ \\
  7					& $ 4.60^{+0.34}_{-0.17} $	& $5.74^{+0.15}_{-0.15}$	& $5.98^{+0.09}_{-0.07}$ \\
  8					& $ 4.66^{+0.35}_{-0.17} $	& $5.90^{+0.15}_{-0.13}$	& $6.13^{+0.10}_{-0.08}$ \\
  \midrule
 \rowfont[r]{} {}				& \multicolumn{3}{c}{\footnotesize{${}^{\dagger}$Data taken from \cite{aamodt2010charged09}}}
 \end{tabu}}%
  \caption{Summary of ALICE measurements of $\dndetainline$ at $\eta = 0$ (integral of the data over $|\eta| < 0.5$), for
centre-of-mass energies and event classes considered in this study. The errors shown are systematic errors. Statistical errors are negligible.\label{tab:dndeta-results-at-0}}%
\end{table}%

The results of the measurements of $\dndetainline$ at $\eta = 0$ are given in \cref{tab:dndeta-results-at-0}. The energy dependence of $\dndetainline$ at $\eta = 0$ is of interest not only because it provides information about the basic properties of pp collisions, but also because it is related to the average energy density achieved in the interaction of protons, and constitutes a reference for the comparison with heavy ion collisions. At mid-rapidity, $\dndetainline$ can be parameterized as $\dndetainline \sim s^{\delta}$. Combining the ALICE data with other data at the LHC and at lower energies, we obtain $\delta = 0.102 \pm 0.003$, $0.114 \pm 0.003$ and $0.114 \pm 0.0015$\footnote{The uncertainty on $\delta$ was obtained, assuming that the data errors are independent at different centre-of-mass energies, which is not strictly the case.}, for the \inel, \nsd and \inelgt event classes, respectively, to be compared to $\delta \simeq 0.15$ for central Pb--Pb collisions \cite{aamodt2010charged}. This is clear evidence that the particle pseudorapidity density increases faster with energy in Pb--Pb collisions than in pp collisions. Fits are shown on \cref{fig:dndeta-result-at-0-fit} and \cref{tab:dndeta-results-at-0-extrapolation} gives extrapolations to centre-of-mass energies of 13 and 14~TeV (LHC design energy). While this paper was being prepared, the first measurement at 13~TeV by CMS appeared \cite{Khachatryan:2015jna}, resulting in $\left.\dndetainline\right|_{|\eta|<0.5} = 5.49\pm0.01\ \text{(stat)}\pm0.17\ \text{(syst)}$ for inelastic events, which is consistent with our extrapolation of $5.30\pm0.24$. Over the LHC energy range, from 0.9 to 14~TeV, while the centre-of-mass energy increases by a factor 15.5, extrapolation of present data for $\dndetainline$ at $\eta = 0$ shows an increase by factors $1.75\pm0.03$, $1.87\pm0.03$ and $1.87\pm0.01$, respectively for the three event classes. The multiplicity increase is similar for \nsd and \inelgt classes but slightly lower for the \inel class.

\begin{table}[ht]%
  \centering
 \begin{tabu}{>{\bfseries}llll}%
\rowfont[c]{\bfseries}  $\sqrt{\bm{s}}$ (TeV)	& \inel				& \nsd				& \inelgt \\
  \midrule
  13					& $5.30\pm0.24$			& $6.50\pm0.20$			& $6.86\pm0.10$ \\
  13.5					& $5.33\pm0.25$			& $6.56\pm0.20$			& $6.92\pm0.10$ \\
  14					& $5.37\pm0.25$			& $6.62\pm0.20$			& $6.98\pm0.10$ \\
 \end{tabu}%
 \caption{Extrapolations of $\dndetainline$, at $\eta  = 0$, for the three event classes, to higher energies at the LHC ($\sqrt{s} = 13$ and $14$~TeV), using the fits described in the text.\label{tab:dndeta-results-at-0-extrapolation}}%
\end{table}%
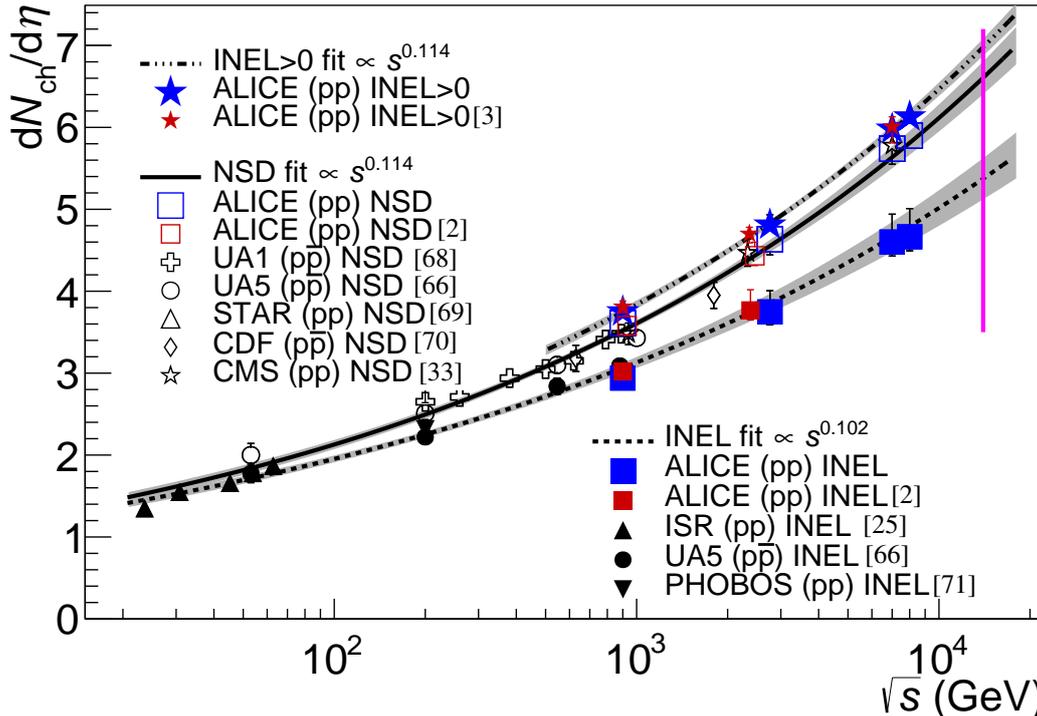
\begin{figure}[t]
 \centering
 \begin{overpic}[width=0.85\textwidth]{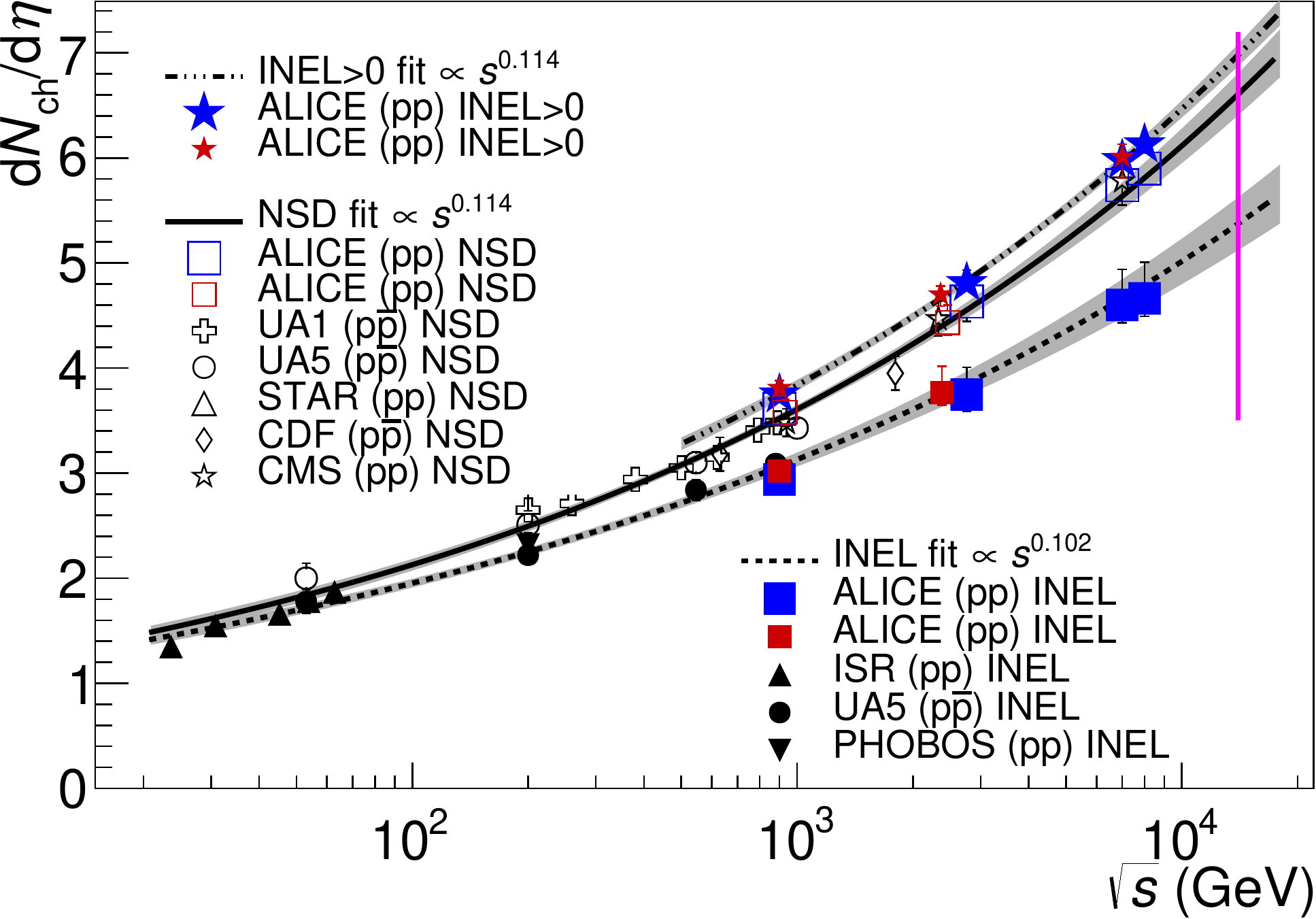}
  \put(447,582){\small{\cite{aamodt2010charged7}}}
  \put(413,472){\small{\cite{aamodt2010charged09}}}
  \put(390,443){\small{\cite{Albajar:1989an}}}
  \put(390,416){\small{\cite{bUA5}}}
  \put(405,390){\small{\cite{Abelev:2008ab}}}
  \put(390,362){\small{\cite{Abe:1989td}}}
  \put(395,332){\small{\cite{bCMS-eta}}}
  \put(852,212){\small{\cite{aamodt2010charged09}}}
  \put(825,183){\small{\cite{breakstone1984charged}}}
  \put(827,154){\small{\cite{bUA5}}}
  \put(892,125){\small{\cite{Nouicer:2004ke}}}
 \end{overpic}\\[-0.35cm]
 \caption{Charged-particle pseudorapidity density in the pseudorapidity region $|\eta| < 0.5$ ($\dndetainline$ at $\eta = 0$ calculated as the integral of the data over $|\eta| < 0.5$) for \inel, \nsd, and \inelgt collisions, as a function of the centre-of-mass energy. Lines indicate fits with a power-law dependence on $\sqrt{s}$. Grey bands represent the one standard deviation range. Data points at the same energy have been shifted horizontally for visibility. The LHC nominal centre-of-mass energy is indicated by a vertical line. Data other than from ALICE used in this figure are taken from references \cite{breakstone1984charged,bUA5,bCMS-eta,Thome:1977ky,Alpgard:1982zx,ambrosio1982total,Nouicer:2004ke,Abelev:2008ab,Alner:1987wb,Albajar:1989an,Abe:1989td}.\label{fig:dndeta-result-at-0-fit}}%
\end{figure}%
\subsection{Multiplicity distributions of primary charged particles: measurements}
\begin{figure}[!p]
 \centering
 \subfigure{\includegraphics[width=0.48\textwidth]{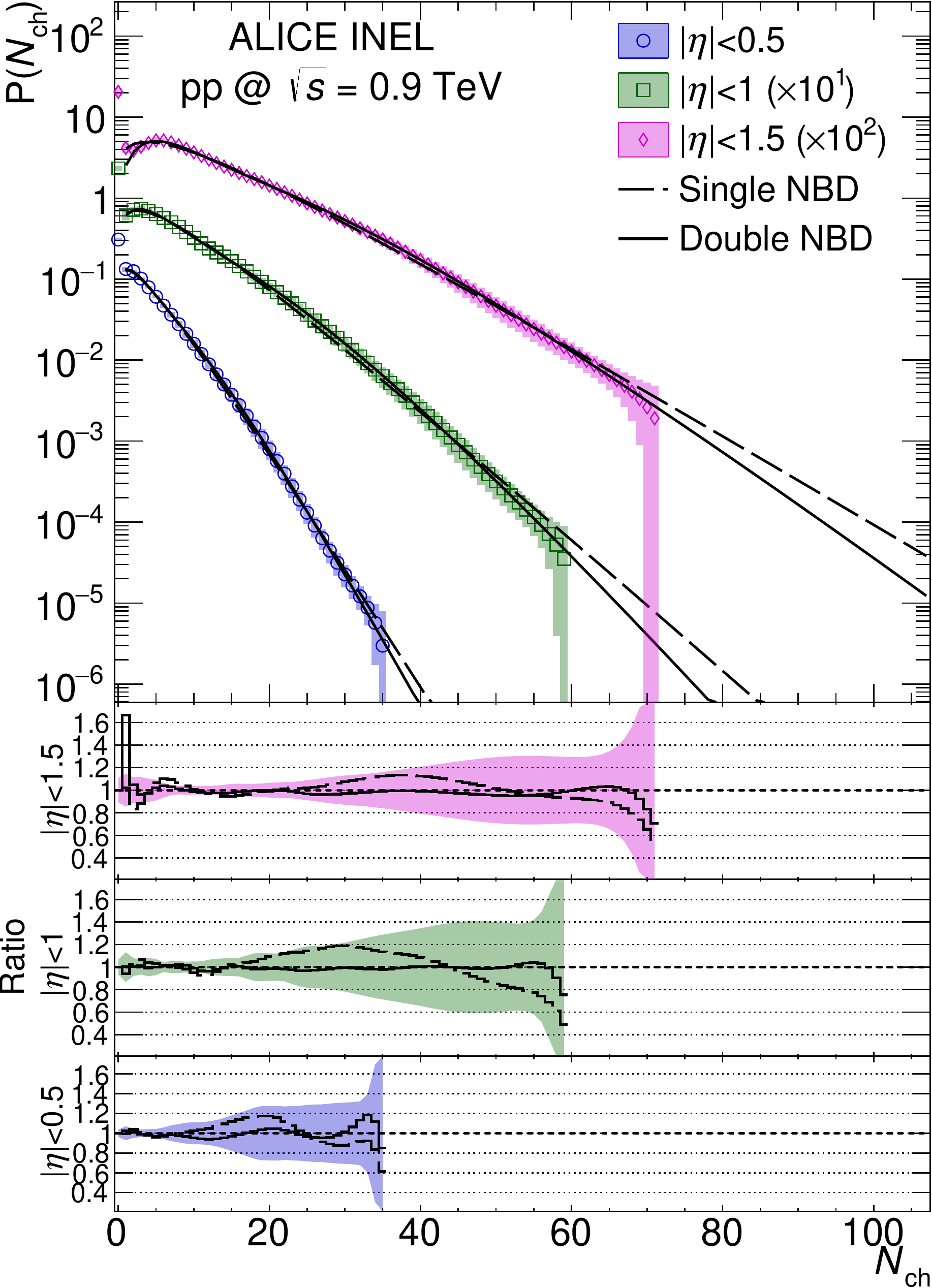}}\,%
 \subfigure{\includegraphics[width=0.48\textwidth]{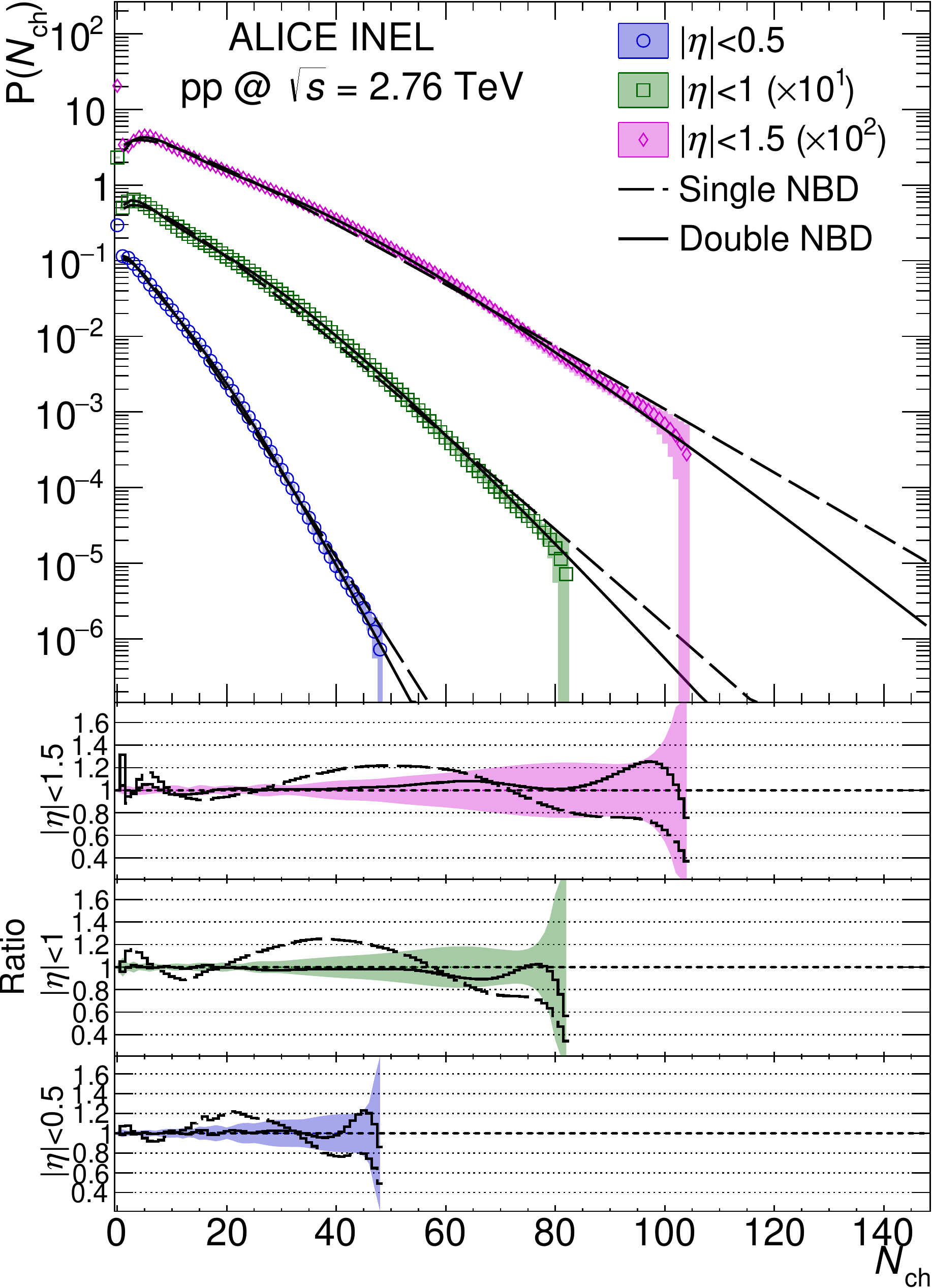}} \\[-0.7cm]
 \subfigure{\includegraphics[width=0.48\textwidth]{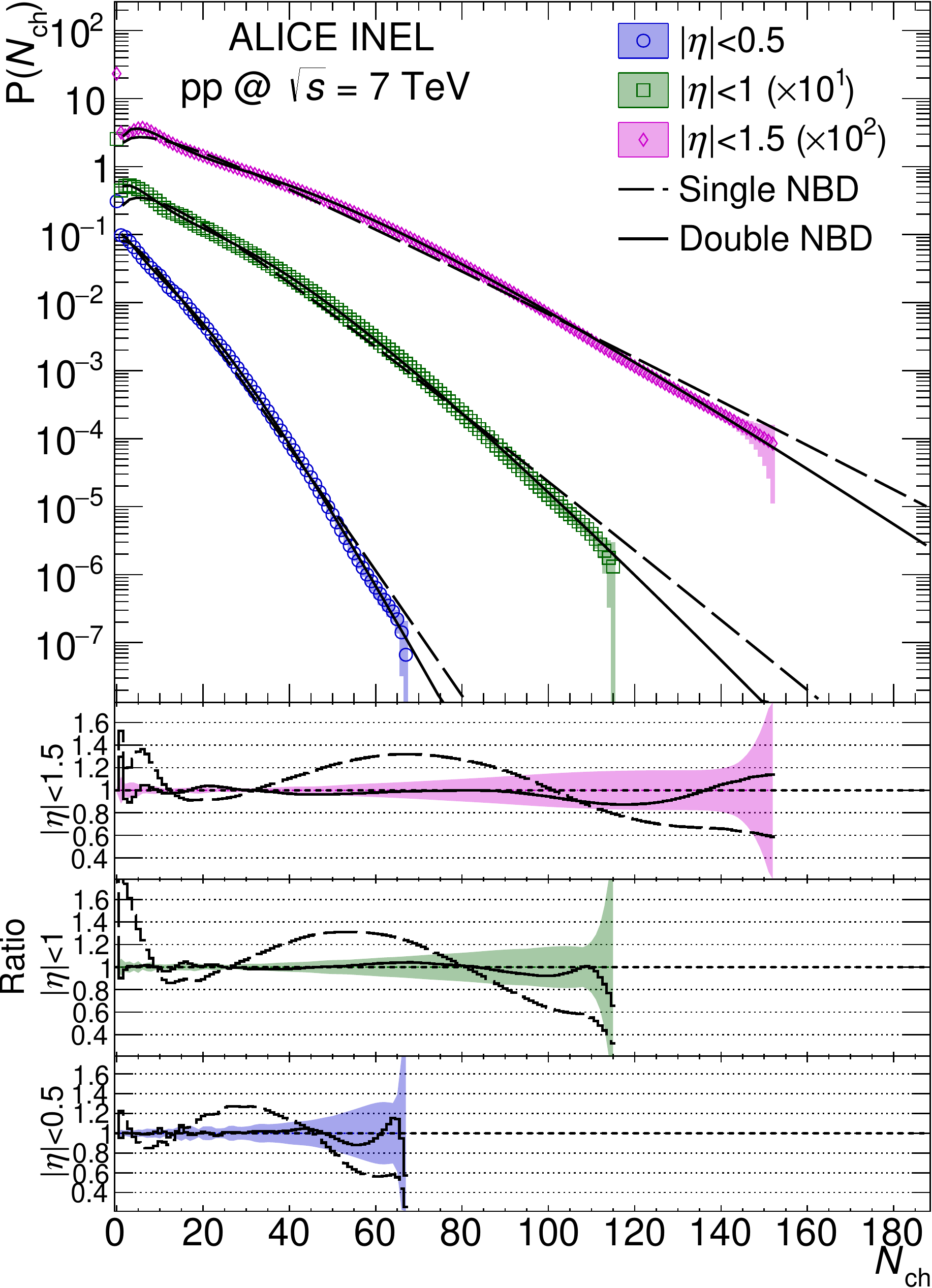}}\,%
 \subfigure{\includegraphics[width=0.48\textwidth]{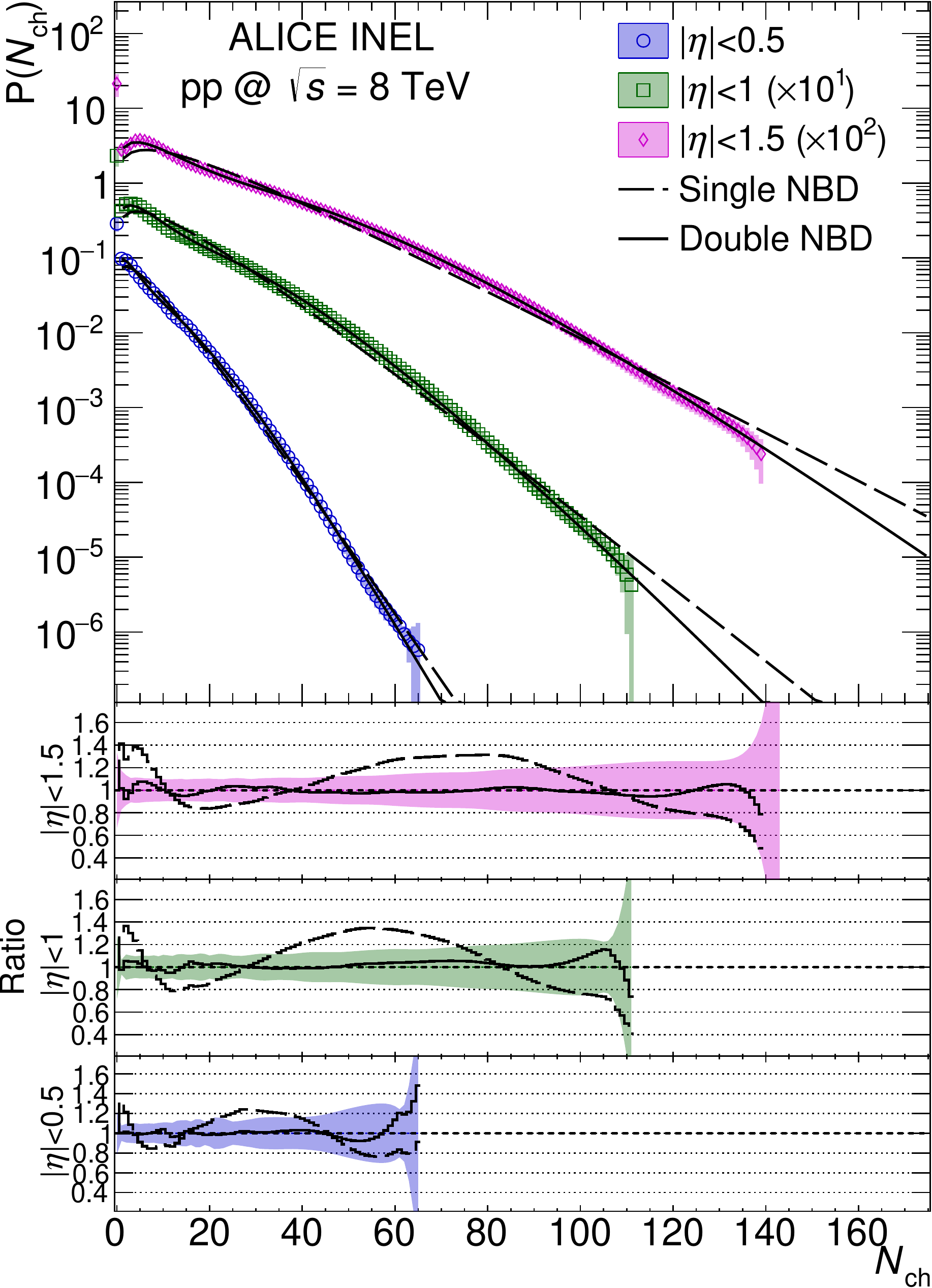}}\\[-0.4cm]%
 \caption{Measured multiplicity distributions in three pseudorapidity ranges for \inel events. The dashed and solid lines show the single and double NBD fits (see \cref{section:experimental-results:subsection:NBD-parameterization}). Shaded areas represent statistical and systematic uncertainties combined: (a) data at $\sqrt{s} = 0.9$~TeV (top left); (b) data at $\sqrt{s} = 2.76$~TeV (top right); (c) data at $\sqrt{s} = 7$~TeV (bottom left); (d) data at $\sqrt{s} = 8$~TeV (bottom right). Ratios of data to the fits are also shown, with shaded areas representing combined systematic and statistical uncertainties.\label{fig:pnch-results-INEL}}
\end{figure}

\begin{figure}[!p]
 \centering
 \subfigure{\includegraphics[width=0.48\textwidth]{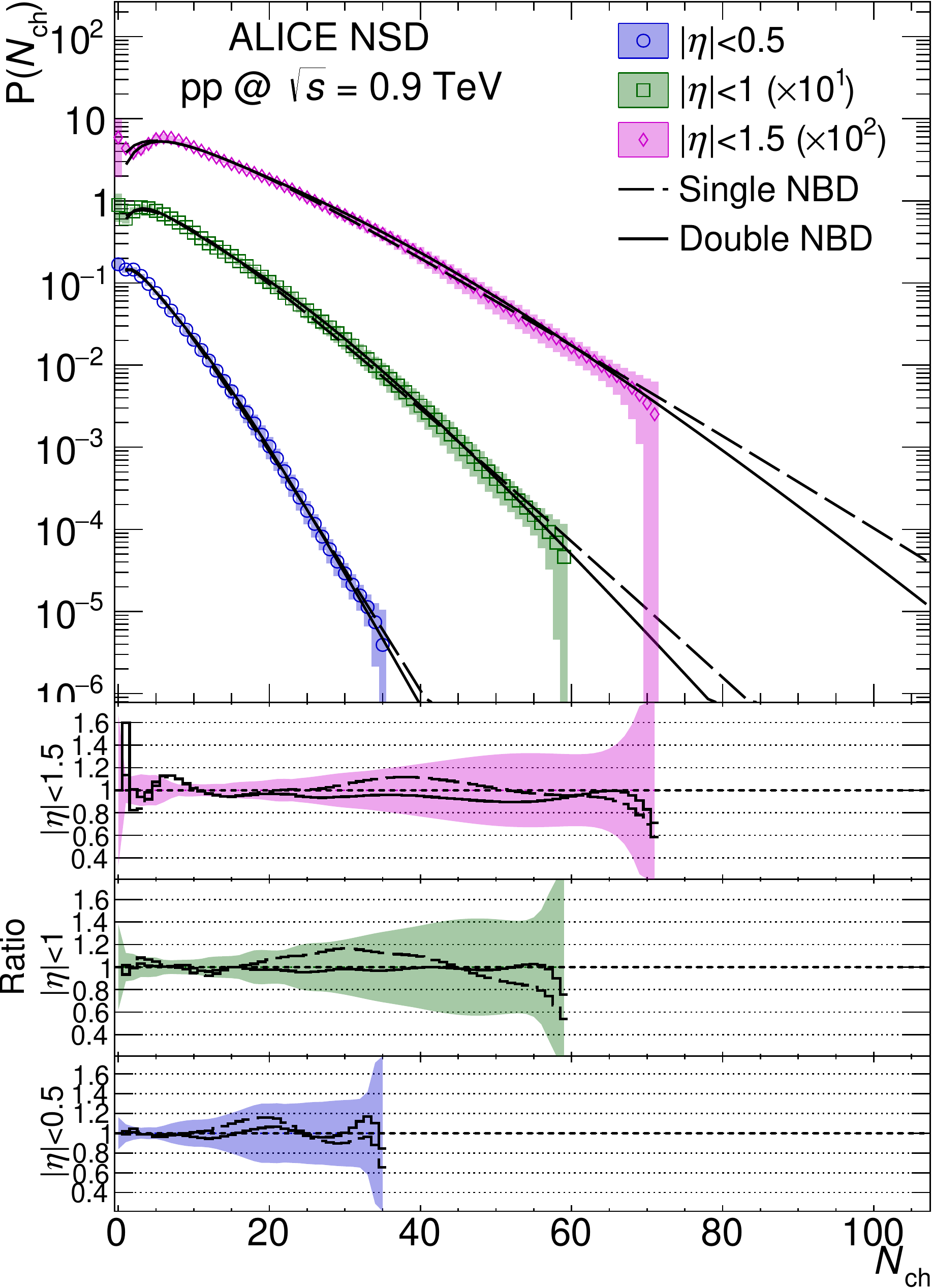}}\,%
 \subfigure{\includegraphics[width=0.48\textwidth]{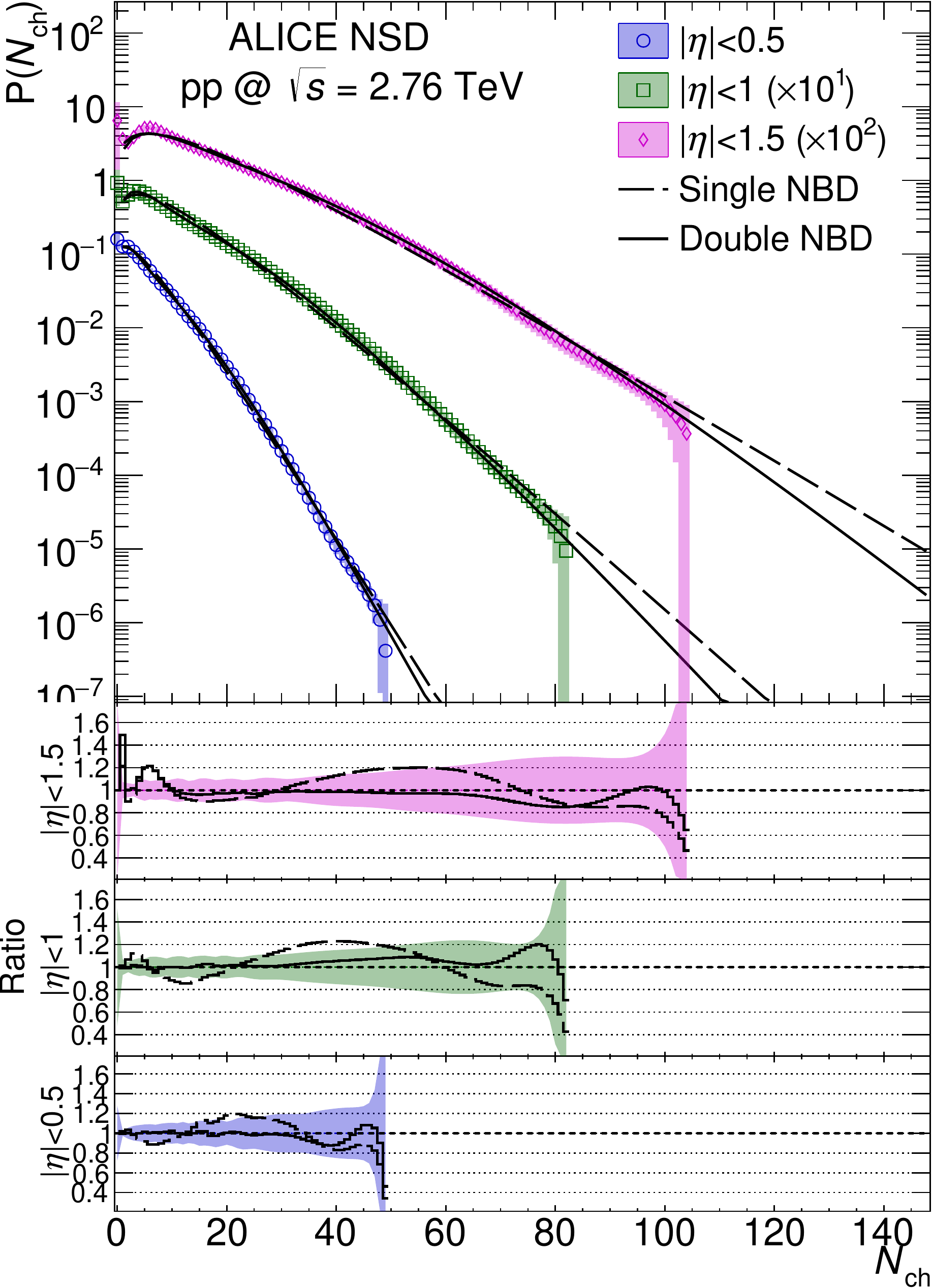}} \\[-0.7cm]
 \subfigure{\includegraphics[width=0.48\textwidth]{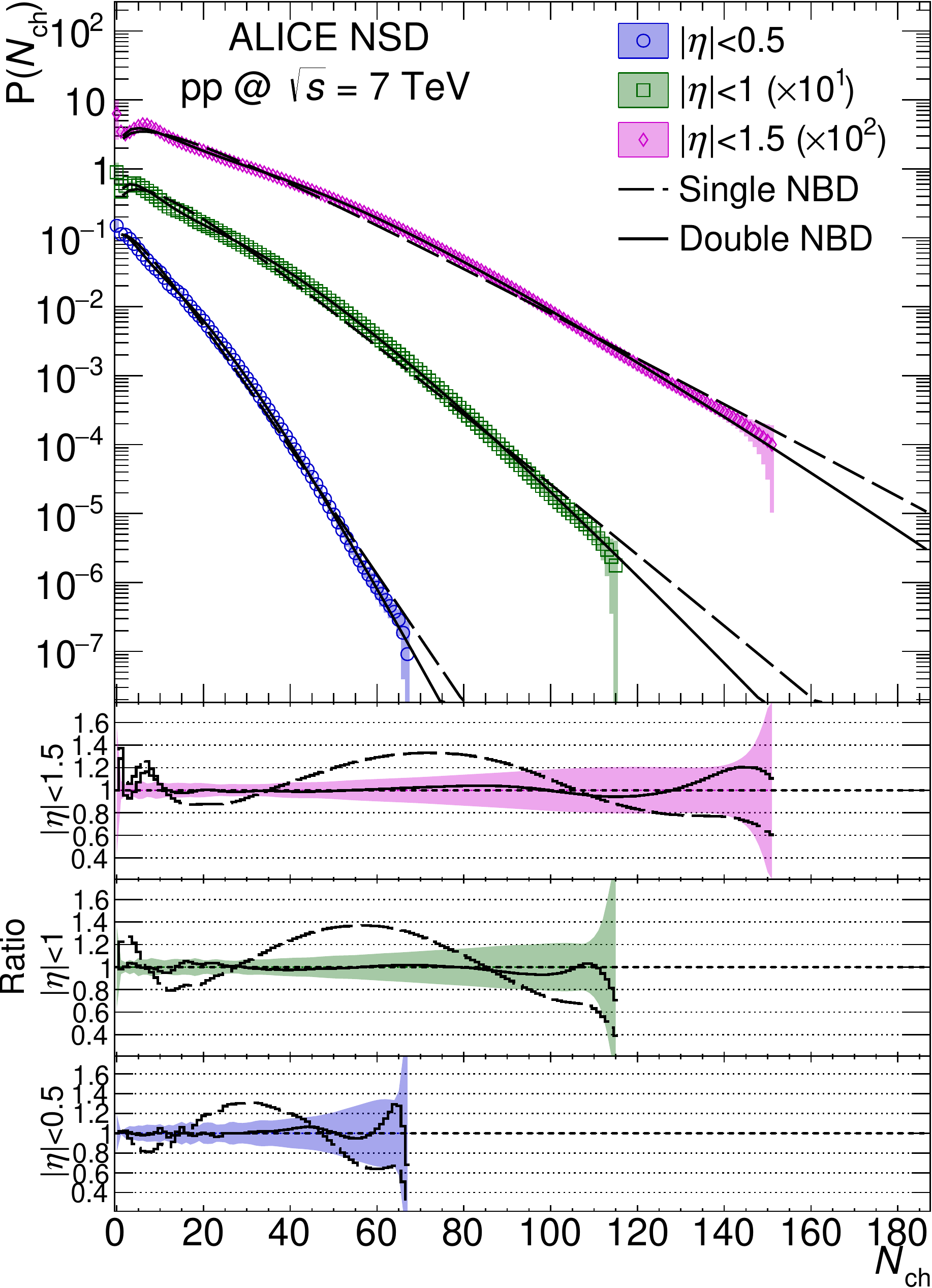}}\,%
 \subfigure{\includegraphics[width=0.48\textwidth]{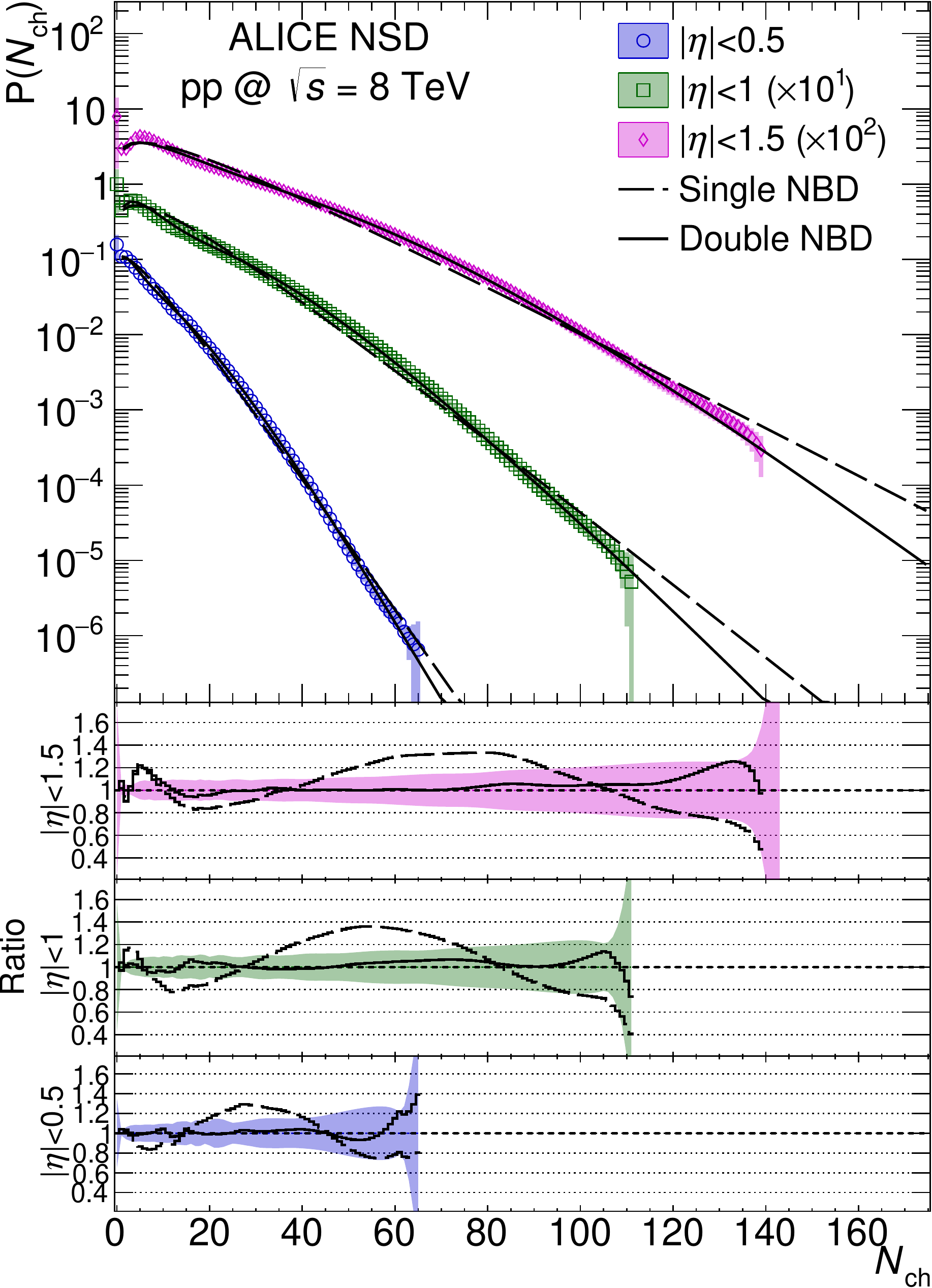}}\\[-0.4cm]%
 \caption{Measured multiplicity distributions in three pseudorapidity ranges for \nsd events. The dashed and solid lines show the single and double NBD fits (see \cref{section:experimental-results:subsection:NBD-parameterization}). Shaded areas represent statistical and systematic uncertainties combined: (a) data at $\sqrt{s} = 0.9$~TeV (top left); (b) data at $\sqrt{s} = 2.76$~TeV (top right); (c) data at $\sqrt{s} = 7$~TeV (bottom left); (d) data at $\sqrt{s} = 8$~TeV (bottom right). Ratios of data to the fits are also shown, with shaded areas representing combined systematic and statistical uncertainties.\label{fig:pnch-results-NSD}}
\end{figure}

\begin{figure}[!ht]
 \centering
 \subfigure{\includegraphics[width=0.48\textwidth]{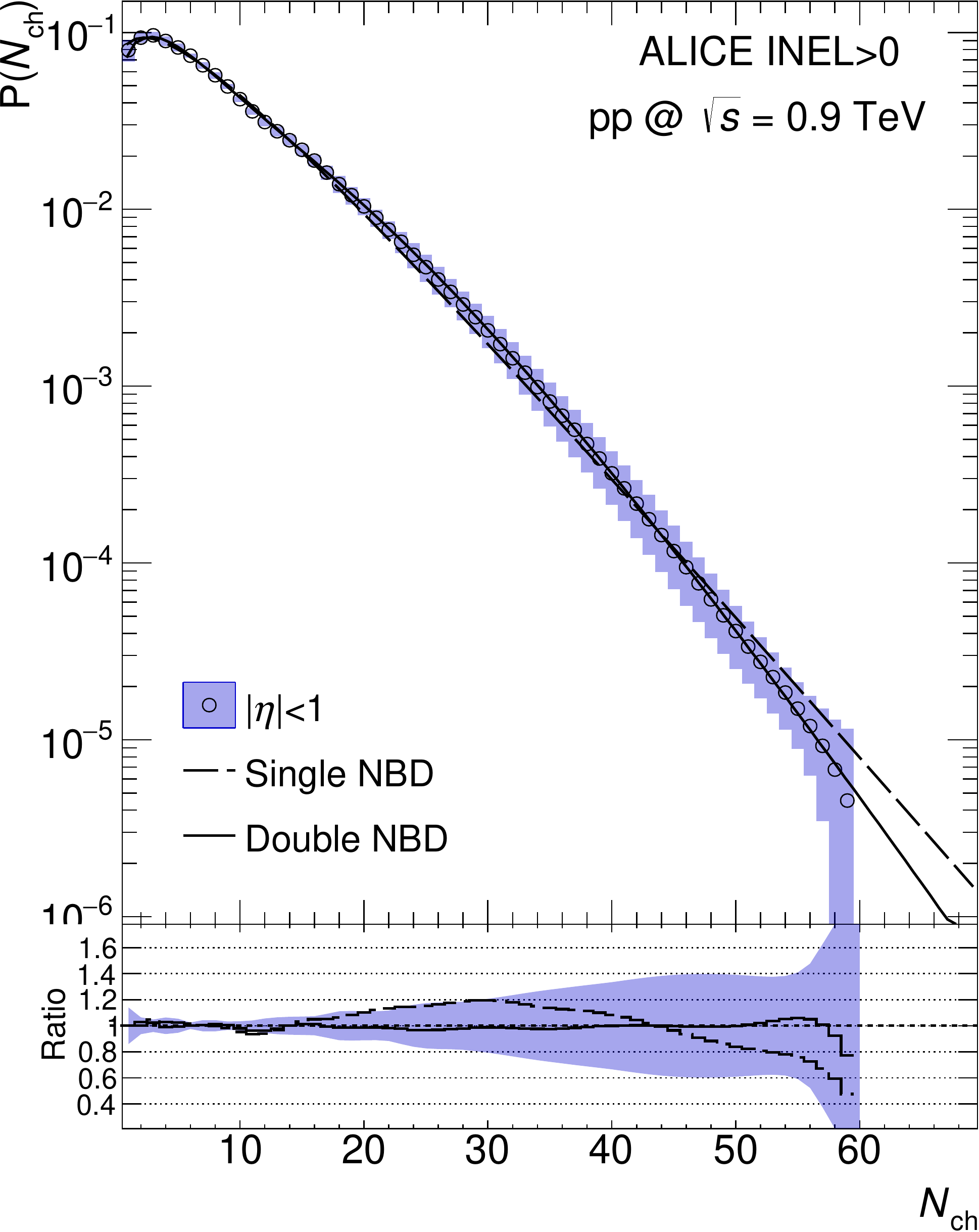}}\,%
 \subfigure{\includegraphics[width=0.48\textwidth]{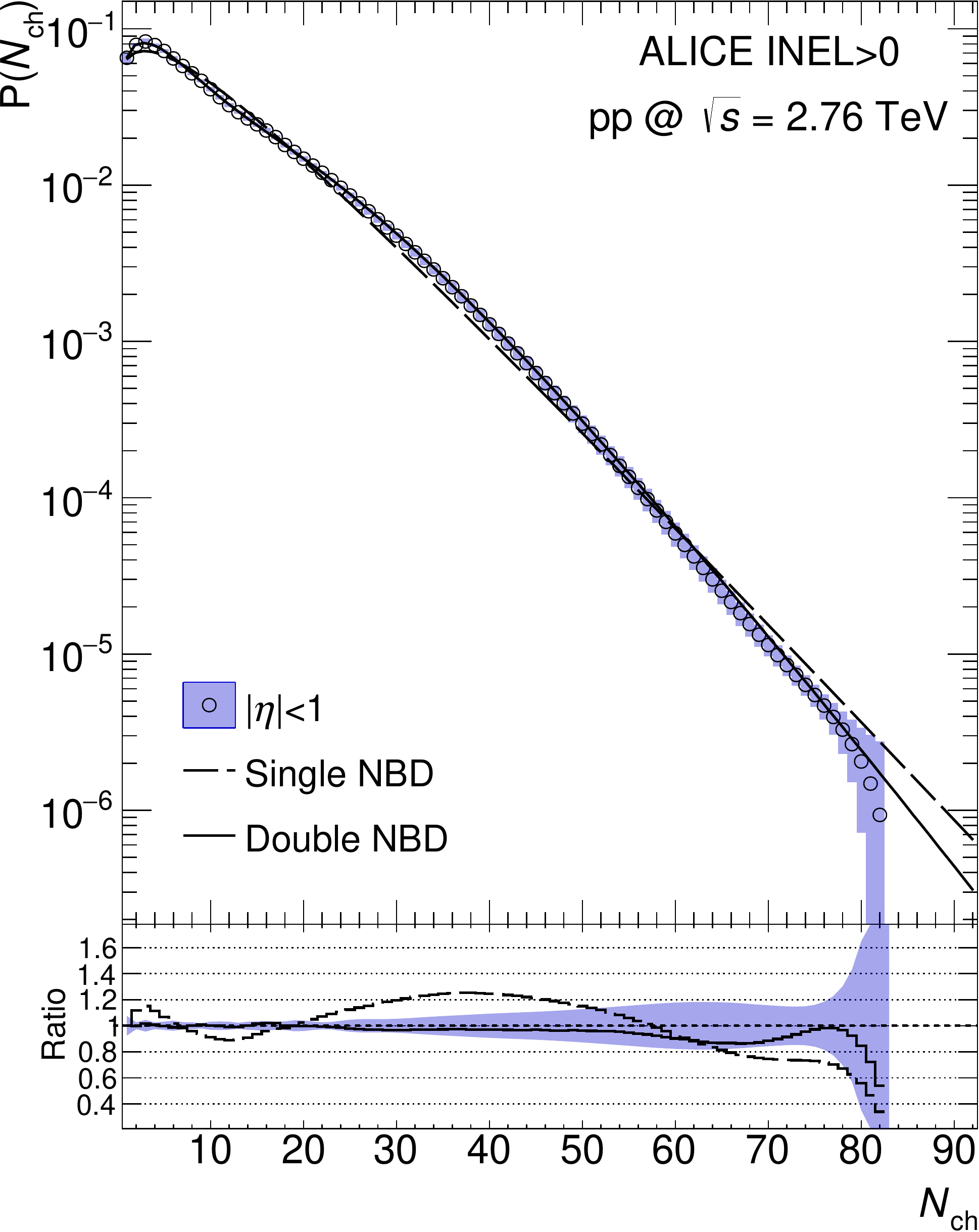}} \\[-0.7cm]
 \subfigure{\includegraphics[width=0.48\textwidth]{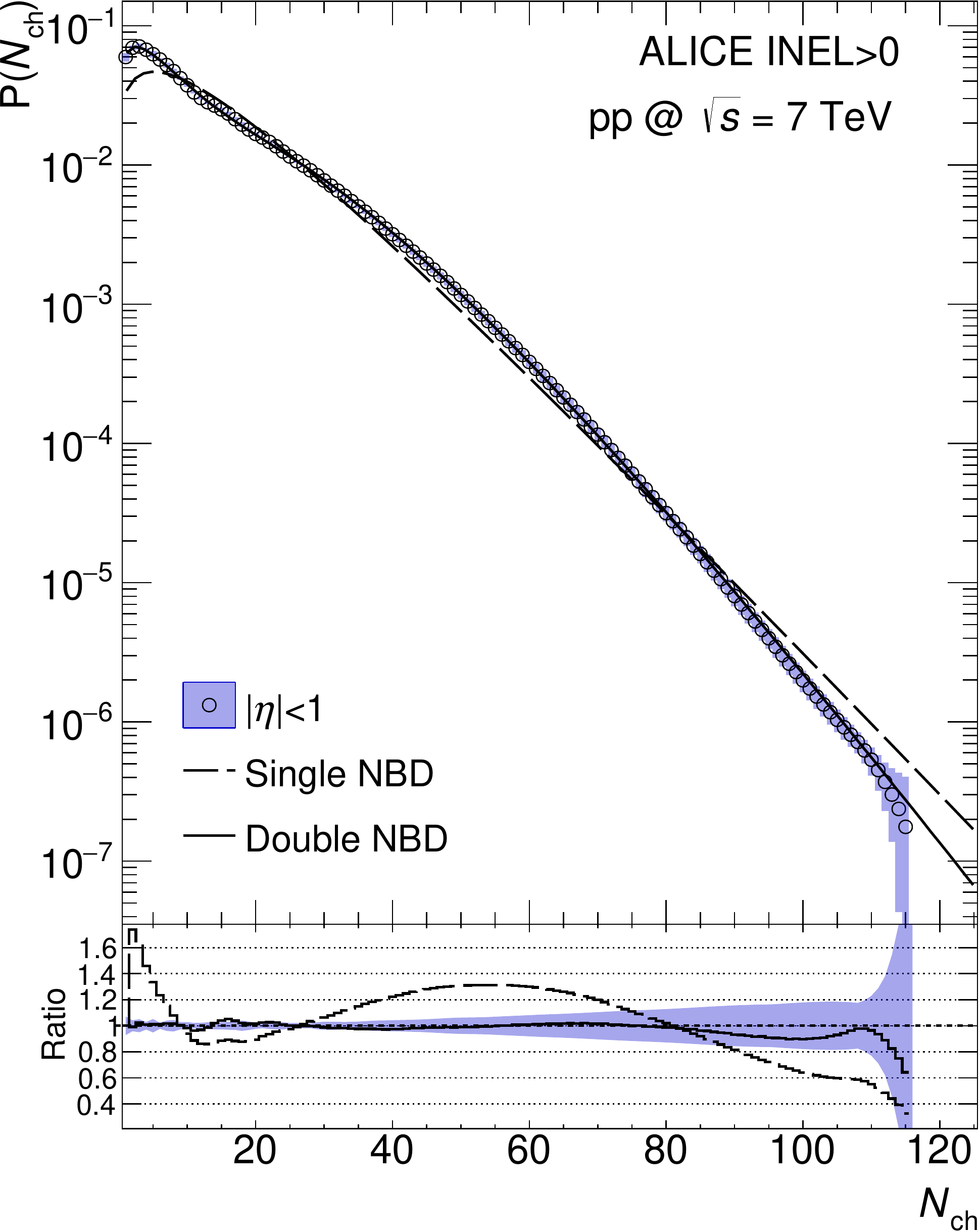}}\,%
 \subfigure{\includegraphics[width=0.48\textwidth]{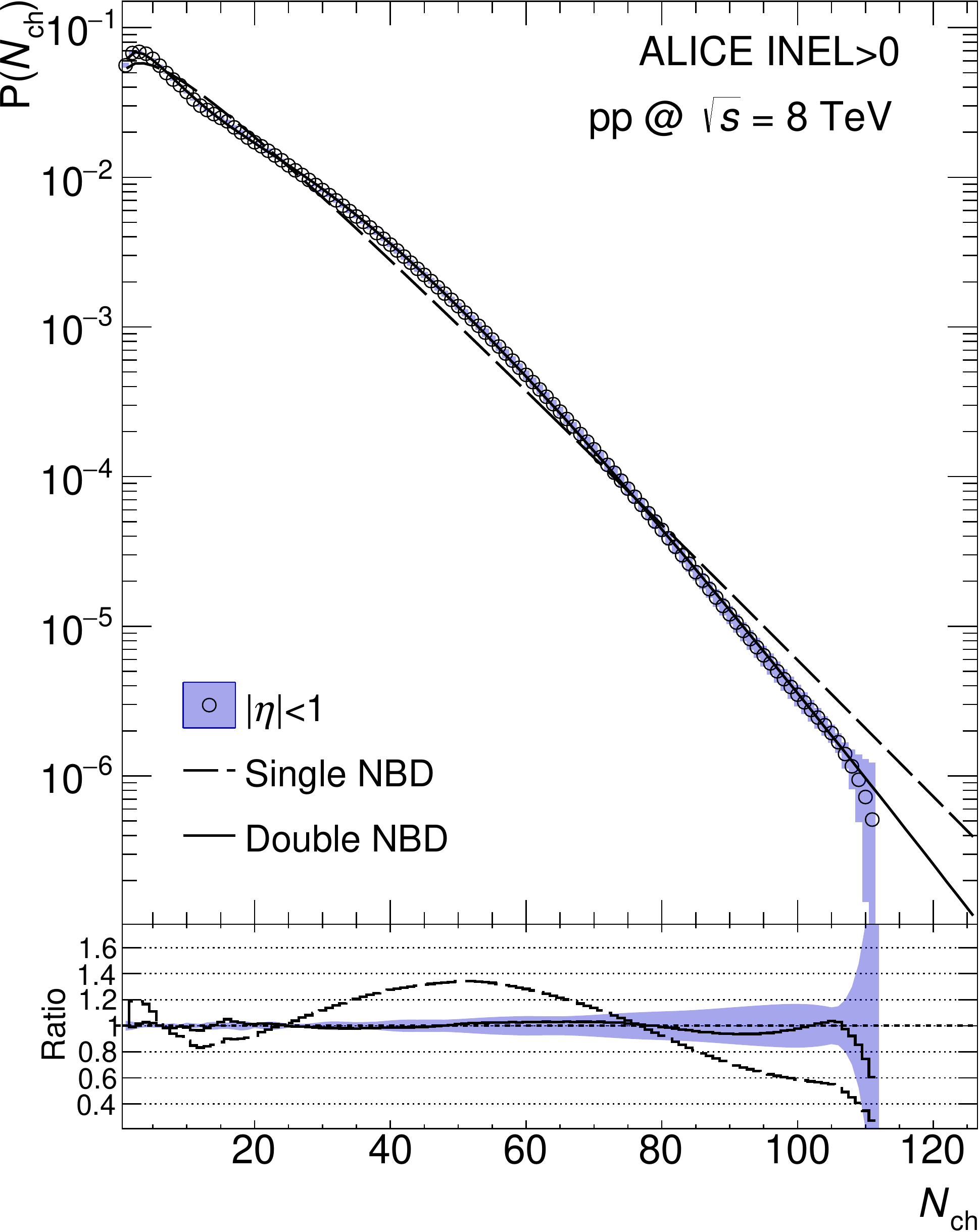}}\\[-0.4cm]%
 \caption{Measured multiplicity distributions in three pseudorapidity ranges for \inelgt events. The dashed and solid lines show the single and double NBD fits (see \cref{section:experimental-results:subsection:NBD-parameterization}). Shaded areas represent statistical and systematic uncertainties combined: (a) data at $\sqrt{s} = 0.9$~TeV (top left); (b) data at $\sqrt{s} = 2.76$~TeV (top right); (c) data at $\sqrt{s} = 7$~TeV (bottom left); (d) data at $\sqrt{s} = 8$~TeV (bottom right). Ratios of data to the fits are also shown, with shaded areas representing combined systematic and statistical uncertainties.\label{fig:pnch-results-INELgt}}
\end{figure}

The results of ALICE measurements of multiplicity distributions of charged primary particles are displayed as probability distributions ($\pnch$) in Figures \ref{fig:pnch-results-INEL} (\inel), \ref{fig:pnch-results-NSD} (\nsd) and \ref{fig:pnch-results-INELgt} (\inelgt). For the first two event classes the measurements were obtained in three pseudorapidity intervals $|\eta| <$ 0.5, 1 and 1.5, and for \inelgt in $|\eta| < 1$. At $\sqrt{s} = 7$~TeV, $\pnch$ varies over 6 to 7 orders of magnitude and the multiplicity range reaches up to 160 in $|\eta| < 1.5$ for both \inel and \nsd event classes. In $|\eta| < 0.5$ and $|\eta| < 1$, the observed multiplicity reaches 10 times the mean multiplicity. It is expected that the average energy density in proton collisions at the LHC, at $\sqrt{s} = 14$~TeV, is about 5 to 15 times smaller than energy densities reached in gold ions at RHIC \cite{Csanad:2013lba}. Therefore, in proton-proton collisions of multiplicity exceeding 10 times the average multiplicity, energy densities should overlap with those of heavy ion collisions at RHIC, allowing to compare properties of systems with very different collision volumes (two to three orders of magnitude) but the same energy density. Future runs of the LHC should allow extending much further the range of multiplicities probed so far.

\begin{figure}[!ht]
 \centering
 \subfigure{\includegraphics[width=0.5\textwidth]{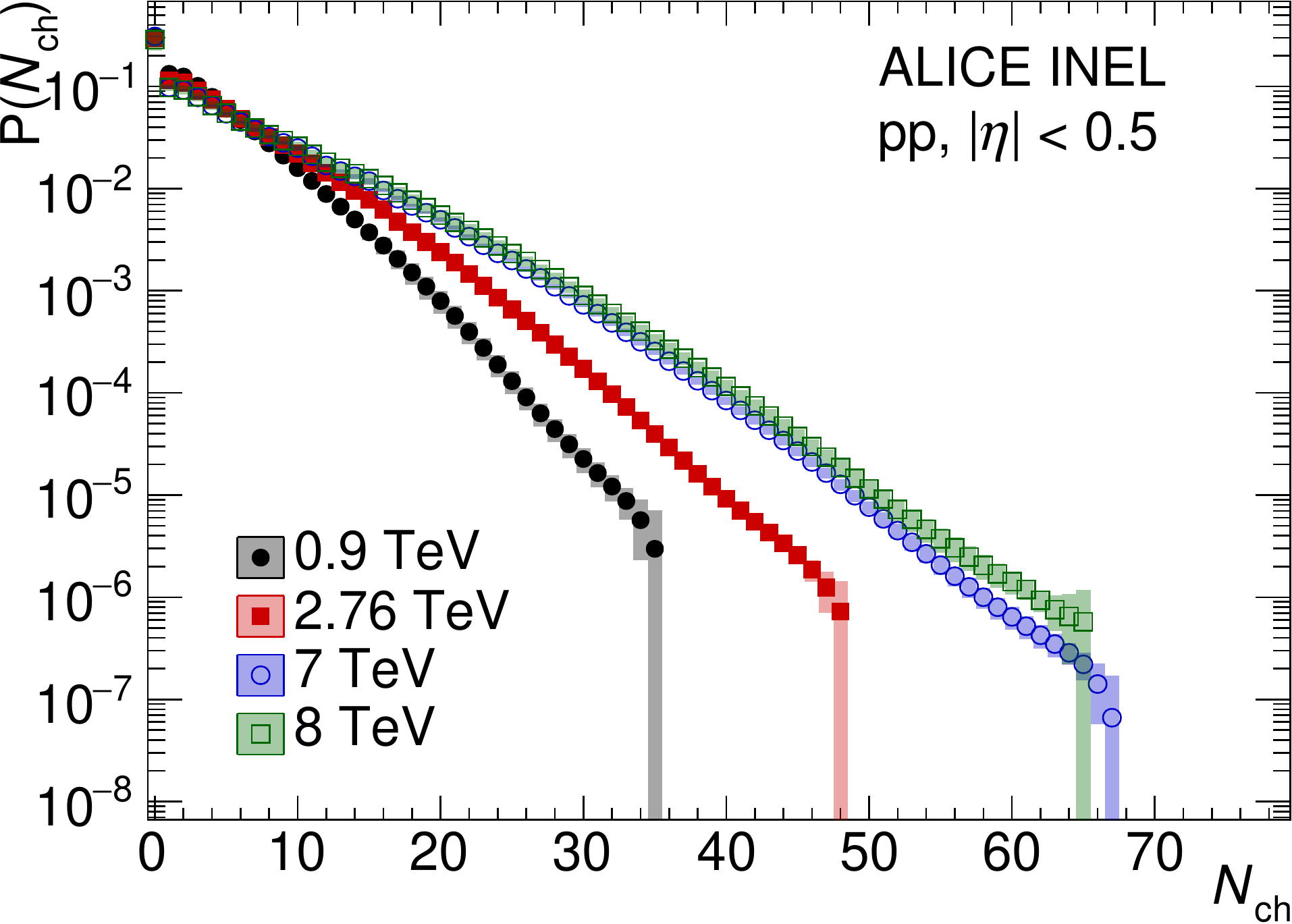}}\,%
 \subfigure{\includegraphics[width=0.5\textwidth]{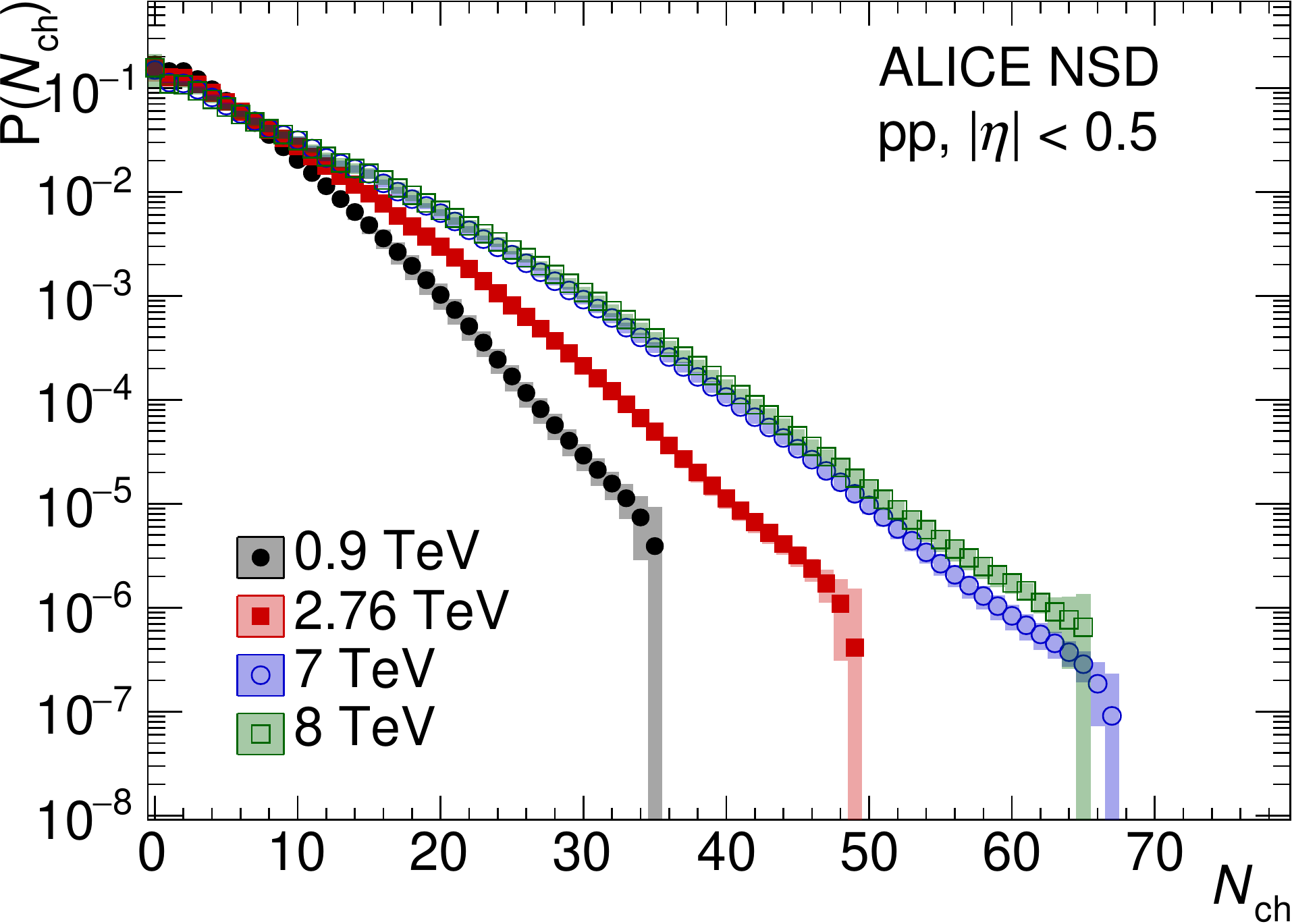}} \\[-0.3cm]
 \subfigure{\includegraphics[width=0.5\textwidth]{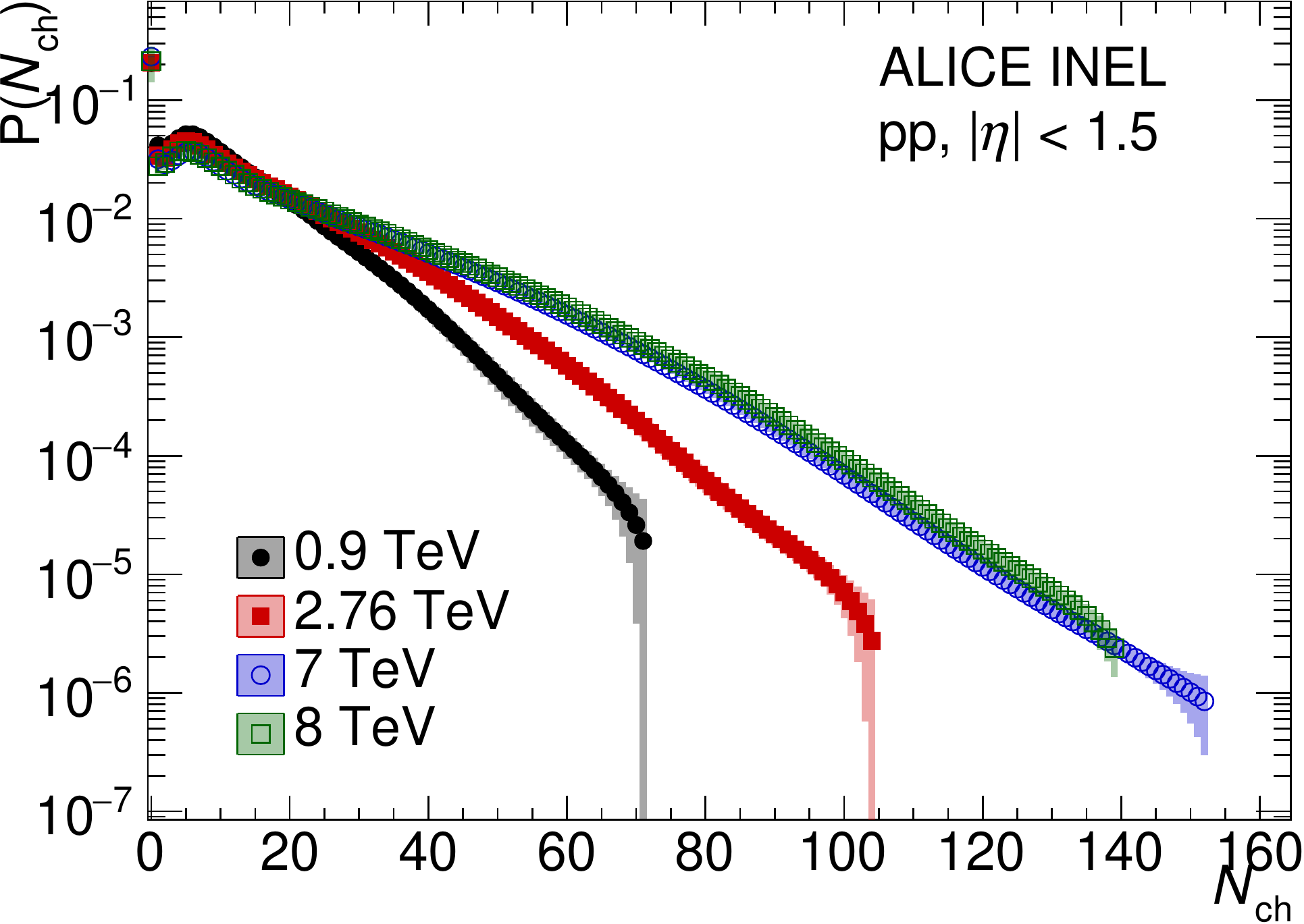}}\,%
 \subfigure{\includegraphics[width=0.5\textwidth]{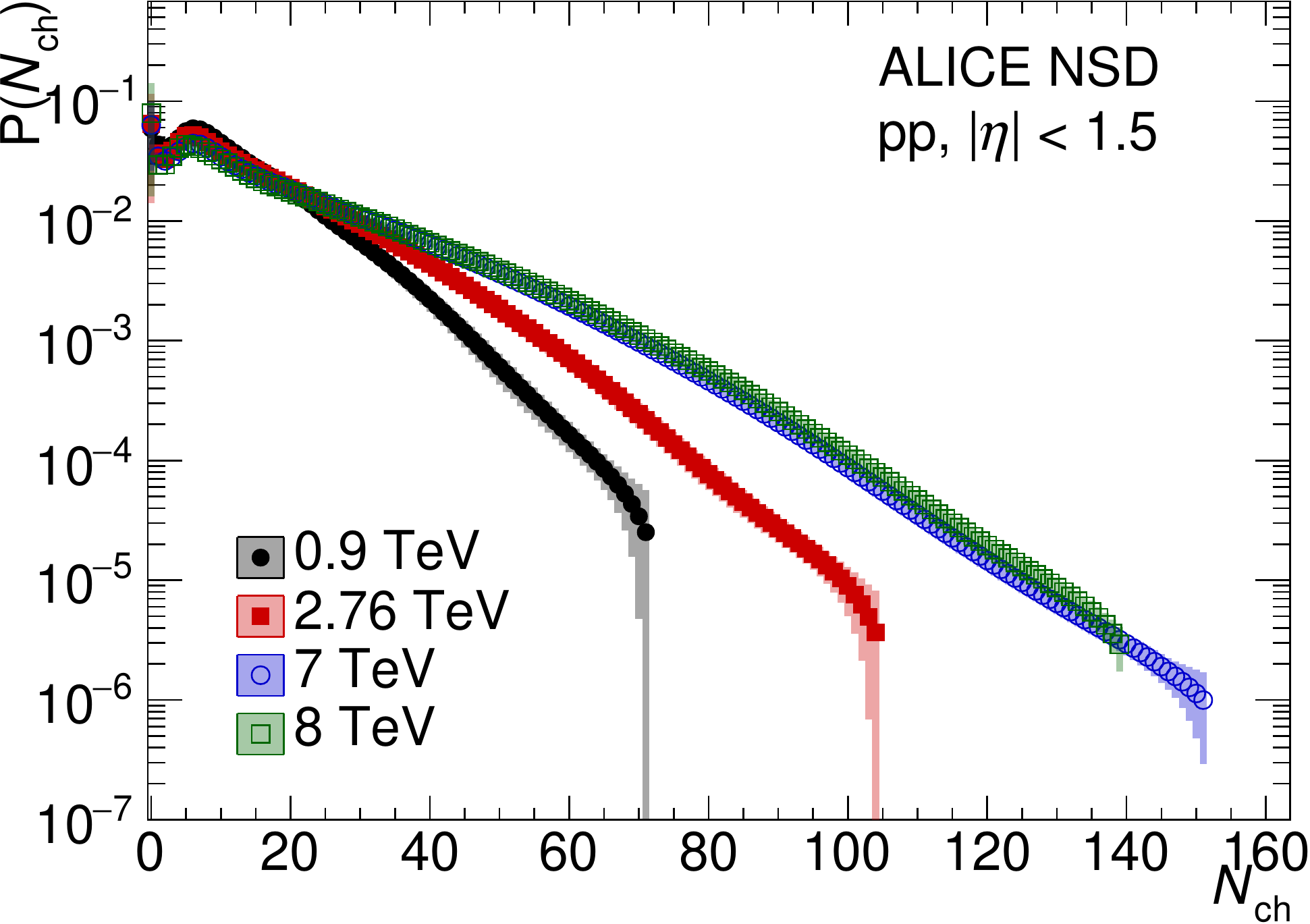}}\\[-0.4cm]%
 \caption{Evolution of measured multiplicity distributions as a function of centre-of-mass energy (from 0.9 to 8~TeV), for \inel and \nsd event classes and for $|\eta| < 0.5$ (top row) and $|\eta| < 1.5$ (bottom row).\label{fig:pnch-results-energy-evolution}}
\end{figure}

The high-multiplicity tail of the distribution increases as expected with increasing energy (\cref{fig:pnch-results-energy-evolution}). This behaviour is studied quantitatively in \cref{section:experimental-results:subsection:KNO-studies} on KNO scaling and $q$-moment analysis.

The measurements presented in this publication are consistent with previous ALICE data, for INEL \cite{aamodt2010charged09} at $\sqrt{s} = 0.9$~TeV and \inelgt \cite{aamodt2010charged7} at $\sqrt{s} = 7$~TeV, in the multiplicity range where they overlap (\cref{fig:pnch-results-experiments-comparison}). The wavy structure already observed by ALICE in \cite{aamodt2010charged09,aamodt2010charged7}, for multiplicities above $\nch = 25$, and previously by UA5 \cite{Ansorge:1988kn}, is still hardly significant, and it is not present in the raw data. This feature was also observed in a study of CMS data \cite{bPremGosh}. However, applying the same procedure to Monte Carlo data produces similar structures in the unfolded distribution, while no oscillation was present at particle level. We conclude that this structure is an unfolding procedure artifact. The period of the structure is related to response matrix width.
\subsection{Comparison of multiplicity distributions with other experiments and models}
\label{section:experimental-results:subsection:subsection:comparison-to-other}
\begin{figure}[!p]
 \centering
 \subfigure{
 \begin{overpic}[width=0.47\textwidth]{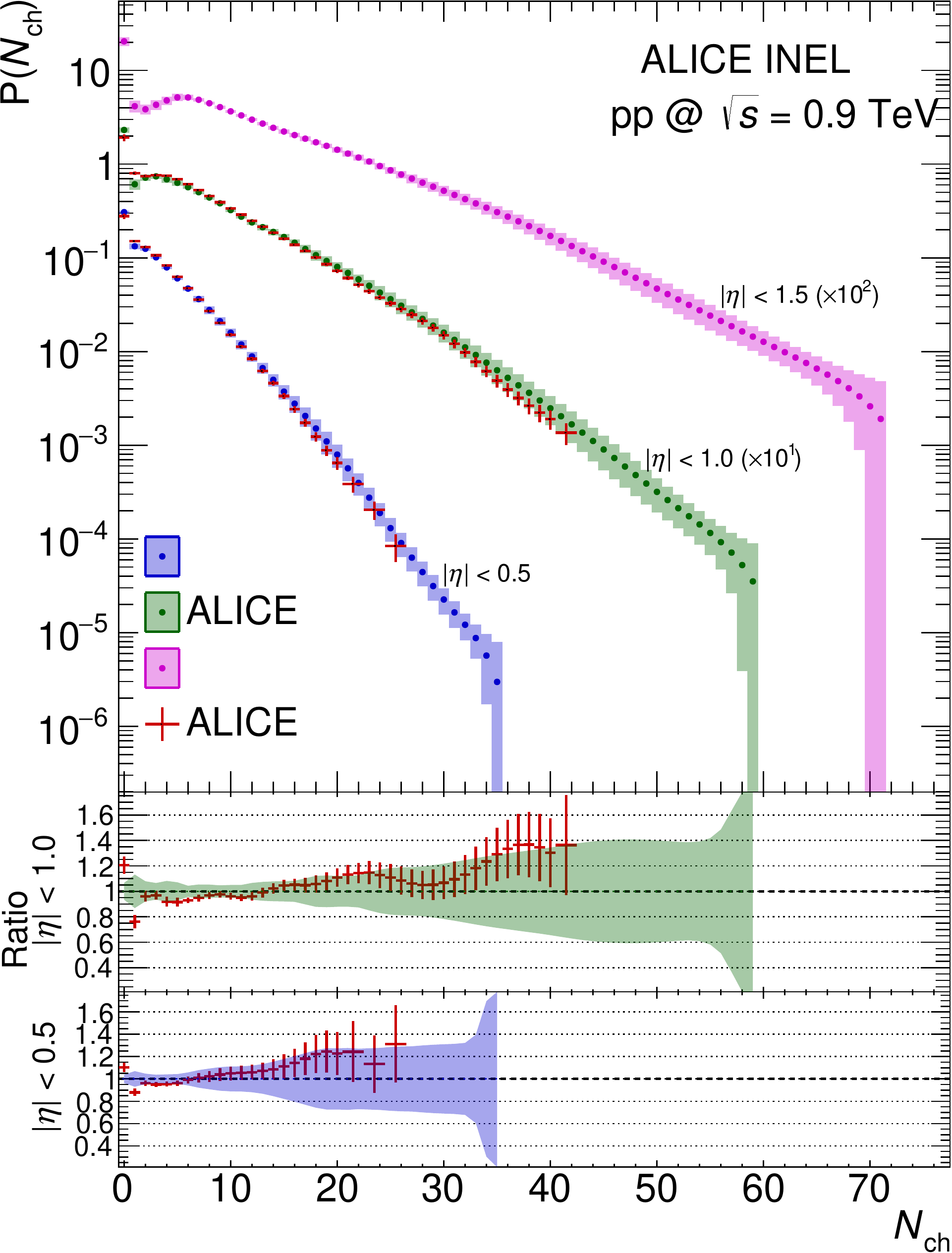}
  \put(240,413){{\cite{aamodt2010charged09}}}
 \end{overpic}
 }%
 \subfigure{
 \begin{overpic}[width=0.47\textwidth]{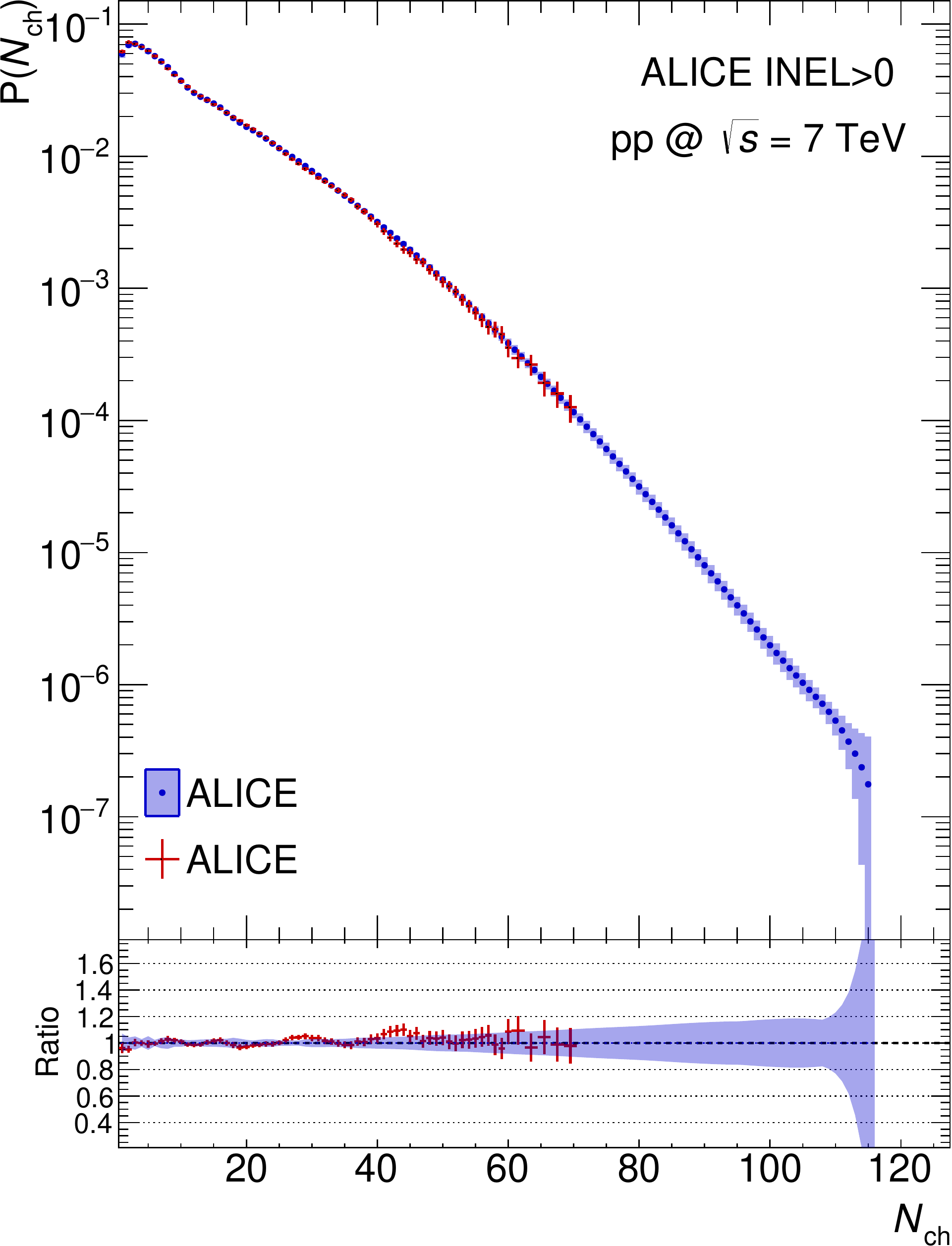}
  \put(242,301){{\cite{aamodt2010charged7}}}
 \end{overpic}
 } \\[-0.7cm]
 \subfigure{
 \begin{overpic}[width=0.47\textwidth]{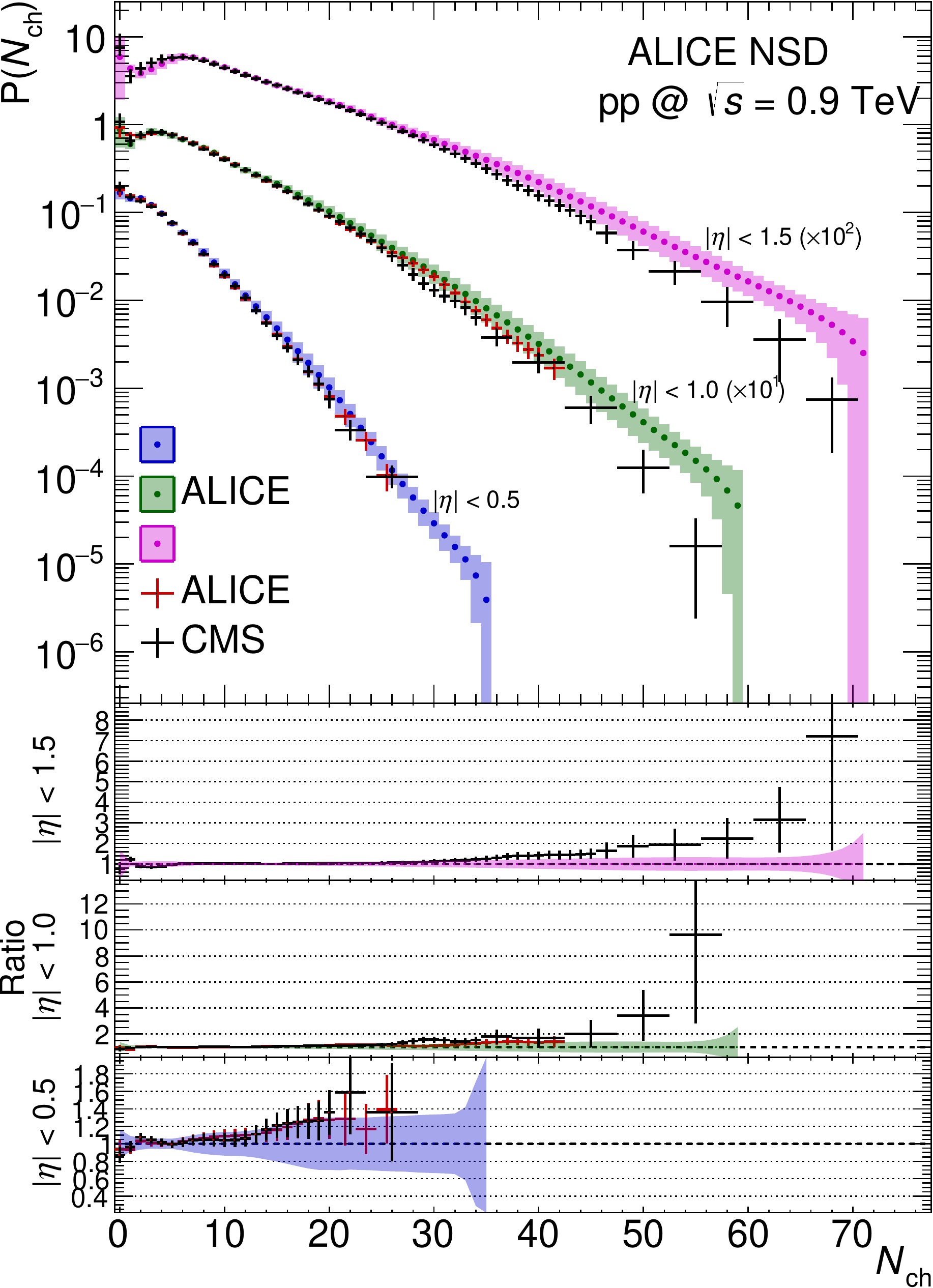}
  \put(227,528){{\cite{aamodt2010charged09}}}
  \put(210,492){{\cite{Khachatryan:2010nk}}}
 \end{overpic}
 }%
 \subfigure{
 \begin{overpic}[width=0.47\textwidth]{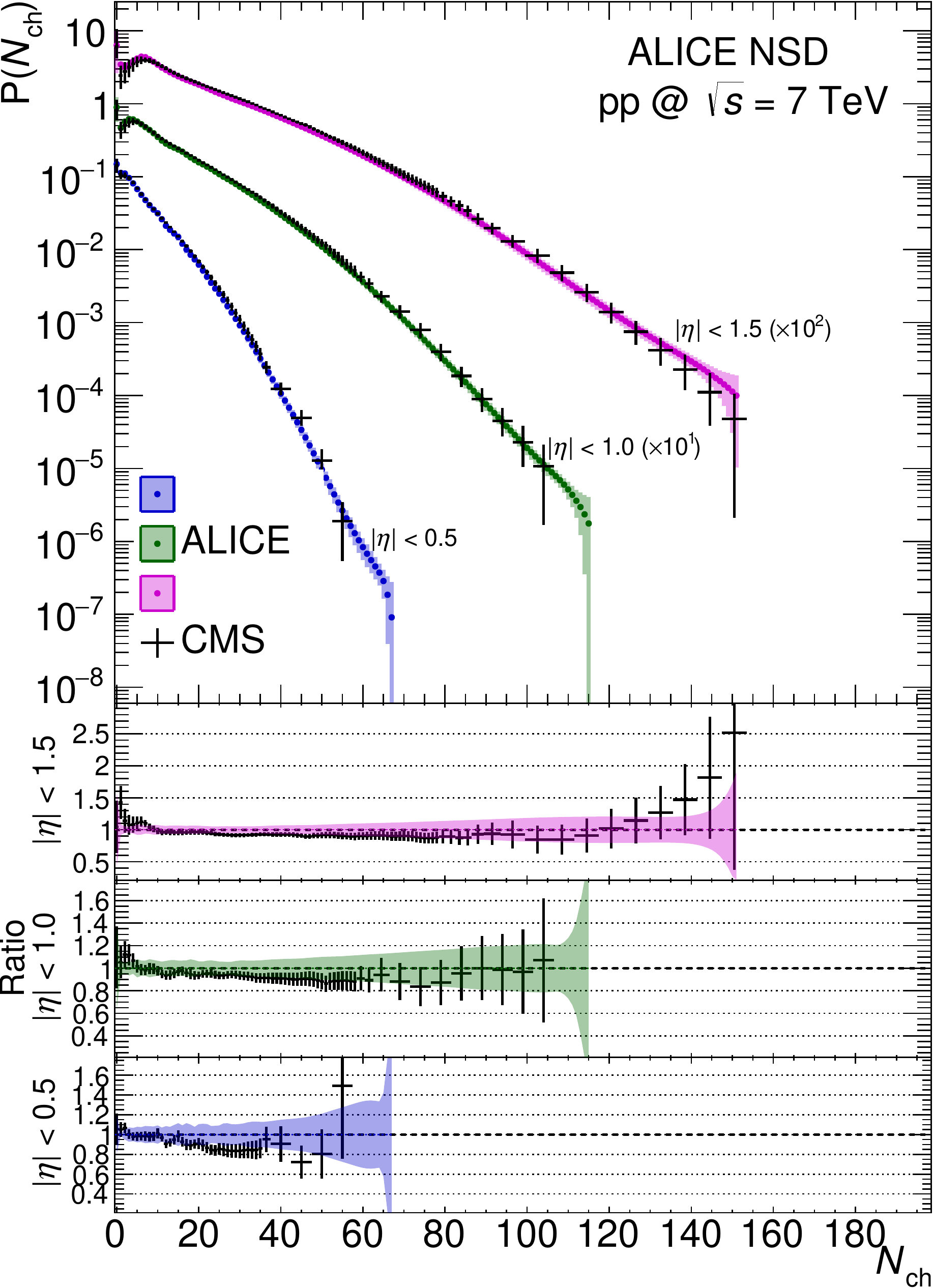}
   \put(208,495){{\cite{Khachatryan:2010nk}}}
 \end{overpic}
 }\\[-0.4cm]%
 \caption{Comparison with ALICE previous publications \cite{aamodt2010charged09,aamodt2010charged7}, for the INEL event class, at $\sqrt{s} = 0.9$~TeV (top left) and the \inelgt event class at $\sqrt{s} = 7$~TeV (top right); in both cases, ratios between ALICE new data and ALICE previous data are also shown. The total uncertainties are shown as error bars for the previous data and as a band for the present measurement. For the \nsd event class, comparison with ALICE previous  publication \cite{aamodt2010charged09} and with CMS data at $\sqrt{s} = 0.9$~TeV \cite{Khachatryan:2010nk} (bottom left), and comparison with CMS data at  $\sqrt{s} = 7$~TeV \cite{Khachatryan:2010nk} (bottom right); ratios of the NBD fits of ALICE data taken without errors to CMS data, for the various $\eta$ intervals (indicated on the figure), are also shown. Error bars represent the contributions of the CMS errors to the ratios, the bands represent the ALICE total uncertainty assigned to the ratio of 1.\label{fig:pnch-results-experiments-comparison}}
\end{figure}

CMS data are available for the NSD normalization only. At $\sqrt{s} =$ 0.9 and 7~TeV, in the three pseudorapidity intervals where they could be compared, the ALICE multiplicity measurements are generally in agreement with CMS data \cite{Khachatryan:2010nk} (\cref{fig:pnch-results-experiments-comparison}). However, at $\sqrt{s} = 0.9$~TeV, above $\nch \approx 40$ in $|\eta| < 1$ and above $\nch \approx 50$ in $|\eta|< 1.5$, the ALICE multiplicity distributions tend to be higher than CMS data. The difference is not significant in each isolated bin, however the general trend seems to be different. As the bin-to-bin correlations in CMS{\textquoteright}s results are unknown, it is very difficult to obtain reliable unbiased quantitative comparison. A possible source of the discrepancy is the different treatment of single-diffractive events in the two analyses making the NSD event sample definitions not strictly compatible. More precisely, there was no diffraction tuning in the simulations used by CMS and, moreover, ALICE{\textquoteright}s criteria for events to be considered single-diffractive (see \cite{alice-crosssection-diffraction}) include a fixed cut on diffractive mass that differs from the value that CMS used. This explanation is supported by the fact that the difference is insignificant at $\sqrt{s} = 7$~TeV, where single-diffractive contribution is negligible. The comparison with CMS data is further discussed in the \cref{section:experimental-results:subsection:NBD-parameterization}, where NBD fits to the  data are reported.

\begin{figure}[!ht]
 \centering
 \includegraphics[width=\textwidth]{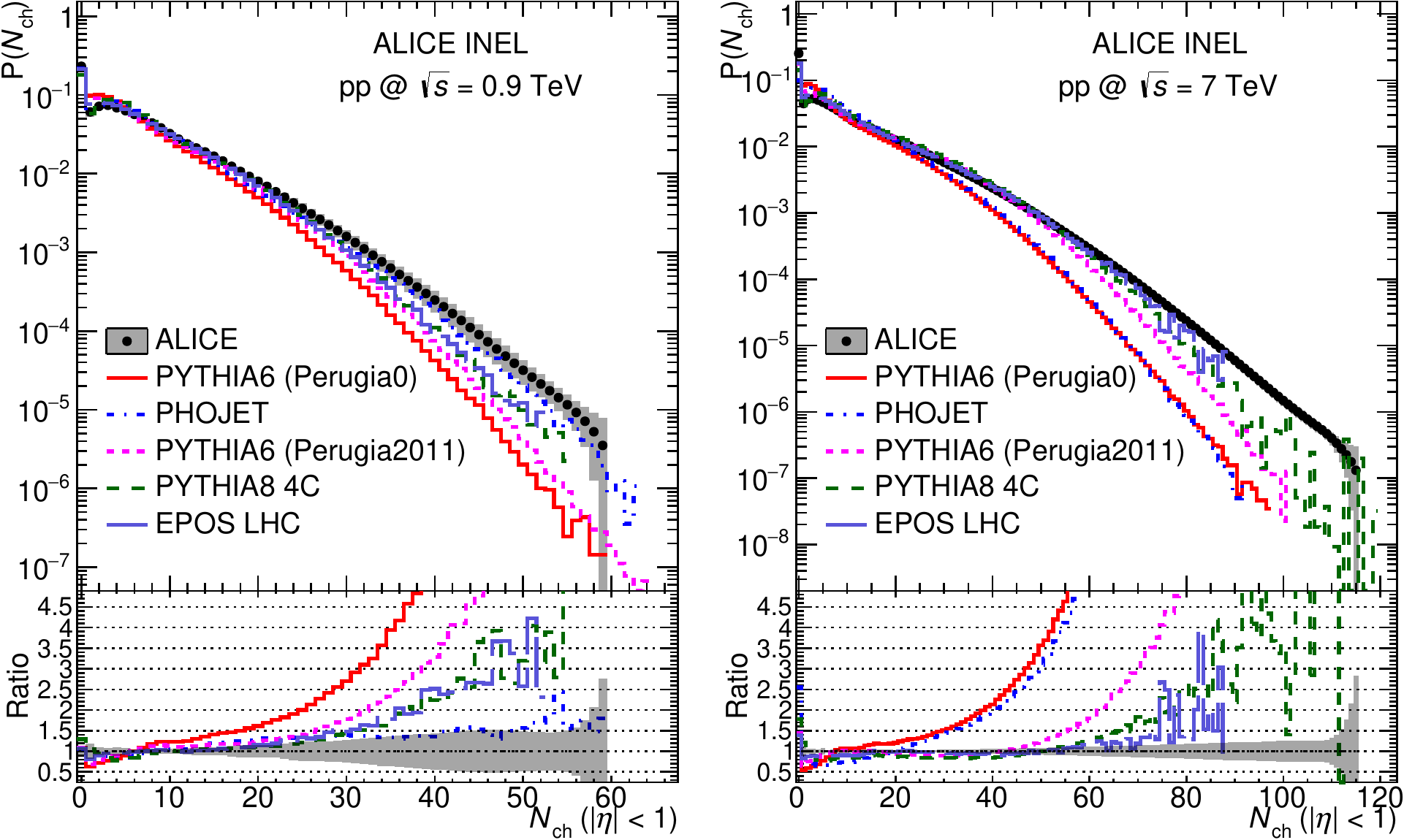}
 \caption{Comparison with models of measured multiplicity distributions for the \inel event class in the pseudorapidity range $|\eta| < 1.0$: ALICE data (black circles with grey bands), \pythia{6} \cite{Skands:2009zm} tune Perugia0 (red continuous line), \phojet \cite{Engel:1995sb} (blue dot-dashed line), \pythia{6} Perugia 2011 (pink dashed line), \pythia{8} 4C \cite{Sjostrand:2007gs,Corke:2010yf} (green dashed line), and EPOS LHC \cite{pierog2013epos} (long dashed light blue line). The shaded areas represent total uncertainties: comparison at 0.9 (left) and at 7~TeV (right). The ratios between measured values and model calculations are shown in the lower parts of the figures with the same convention. \label{fig:pnch-results-model-comparison}}
\end{figure}

Measured multiplicity distributions are compared to models (\cref{fig:pnch-results-model-comparison}), for two energies, the lowest one ($\sqrt{s} = 0.9$~TeV), and the energy at which there is the largest event sample ($\sqrt{s} = 7$~TeV), for the \inel event class, in the pseudorapidity range $|\eta| < 1$.

At $\sqrt{s} = 0.9$~TeV, \pythia{6} Perugia0 fails to reproduce the data. \pythia{6} Perugia 2011, \pythia{8} 4C and EPOS LHC underestimate the multiplicity distribution at high  multiplicities ($\nch \gtrsim 25$). \phojet provides the best match to the data over the observed multiplicity range, up to $\nch \approx 60$.

At $\sqrt{s} = 7$~TeV, both \pythia{6} Perugia0 and \phojet fail to reproduce the data. They severely underestimate the high multiplicity part of the distribution. \pythia{6} Perugia 2011, EPOS LHC and \pythia{8} 4C give a reasonable fit of the low multiplicity region but underestimate the data above $\nch \approx$ 60.%
\subsection{Parameterization of multiplicity distributions with NBDs}
\label{section:experimental-results:subsection:NBD-parameterization}
Single Negative Binomial Distributions (NBD) have been traditionally used to parameterize particle multiplicity distributions in hadron collisions
\begin{equation}
 P_{\rm NBD}\left( n, \left< n \right>, k \right) = \frac{\Gamma\left( n + k \right)}{\Gamma\left( k \right)\Gamma\left( n+1 \right)} \left[ \frac{\left< n \right>}{\left< n \right> + k} \right]^n \times \left[ \frac{k}{\left< n \right> + k} \right]^k
\end{equation}
where $\left< n \right>$ is the average multiplicity and the variance is given by
\begin{equation}
D^2 = \left< n^2 \right> - \left< n \right>^2 = \left< n \right> + \frac{\left< n \right>^2}{k}
\end{equation}
The parameter $k$ is related to the two-particle correlation function, in the pseudorapidity interval considered \cite{Giovannini:1985mz}. In the limit $k\rightarrow\infty$, the NBD becomes a Poisson distribution.

In previous ALICE data, for the NSD event class, no strong deviation from a single NBD fit was observed at $\sqrt{s} =$ 0.9 and 2.36~TeV, for $|\eta| \leq 1$, while a hint of a substructure appears at $|\eta| < 1.3$ \cite{aamodt2010charged09}. At $\sqrt{s} = 7$~TeV, for the \inelgt event class, the single NBD fit slightly underestimated the data at low multiplicity ($\nch < 5$), and slightly overestimated the data at high multiplicities ($\nch > 55$) \cite{aamodt2010charged7}.

In the present data, for all event classes, already at $\sqrt{s} = 0.9$~TeV, there is a hint that single NBD fits start diverging from the data at the higher multiplicity, for $|\eta| <$ 0.5 and 1.0. More significant departure from single NBD starts at $\sqrt{s} = 2.76$~TeV. The values of $\slfrac{\chi^2}{\mathrm{dof}}$ reflect this, as they are rather high and increase further with increasing centre-of-mass energies.

The appearance of substructures in multiplicity distributions attributed to the occurrence of several sources in the process of particle production \cite{Fowler:1986nv,Giovannini:1998zb,Alexopoulos:1998bi,Walker:2004tx}, can be parameterized by fitting the data with two NBDs. Indeed, a much better fit to the data is obtained by using a weighted sum of two NBD functions
\begin{equation}
\label{eq:double-weighted-NBD}
P\left( n \right) = \lambda \left[ \alpha P_{\rm NBD}\left( n, \left< n \right>_1, k_1 \right) + \left( 1 - \alpha \right) P_{\rm NBD}\left( n, \left< n \right>_2, k_2 \right)  \right]
\end{equation}
This type of function, however, is not meant to describe the value $P(0)$ for \inel and \nsd distributions, which occurs when the $\eta$ acceptance is limited, therefore the bin $n = 0$ was excluded from the fit and an overall normalization factor ($\lambda$) was introduced, as a free parameter, to account for this. At $\sqrt{s} = 7$~TeV, P. Gosh et al. in \cite{bPremGosh} perform a similar fit for CMS data, however without using an overall scale factor ($\lambda$) and with no account for the effect of bin-to-bin correlations.

As was described in \cref{section:systematics:subsection:unfolding-systematics:correlation}, 18 multiplicity distributions, corresponding to main sources of correlated uncertainty, were fitted independently.
For each of these distributions systematic uncertainties corresponding to the non-leading sources were found to have negligible correlations, as expected. 
Thus these residual systematic uncertainties were added to the diagonals of the statistical covariance matrices for these distributions, that were produced in the unfolding procedure. 
In \crefrange{tab:pnch-results-INEL-NBD-proper}{tab:pnch-results-INELgt-NBD-proper} we present the central values of the fit parameters, i.e. center of the maximum spread of each parameter between the 18 independent fits. 
Half of the spread is used to estimate the systematic uncertainty of the parameters (the statistical uncertainty being negligible). 
However these estimates should be treated with caution due to the significant correlations between parameters. 
The multiplicity distributions for each of the 18 cases differ by their slope and intersect at low-to-middle values of multiplicity ($\nch \approx 10$ to $30$ depending on $\eta$ window and energy). 
There is no single point of intersection for a given set, but rather a small interval. 
Therefore extreme cases were selected as the curves with highest and lowest values at high multiplicities (and consequently lowest and highest values at low multiplicity). 
To indicate the change in shape of the parametrization allowed by the correlations present in systematic uncertainties, the parameters for these extreme cases are given in \crefrange{tab:pnch-results-INEL-NBD-proper}{tab:pnch-results-INELgt-NBD-proper} as well.

\begin{table}[ht]
\centering
\scriptsize{\tabulinesep=0.6mm
\begin{tabu}{>{\bfseries}p{0.5cm}cllllll}
\toprule
\rowfont[c]{\bfseries} $\sqrt{\mathbf{s}}$ (TeV) & $\Delta\mathbf{\eta}$ & $\mathbf{\alpha}$ & $\mathbf{\left<n\right>_1}$ & $\mathbf{k_1}$ & $\mathbf{\left<n\right>_2}$ & $\mathbf{k_2}$ & $\mathbf{\lambda}$  \\
\midrule
 \multirow{9}{*}{0.9}	& \multirow{3}{*}{ $0.5$ } & $ 0.427 $ & $ 1.87 $ & $ 2.59 $ & $ 4.92 $ & $ 2.52 $ & $ 0.805 $ \\
 {} & {} &\cellcolor{gray!50}$ 0.469 \pm 0.118 $ & \cellcolor{gray!50}$ 2.08 \pm 0.30 $ & \cellcolor{gray!50}$ 3.09 \pm 0.67 $ & \cellcolor{gray!50}$ 5.56 \pm 0.82 $ & \cellcolor{gray!50}$ 2.97 \pm 0.60 $ & \cellcolor{gray!50}$ 0.793 \pm 0.013 $ \\
 {} & {} &$ 0.587 $ & $ 2.37 $ & $ 2.64 $ & $ 6.38 $ & $ 3.48 $ & $ 0.788 $ \\
 \cmidrule{2-8}
 {}	& \multirow{3}{*}{ $1$ } & $ 0.615 $ & $ 4.46 $ & $ 2.38 $ & $ 11.98 $ & $ 3.76 $ & $ 0.813 $ \\
 {} & {} &\cellcolor{gray!50}$ 0.627 \pm 0.048 $ & \cellcolor{gray!50}$ 4.77 \pm 0.45 $ & \cellcolor{gray!50}$ 2.45 \pm 0.31 $ & \cellcolor{gray!50}$ 12.93 \pm 1.20 $ & \cellcolor{gray!50}$ 3.95 \pm 0.38 $ & \cellcolor{gray!50}$ 0.805 \pm 0.012 $ \\
 {} & {} &$ 0.672 $ & $ 5.22 $ & $ 2.49 $ & $ 14.13 $ & $ 4.33 $ & $ 0.798 $ \\
 \cmidrule{2-8}
 {}	& \multirow{3}{*}{ $1.5$ } & $ 0.444 $ & $ 6.18 $ & $ 3.54 $ & $ 15.63 $ & $ 3.04 $ & $ 0.793 $ \\
 {} & {} &\cellcolor{gray!50}$ 0.502 \pm 0.069 $ & \cellcolor{gray!50}$ 6.81 \pm 0.66 $ & \cellcolor{gray!50}$ 3.22 \pm 0.32 $ & \cellcolor{gray!50}$ 17.36 \pm 1.74 $ & \cellcolor{gray!50}$ 3.34 \pm 0.40 $ & \cellcolor{gray!50}$ 0.794 \pm 0.008 $ \\
 {} & {} &$ 0.553 $ & $ 7.25 $ & $ 2.90 $ & $ 18.79 $ & $ 3.60 $ & $ 0.791 $ \\
\midrule
 \multirow{9}{*}{2.76}	& \multirow{3}{*}{ $0.5$ } & $ 0.546 $ & $ 2.39 $ & $ 2.34 $ & $ 7.46 $ & $ 3.12 $ & $ 0.799 $ \\
 {} & {} &\cellcolor{gray!50}$ 0.588 \pm 0.052 $ & \cellcolor{gray!50}$ 2.54 \pm 0.15 $ & \cellcolor{gray!50}$ 2.20 \pm 0.36 $ & \cellcolor{gray!50}$ 7.87 \pm 0.41 $ & \cellcolor{gray!50}$ 3.32 \pm 0.24 $ & \cellcolor{gray!50}$ 0.800 \pm 0.009 $ \\
 {} & {} &$ 0.578 $ & $ 2.54 $ & $ 2.13 $ & $ 7.89 $ & $ 3.24 $ & $ 0.799 $ \\
 \cmidrule{2-8}
 {}	& \multirow{3}{*}{ $1$ } & $ 0.574 $ & $ 5.05 $ & $ 2.40 $ & $ 15.54 $ & $ 3.53 $ & $ 0.802 $ \\
 {} & {} &\cellcolor{gray!50}$ 0.591 \pm 0.025 $ & \cellcolor{gray!50}$ 5.20 \pm 0.20 $ & \cellcolor{gray!50}$ 2.26 \pm 0.18 $ & \cellcolor{gray!50}$ 16.16 \pm 0.63 $ & \cellcolor{gray!50}$ 3.63 \pm 0.19 $ & \cellcolor{gray!50}$ 0.805 \pm 0.006 $ \\
 {} & {} &$ 0.591 $ & $ 5.15 $ & $ 2.08 $ & $ 16.07 $ & $ 3.52 $ & $ 0.811 $ \\
 \cmidrule{2-8}
 {}	& \multirow{3}{*}{ $1.5$ } & $ 0.580 $ & $ 7.86 $ & $ 2.55 $ & $ 23.04 $ & $ 3.69 $ & $ 0.797 $ \\
 {} & {} &\cellcolor{gray!50}$ 0.601 \pm 0.021 $ & \cellcolor{gray!50}$ 8.15 \pm 0.29 $ & \cellcolor{gray!50}$ 2.42 \pm 0.13 $ & \cellcolor{gray!50}$ 23.94 \pm 0.90 $ & \cellcolor{gray!50}$ 3.81 \pm 0.12 $ & \cellcolor{gray!50}$ 0.800 \pm 0.003 $ \\
 {} & {} &$ 0.593 $ & $ 8.15 $ & $ 2.42 $ & $ 24.68 $ & $ 3.79 $ & $ 0.799 $ \\
\midrule
 \multirow{9}{*}{7}	& \multirow{3}{*}{ $0.5$ } & $ 0.713 $ & $ 3.39 $ & $ 1.54 $ & $ 11.52 $ & $ 4.10 $ & $ 0.787 $ \\
 {} & {} &\cellcolor{gray!50}$ 0.716 \pm 0.046 $ & \cellcolor{gray!50}$ 3.42 \pm 0.24 $ & \cellcolor{gray!50}$ 1.49 \pm 0.14 $ & \cellcolor{gray!50}$ 11.54 \pm 0.61 $ & \cellcolor{gray!50}$ 4.01 \pm 0.28 $ & \cellcolor{gray!50}$ 0.792 \pm 0.007 $ \\
 {} & {} &$ 0.733 $ & $ 3.55 $ & $ 1.42 $ & $ 11.92 $ & $ 4.13 $ & $ 0.795 $ \\
 \cmidrule{2-8}
 {}	& \multirow{3}{*}{ $1$ } & $ 0.668 $ & $ 6.46 $ & $ 1.70 $ & $ 22.07 $ & $ 4.04 $ & $ 0.789 $ \\
 {} & {} &\cellcolor{gray!50}$ 0.679 \pm 0.013 $ & \cellcolor{gray!50}$ 6.62 \pm 0.16 $ & \cellcolor{gray!50}$ 1.68 \pm 0.06 $ & \cellcolor{gray!50}$ 22.66 \pm 0.60 $ & \cellcolor{gray!50}$ 4.14 \pm 0.11 $ & \cellcolor{gray!50}$ 0.788 \pm 0.005 $ \\
 {} & {} &$ 0.680 $ & $ 6.69 $ & $ 1.66 $ & $ 22.94 $ & $ 4.14 $ & $ 0.790 $ \\
 \cmidrule{2-8}
 {}	& \multirow{3}{*}{ $1.5$ } & $ 0.581 $ & $ 8.59 $ & $ 2.06 $ & $ 29.84 $ & $ 3.46 $ & $ 0.785 $ \\
 {} & {} &\cellcolor{gray!50}$ 0.577 \pm 0.030 $ & \cellcolor{gray!50}$ 8.72 \pm 0.43 $ & \cellcolor{gray!50}$ 2.12 \pm 0.29 $ & \cellcolor{gray!50}$ 30.69 \pm 0.98 $ & \cellcolor{gray!50}$ 3.50 \pm 0.13 $ & \cellcolor{gray!50}$ 0.781 \pm 0.014 $ \\
 {} & {} &$ 0.607 $ & $ 8.92 $ & $ 1.83 $ & $ 31.66 $ & $ 3.62 $ & $ 0.795 $ \\
\midrule
 \multirow{9}{*}{8}	& \multirow{3}{*}{ $0.5$ } & $ 0.589 $ & $ 2.95 $ & $ 1.90 $ & $ 10.48 $ & $ 3.29 $ & $ 0.793 $ \\
 {} & {} &\cellcolor{gray!50}$ 0.593 \pm 0.029 $ & \cellcolor{gray!50}$ 3.01 \pm 0.12 $ & \cellcolor{gray!50}$ 1.85 \pm 0.07 $ & \cellcolor{gray!50}$ 10.64 \pm 0.32 $ & \cellcolor{gray!50}$ 3.30 \pm 0.19 $ & \cellcolor{gray!50}$ 0.795 \pm 0.002 $ \\
 {} & {} &$ 0.568 $ & $ 2.92 $ & $ 1.89 $ & $ 10.44 $ & $ 3.13 $ & $ 0.794 $ \\
 \cmidrule{2-8}
 {}	& \multirow{3}{*}{ $1$ } & $ 0.626 $ & $ 6.42 $ & $ 1.93 $ & $ 22.24 $ & $ 3.82 $ & $ 0.794 $ \\
 {} & {} &\cellcolor{gray!50}$ 0.673 \pm 0.047 $ & \cellcolor{gray!50}$ 7.03 \pm 0.61 $ & \cellcolor{gray!50}$ 1.81 \pm 0.12 $ & \cellcolor{gray!50}$ 23.61 \pm 1.37 $ & \cellcolor{gray!50}$ 4.19 \pm 0.37 $ & \cellcolor{gray!50}$ 0.795 \pm 0.001 $ \\
 {} & {} &$ 0.669 $ & $ 7.06 $ & $ 1.79 $ & $ 23.79 $ & $ 4.14 $ & $ 0.795 $ \\
 \cmidrule{2-8}
 {}	& \multirow{3}{*}{ $1.5$ } & $ 0.709 $ & $ 10.88 $ & $ 1.71 $ & $ 36.16 $ & $ 4.64 $ & $ 0.803 $ \\
 {} & {} &\cellcolor{gray!50}$ 0.696 \pm 0.018 $ & \cellcolor{gray!50}$ 10.83 \pm 0.30 $ & \cellcolor{gray!50}$ 1.71 \pm 0.05 $ & \cellcolor{gray!50}$ 36.22 \pm 0.70 $ & \cellcolor{gray!50}$ 4.48 \pm 0.19 $ & \cellcolor{gray!50}$ 0.803 \pm 0.002 $ \\
 {} & {} &$ 0.689 $ & $ 10.84 $ & $ 1.73 $ & $ 36.38 $ & $ 4.44 $ & $ 0.802 $ \\
\bottomrule
\end{tabu}
}
\caption{Double-NBD fit parameters for charged-particle multiplicity distributions of \inel events. The shaded (middle) row for each entry contains central values of the parameters with their estimated uncertainty, and the surrounding rows represent the parameter values of the bounding fits: topmost in the first row and the bottommost in the last one. \label{tab:pnch-results-INEL-NBD-proper}}
\end{table}

\begin{table}[ht]
 \centering
\scriptsize{\tabulinesep=0.6mm
\begin{tabu}{>{\bfseries}p{0.5cm}cllllll}
\toprule
\rowfont[c]{\bfseries} $\sqrt{\mathbf{s}}$ (TeV) & $\Delta\mathbf{\eta}$ & $\mathbf{\alpha}$ & $\mathbf{\left<n\right>_1}$ & $\mathbf{k_1}$ & $\mathbf{\left<n\right>_2}$ & $\mathbf{k_2}$ & $\mathbf{\lambda}$  \\
\midrule
 \multirow{9}{*}{0.9}	& \multirow{3}{*}{ $0.5$ } & $ 0.317 $ & $ 1.90 $ & $ 4.47 $ & $ 4.80 $ & $ 2.45 $ & $ 0.935 $ \\
 {} & {} &\cellcolor{gray!50}$ 0.363 \pm 0.094 $ & \cellcolor{gray!50}$ 2.07 \pm 0.23 $ & \cellcolor{gray!50}$ 4.34 \pm 1.22 $ & \cellcolor{gray!50}$ 5.31 \pm 0.65 $ & \cellcolor{gray!50}$ 2.75 \pm 0.43 $ & \cellcolor{gray!50}$ 0.930 \pm 0.014 $ \\
 {} & {} &$ 0.457 $ & $ 2.30 $ & $ 3.46 $ & $ 5.92 $ & $ 3.09 $ & $ 0.922 $ \\
 \cmidrule{2-8}
 {}	& \multirow{3}{*}{ $1$ } & $ 0.548 $ & $ 4.58 $ & $ 2.80 $ & $ 11.70 $ & $ 3.62 $ & $ 0.952 $ \\
 {} & {} &\cellcolor{gray!50}$ 0.572 \pm 0.074 $ & \cellcolor{gray!50}$ 4.98 \pm 0.57 $ & \cellcolor{gray!50}$ 2.82 \pm 0.29 $ & \cellcolor{gray!50}$ 12.79 \pm 1.41 $ & \cellcolor{gray!50}$ 3.88 \pm 0.50 $ & \cellcolor{gray!50}$ 0.944 \pm 0.012 $ \\
 {} & {} &$ 0.646 $ & $ 5.55 $ & $ 2.77 $ & $ 14.19 $ & $ 4.37 $ & $ 0.937 $ \\
 \cmidrule{2-8}
 {}	& \multirow{3}{*}{ $1.5$ } & $ 0.661 $ & $ 8.28 $ & $ 2.73 $ & $ 19.28 $ & $ 4.28 $ & $ 0.933 $ \\
 {} & {} &\cellcolor{gray!50}$ 0.703 \pm 0.120 $ & \cellcolor{gray!50}$ 9.20 \pm 1.23 $ & \cellcolor{gray!50}$ 2.54 \pm 0.24 $ & \cellcolor{gray!50}$ 21.51 \pm 3.38 $ & \cellcolor{gray!50}$ 4.73 \pm 1.06 $ & \cellcolor{gray!50}$ 0.937 \pm 0.012 $ \\
 {} & {} &$ 0.823 $ & $ 10.43 $ & $ 2.30 $ & $ 24.89 $ & $ 5.67 $ & $ 0.936 $ \\
\midrule
 \multirow{9}{*}{2.76}	& \multirow{3}{*}{ $0.5$ } & $ 0.513 $ & $ 2.49 $ & $ 2.73 $ & $ 7.42 $ & $ 3.11 $ & $ 0.934 $ \\
 {} & {} &\cellcolor{gray!50}$ 0.509 \pm 0.019 $ & \cellcolor{gray!50}$ 2.50 \pm 0.05 $ & \cellcolor{gray!50}$ 2.77 \pm 0.28 $ & \cellcolor{gray!50}$ 7.53 \pm 0.13 $ & \cellcolor{gray!50}$ 3.04 \pm 0.09 $ & \cellcolor{gray!50}$ 0.931 \pm 0.006 $ \\
 {} & {} &$ 0.493 $ & $ 2.48 $ & $ 2.66 $ & $ 7.45 $ & $ 2.95 $ & $ 0.931 $ \\
 \cmidrule{2-8}
 {}	& \multirow{3}{*}{ $1$ } & $ 0.605 $ & $ 5.64 $ & $ 2.48 $ & $ 16.36 $ & $ 3.87 $ & $ 0.941 $ \\
 {} & {} &\cellcolor{gray!50}$ 0.545 \pm 0.060 $ & \cellcolor{gray!50}$ 5.32 \pm 0.34 $ & \cellcolor{gray!50}$ 2.71 \pm 0.24 $ & \cellcolor{gray!50}$ 15.65 \pm 0.75 $ & \cellcolor{gray!50}$ 3.52 \pm 0.35 $ & \cellcolor{gray!50}$ 0.935 \pm 0.006 $ \\
 {} & {} &$ 0.498 $ & $ 5.11 $ & $ 2.87 $ & $ 15.26 $ & $ 3.24 $ & $ 0.930 $ \\
 \cmidrule{2-8}
 {}	& \multirow{3}{*}{ $1.5$ } & $ 0.703 $ & $ 10.31 $ & $ 2.38 $ & $ 26.50 $ & $ 4.63 $ & $ 0.908 $ \\
 {} & {} &\cellcolor{gray!50}$ 0.703 \pm 0.067 $ & \cellcolor{gray!50}$ 10.63 \pm 0.92 $ & \cellcolor{gray!50}$ 2.18 \pm 0.21 $ & \cellcolor{gray!50}$ 27.39 \pm 1.77 $ & \cellcolor{gray!50}$ 4.52 \pm 0.49 $ & \cellcolor{gray!50}$ 0.913 \pm 0.006 $ \\
 {} & {} &$ 0.655 $ & $ 10.04 $ & $ 2.20 $ & $ 26.47 $ & $ 4.19 $ & $ 0.915 $ \\
\midrule
 \multirow{9}{*}{7}	& \multirow{3}{*}{ $0.5$ } & $ 0.748 $ & $ 3.87 $ & $ 1.66 $ & $ 12.24 $ & $ 4.54 $ & $ 0.944 $ \\
 {} & {} &\cellcolor{gray!50}$ 0.688 \pm 0.060 $ & \cellcolor{gray!50}$ 3.58 \pm 0.32 $ & \cellcolor{gray!50}$ 1.81 \pm 0.15 $ & \cellcolor{gray!50}$ 11.54 \pm 0.80 $ & \cellcolor{gray!50}$ 4.07 \pm 0.46 $ & \cellcolor{gray!50}$ 0.942 \pm 0.002 $ \\
 {} & {} &$ 0.649 $ & $ 3.38 $ & $ 1.90 $ & $ 11.16 $ & $ 3.75 $ & $ 0.940 $ \\
 \cmidrule{2-8}
 {}	& \multirow{3}{*}{ $1$ } & $ 0.673 $ & $ 7.00 $ & $ 2.02 $ & $ 22.77 $ & $ 4.29 $ & $ 0.948 $ \\
 {} & {} &\cellcolor{gray!50}$ 0.671 \pm 0.020 $ & \cellcolor{gray!50}$ 7.07 \pm 0.27 $ & \cellcolor{gray!50}$ 2.02 \pm 0.11 $ & \cellcolor{gray!50}$ 23.06 \pm 0.74 $ & \cellcolor{gray!50}$ 4.27 \pm 0.18 $ & \cellcolor{gray!50}$ 0.942 \pm 0.006 $ \\
 {} & {} &$ 0.679 $ & $ 7.25 $ & $ 1.95 $ & $ 23.49 $ & $ 4.34 $ & $ 0.943 $ \\
 \cmidrule{2-8}
 {}	& \multirow{3}{*}{ $1.5$ } & $ 0.615 $ & $ 10.17 $ & $ 1.94 $ & $ 31.42 $ & $ 3.75 $ & $ 0.931 $ \\
 {} & {} &\cellcolor{gray!50}$ 0.622 \pm 0.019 $ & \cellcolor{gray!50}$ 10.43 \pm 0.43 $ & \cellcolor{gray!50}$ 1.89 \pm 0.07 $ & \cellcolor{gray!50}$ 32.31 \pm 0.89 $ & \cellcolor{gray!50}$ 3.84 \pm 0.13 $ & \cellcolor{gray!50}$ 0.930 \pm 0.004 $ \\
 {} & {} &$ 0.618 $ & $ 10.36 $ & $ 1.89 $ & $ 32.76 $ & $ 3.82 $ & $ 0.932 $ \\
\midrule
 \multirow{9}{*}{8}	& \multirow{3}{*}{ $0.5$ } & $ 0.557 $ & $ 2.95 $ & $ 2.10 $ & $ 10.28 $ & $ 3.18 $ & $ 0.927 $ \\
 {} & {} &\cellcolor{gray!50}$ 0.567 \pm 0.018 $ & \cellcolor{gray!50}$ 3.03 \pm 0.08 $ & \cellcolor{gray!50}$ 2.04 \pm 0.06 $ & \cellcolor{gray!50}$ 10.48 \pm 0.20 $ & \cellcolor{gray!50}$ 3.22 \pm 0.12 $ & \cellcolor{gray!50}$ 0.929 \pm 0.002 $ \\
 {} & {} &$ 0.555 $ & $ 3.01 $ & $ 2.02 $ & $ 10.44 $ & $ 3.13 $ & $ 0.929 $ \\
 \cmidrule{2-8}
 {}	& \multirow{3}{*}{ $1$ } & $ 0.589 $ & $ 6.25 $ & $ 2.25 $ & $ 21.71 $ & $ 3.68 $ & $ 0.932 $ \\
 {} & {} &\cellcolor{gray!50}$ 0.625 \pm 0.035 $ & \cellcolor{gray!50}$ 6.70 \pm 0.44 $ & \cellcolor{gray!50}$ 2.13 \pm 0.12 $ & \cellcolor{gray!50}$ 22.72 \pm 1.00 $ & \cellcolor{gray!50}$ 3.94 \pm 0.26 $ & \cellcolor{gray!50}$ 0.932 \pm 0.001 $ \\
 {} & {} &$ 0.631 $ & $ 6.88 $ & $ 2.07 $ & $ 23.18 $ & $ 3.98 $ & $ 0.931 $ \\
 \cmidrule{2-8}
 {}	& \multirow{3}{*}{ $1.5$ } & $ 0.817 $ & $ 14.18 $ & $ 1.56 $ & $ 40.81 $ & $ 5.60 $ & $ 0.917 $ \\
 {} & {} &\cellcolor{gray!50}$ 0.857 \pm 0.042 $ & \cellcolor{gray!50}$ 15.31 \pm 1.14 $ & \cellcolor{gray!50}$ 1.49 \pm 0.07 $ & \cellcolor{gray!50}$ 43.06 \pm 2.25 $ & \cellcolor{gray!50}$ 6.25 \pm 0.73 $ & \cellcolor{gray!50}$ 0.918 \pm 0.003 $ \\
 {} & {} &$ 0.839 $ & $ 15.18 $ & $ 1.51 $ & $ 42.89 $ & $ 5.83 $ & $ 0.915 $ \\
\bottomrule
\end{tabu}
}
\caption{Double-NBD fit parameters for charged-particle multiplicity distributions of \nsd events. The shaded (middle) row for each entry contains central values of the parameters with their estimated uncertainty, and the surrounding rows represent the parameter values of the bounding fits: topmost in the first row and the bottommost in the last one.\label{tab:pnch-results-NSD-NBD-proper}}
\end{table}

\begin{table}[hb]
 \centering
\scriptsize{\tabulinesep=0.6mm
\begin{tabu}{>{\bfseries}p{0.5cm}llllll}
\toprule
\rowfont[c]{\bfseries} $\sqrt{\mathbf{s}}$ (TeV) & $\mathbf{\alpha}$ & $\mathbf{\left<n\right>_1}$ & $\mathbf{k_1}$ & $\mathbf{\left<n\right>_2}$ & $\mathbf{k_2}$ & $\mathbf{\lambda}$  \\
\midrule
  \multirow{3}{*} { $ 0.9 $ }	& $ 0.595 $ & $ 4.40 $ & $ 2.56 $ & $ 11.82 $ & $ 3.69 $ & $ 1.047 $ \\
 {} &\cellcolor{gray!50}$ 0.661 \pm 0.069 $ & \cellcolor{gray!50}$ 4.97 \pm 0.58 $ & \cellcolor{gray!50}$ 2.45 \pm 0.20 $ & \cellcolor{gray!50}$ 13.39 \pm 1.57 $ & \cellcolor{gray!50}$ 4.21 \pm 0.55 $ & \cellcolor{gray!50}$ 1.046 \pm 0.004 $ \\
 {} &$ 0.731 $ & $ 5.55 $ & $ 2.25 $ & $ 14.96 $ & $ 4.76 $ & $ 1.048 $ \\
\midrule
  \multirow{3}{*} { $ 2.76 $ }	& $ 0.581 $ & $ 5.12 $ & $ 2.37 $ & $ 15.64 $ & $ 3.57 $ & $ 1.038 $ \\
 {} &\cellcolor{gray!50}$ 0.588 \pm 0.028 $ & \cellcolor{gray!50}$ 5.15 \pm 0.29 $ & \cellcolor{gray!50}$ 2.30 \pm 0.11 $ & \cellcolor{gray!50}$ 16.12 \pm 0.69 $ & \cellcolor{gray!50}$ 3.59 \pm 0.24 $ & \cellcolor{gray!50}$ 1.047 \pm 0.012 $ \\
 {} &$ 0.572 $ & $ 4.98 $ & $ 2.19 $ & $ 15.79 $ & $ 3.42 $ & $ 1.060 $ \\
\midrule
  \multirow{3}{*} { $ 7 $ }	& $ 0.677 $ & $ 6.63 $ & $ 1.72 $ & $ 22.36 $ & $ 4.13 $ & $ 1.049 $ \\
 {} &\cellcolor{gray!50}$ 0.683 \pm 0.012 $ & \cellcolor{gray!50}$ 6.65 \pm 0.24 $ & \cellcolor{gray!50}$ 1.68 \pm 0.05 $ & \cellcolor{gray!50}$ 22.86 \pm 0.55 $ & \cellcolor{gray!50}$ 4.19 \pm 0.12 $ & \cellcolor{gray!50}$ 1.053 \pm 0.010 $ \\
 {} &$ 0.684 $ & $ 6.78 $ & $ 1.69 $ & $ 23.08 $ & $ 4.19 $ & $ 1.052 $ \\
\midrule
  \multirow{3}{*} { $ 8 $ }	& $ 0.631 $ & $ 6.26 $ & $ 1.74 $ & $ 22.07 $ & $ 3.75 $ & $ 1.056 $ \\
 {} &\cellcolor{gray!50}$ 0.640 \pm 0.009 $ & \cellcolor{gray!50}$ 6.44 \pm 0.18 $ & \cellcolor{gray!50}$ 1.73 \pm 0.04 $ & \cellcolor{gray!50}$ 22.59 \pm 0.51 $ & \cellcolor{gray!50}$ 3.84 \pm 0.09 $ & \cellcolor{gray!50}$ 1.053 \pm 0.004 $ \\
 {} &$ 0.641 $ & $ 6.57 $ & $ 1.76 $ & $ 22.86 $ & $ 3.85 $ & $ 1.049 $ \\
\bottomrule
\end{tabu}
}
 \caption{Double-NBD fit parameters for charged-particle multiplicity distributions of \inelgt events. The shaded (middle) row for each entry contains central values of the parameters with their estimated uncertainty, and the surrounding rows represent the parameter values of the bounding fits: topmost in the first row and the bottommost in the last one.\label{tab:pnch-results-INELgt-NBD-proper}}
\end{table}

The shape evolution is quantified by the parameter $\left<n\right>_2$, which tends to increase with increasing $\eta$ range and with increasing centre-of-mass energy. The observed relation $\left<n\right>_2 \approx 3\times\left<n\right>_1$, is consistent with the analysis of CMS data reported in \cite{bPremGosh}, despite the fact that the scaling parameter $\lambda$ was not used in their fit function.

In the bins $\left|\eta\right| < 1$ and $\left|\eta\right| < 1.5$, the CMS NSD data at $\sqrt{s} = 0.9$~TeV (\cref{fig:pnch-results-experiments-comparison}) showed a different trend at high multiplicity as compared to the ALICE data (\cref{section:experimental-results:subsection:subsection:comparison-to-other}). We have made our own fit of CMS data excluding the bin $\nch = 0$ (\cref{tab:pnch-results-NBD-NSD}) and using the overall scaling parameter ($\lambda$). As we cannot take into account correlations within CMS data, we compare these fits with fits of our final distributions, ignoring the correlations. We find that the relative weight of the second NBD component in the CMS data is smaller than in the ALICE data, while other parameters are compatible within their respective uncertainties. This confirms the different trends observed in distribution tails.

In conclusion, a double-NBD function provides a precise description of the entire set of multiplicity distributions measured in this experiment.

\begin{table}[h]
\centering
 \footnotesize{
 \begin{tabu} to \textwidth {>{\bfseries}p{0.5cm}cllllllc}
	\toprule
	\rowfont[c]{\bfseries} $\sqrt{\bm{s}}$ (TeV)	& $\Delta\bm{\eta}$	& $\bm{\lambda}$	& $\bm{\alpha}$	& $\bm{\left<n\right>_1}$	& $\bm{k_1}$		& $\bm{\left<n\right>_2}$	& $\bm{k_2}$	& $\bm{\slfrac{\chi^2}{n_\mathrm{dof}}}$ \\
	\midrule
	\rowfont{\bfseries} {}	& \multicolumn{8}{c}{ALICE} \\
	\midrule
	\multirow{2}{*}{0.9}		& 1	& $0.95\pm0.02$	& $0.56\pm0.10$	& $5.0\pm0.8$	& $2.6\pm0.8$	& $12\pm2$	& $3.4\pm1.1$	& 1.5 / 53  \\
					& 1.5	& $0.96\pm0.02$	& $0.70\pm0.18$	& $8.4\pm3.0$	& $2.5\pm1.1$	& $21\pm8$	& $4.7\pm3.1$	& 14.0 / 65  \\
	\midrule
	\rowfont{\bfseries} {}	& \multicolumn{8}{c}{CMS} \\
	\midrule
	\multirow{2}{*}{0.9}		& 1	& $0.93\pm0.03$	& $0.72\pm0.08$	& $5.4\pm1.0$	& $2.6\pm0.1$	& $14\pm2$	& $5.8\pm2.3$	& 2.3 / 33 \\
					& 1.5	& $0.94\pm0.02$	& $0.74\pm0.05$	& $8.4\pm1.0$	& $2.6\pm0.1$	& $21\pm2$	& $6.4\pm1.8$	& 1.8 / 45  \\
	\bottomrule
 \end{tabu}
 }
 \caption{Comparison of double-NBD fit parameters for charged-particle multiplicity distributions of \nsd events by ALICE and CMS, obtained without accounting for correlations using the function defined by \cref{eq:double-weighted-NBD}. CMS data are taken from \cite{Khachatryan:2010nk}. The last column gives the $\chi^2$ value and the number of degrees of freedom ($n_\mathrm{dof}$).\label{tab:pnch-results-NBD-NSD}}
\end{table}

\subsection{KNO studies}
\label{section:experimental-results:subsection:KNO-studies}
The KNO variable $\slfrac{\nch}{\left<\nch\right>}$ provides another way to study the evolution of the shape of multiplicity distributions with varying centre-of-mass energies and varying pseudorapidity intervals. This study is carried out for the NSD event class only so that SD events, which may have a different behaviour, are not included in the data samples.

The quantities of interest are derived from the original set of 72 multiplicity distributions and the resulting spread is used to estimate the various uncertainties presented in this section.
\subsubsection{Evolution of the shape of multiplicity distributions with $\sqrt{s}$}
\begin{figure}[!ht]
 \centering
 \includegraphics[width=\textwidth]{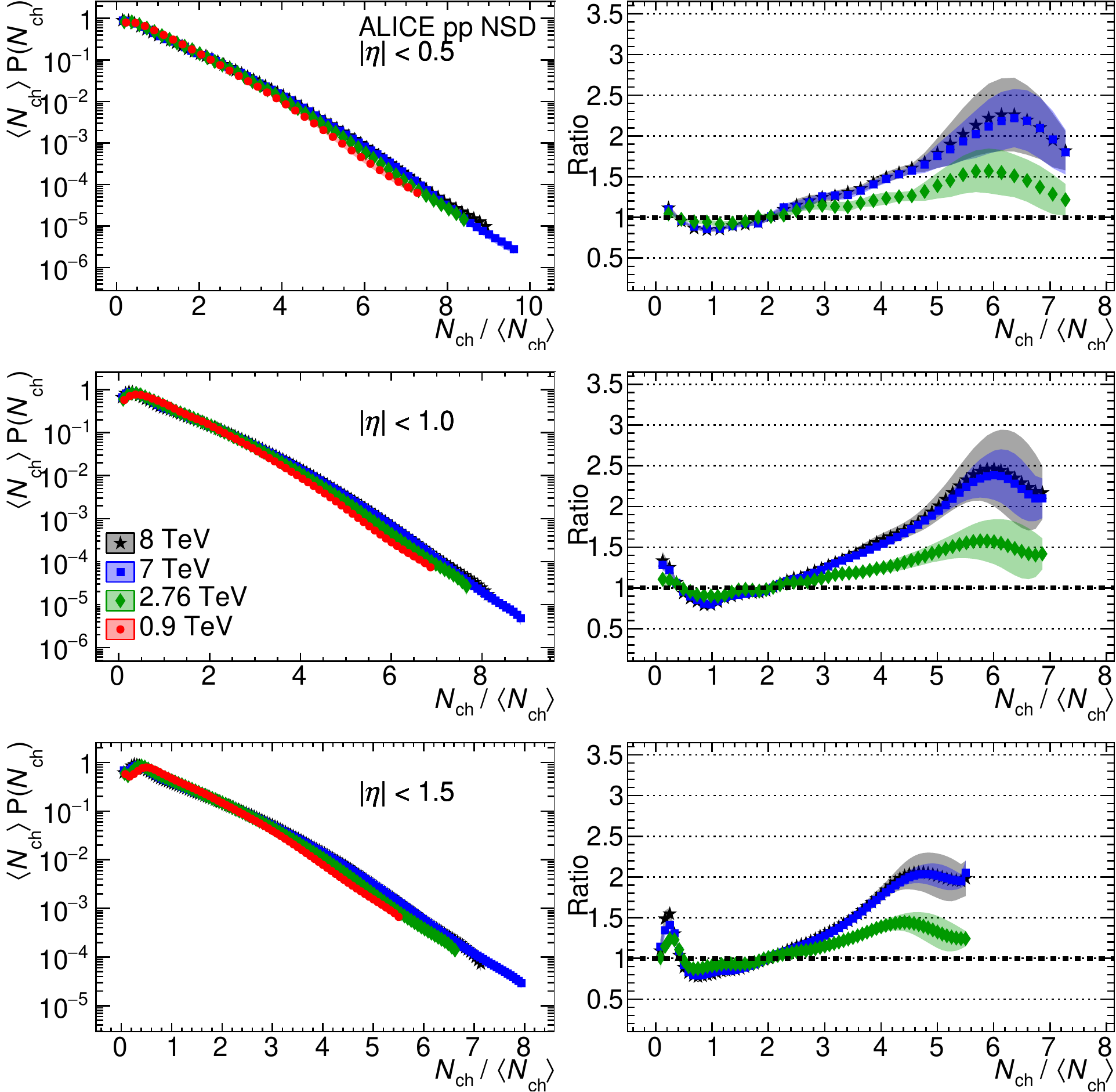}
 \caption{KNO-scaled distribution $\left<\nch\right>\pnch$ versus the KNO variable $\slfrac{\nch}{\left<\nch\right>}$ at $\sqrt{s} =$ 0.9, 2.76, 7 and 8~TeV, for three pseudorapidity intervals: $|\eta| <$ 0.5 (top), 1.0 (middle) and 1.5 (bottom). In each case, ratios to the distribution at $\sqrt{s} = 0.9$~TeV are shown, on the right-hand side parts of the figures. As $\slfrac{\nch}{\left<\nch\right>}$ takes different values at different centre-of-mass energies, ratios were obtained by interpolating the KNO-scaled distributions, and uncertainties were taken from the nearest data point. Bands represent the total uncertainties.\label{fig:pnch-results-KNO-scaled}}
\end{figure}

The KNO test in the range 0.9 to 8~TeV is limited by the multiplicity reach at 0.9~TeV. KNO-scaled distributions and their ratios were obtained for each of the available combinations of corrections with the same procedure used for multiplicity distribution measurements (averaging the 72 cases listed in \cref{section:systematics:subsection:unfolding-systematics} and using the spread as a measure of the systematic uncertainty). Bin to bin correlations were ignored when comparing KNO distributions and $q$-moments at various centre-of-mass energies. Consequently, the relative errors obtained on the ratios are somewhat overestimated. Ratios between the two highest energies and 0.9~TeV exceed the value 2 at $\slfrac{\nch}{\left<\nch\right>}$ larger than 5.5, 5 and 4.5, for $|\eta| < 0.5$, $|\eta| < 1$ and $|\eta| < 1.5$, respectively (\cref{fig:pnch-results-KNO-scaled}). This confirms that KNO scaling violation increases with increasing pseudorapidity intervals. The shape of the KNO scaling violation reflects the fact that the high-multiplicity tail of the distribution increases faster with increasing energy and with increasing pseudorapidity interval than the low ($\nch \leq 20$) multiplicity part as already noted in \cref{section:experimental-results:subsection:NBD-parameterization}.
\subsubsection{Quantitative KNO study with normalized $q$-moments}
\begin{figure}[t]
 \centering
 \begin{overpic}[width=0.55\textwidth]{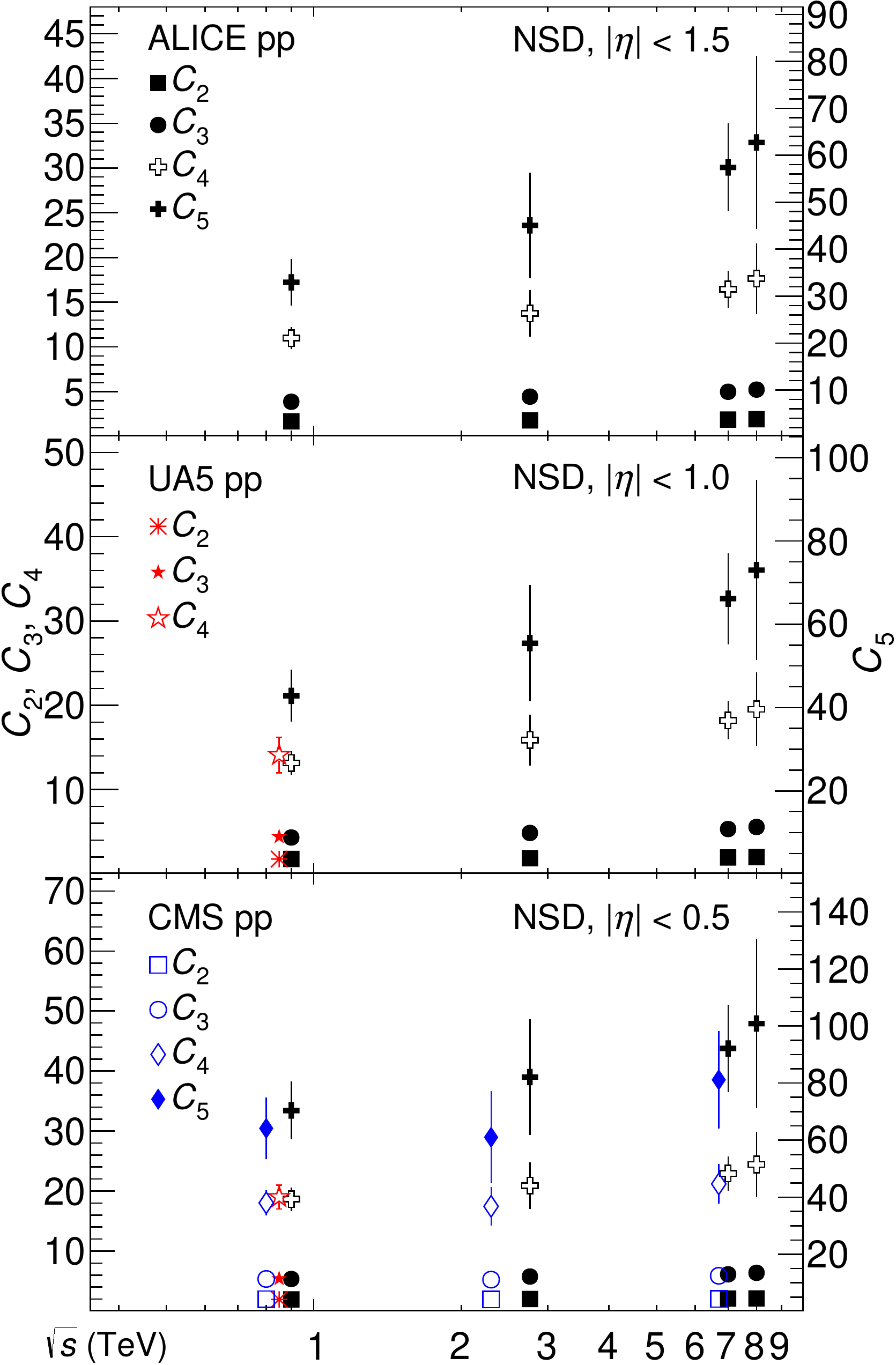}
  \put(208,641){{\cite{Ansorge:1988kn}}}
  \put(217,321){{\cite{Khachatryan:2010nk}}}
 \end{overpic}\\[-0.3cm]
 \caption{Centre-of-mass energy dependence of the $q$-moments ($q =$ 2 to 4, left-hand scale, and $q = 5$, right-hand scale) of the multiplicity distributions for \nsd events in three different pseudorapidity intervals ($|\eta| < 1.5$ top, $|\eta| < 1.0$ middle and $|\eta| < 0.5$ bottom). ALICE data (black) are compared to UA5 \cite{Ansorge:1988kn} (red) for $|\eta| < 0.5$ and $|\eta| < 1$, at $\sqrt{s} = 0.9$~TeV, and with CMS \cite{Khachatryan:2010nk} (blue) at $\sqrt{s} =$ 0.9 and 7~TeV for $|\eta| < 0.5$. The error bars represent the combined statistical and systematic uncertainties. The data at 0.9 and 7~TeV are slightly displaced horizontally for visibility. \label{fig:pnch-results-moments}}
\end{figure}

Multiplicity distributions may be characterized by their normalized $q$-moments defined as
\begin{equation}
 C_q \equiv \frac{\left< n^q \right>}{\left< n \right>^q}
\end{equation}
where $q$ is a positive integer studied here for values 2, 3, 4 and 5, for NSD events (\cref{fig:pnch-results-moments} and \cref{tab:pnch-results-moments}). For the three pseudorapidity intervals studied here ($|\eta| <$ 0.5, 1 and 1.5), $C_2$ remains constant over the energy range, $C_3$ shows a small increase with increasing energy for the two largest $\eta$ intervals, $C_4$ and $C_5$ show an increase with increasing energy, which becomes stronger for larger $\eta$ intervals. These new data are in agreement with UA5 \cite{Ansorge:1988kn}, CMS \cite{Khachatryan:2010nk} and ALICE{\textquoteright}s previous results \cite{aamodt2010charged09,aamodt2010charged7} in all cases where a comparison was possible (\cref{fig:pnch-results-moments}).
\begin{table}[b]
 \centering
 \begin{tabu}{>{\bfseries}c>{\bfseries}cllll}
	\cmidrule[\heavyrulewidth]{3-6}
	\rowfont{\bfseries} {}	& {}	& \multicolumn{4}{c}{$\sqrt{\bm{s}}$ (TeV)} \\
	\midrule
	\rowfont[c]{\bfseries} {$\Delta\bm{\eta}$}	& {Moment}	& 0.9	& 2.76	& 7	& 8 \\
	\midrule
	\multirow{5}{*}{0.5}	& $\left<\nch\right>$	& $3.8\pm0.1$	& $4.6\pm0.2$	& $5.7\pm0.2$	& $5.8\pm0.4$ \\
	{} 			& $C_2$			& $2.0\pm0.1$	& $2.0\pm0.1$	& $2.1\pm0.1$	& $2.1\pm0.2$ \\
	{} 			& $C_3$			& $5.3\pm0.4$	& $5.8\pm0.7$	& $6.1\pm0.5$	& $6.4\pm1.0$ \\
	{} 			& $C_4$			& $19\pm2$	& $21\pm4$	& $23\pm3$	& $24\pm5$    \\
	{} 			& $C_5$			& $78\pm11$	& $91\pm23$	& $102\pm17$	& $112\pm33$  \\
	\midrule
	\multirow{5}{*}{1}	& $\left<\nch\right>$	& $7.8\pm0.3$	& $9.4\pm0.4$	& $11.6\pm0.4$	& $11.9\pm0.7$ \\
	{} 			& $C_2$			& $1.8\pm0.1$	& $1.9\pm0.1$	& $1.9\pm0.1$	& $2.0\pm0.1$  \\
	{} 			& $C_3$			& $4.3\pm0.3$	& $4.9\pm0.6$	& $5.3\pm0.4$	& $5.6\pm0.8$  \\
	{} 			& $C_4$			& $13\pm1$	& $16\pm3$	& $18\pm2$	& $20\pm4$     \\
	{} 			& $C_5$			& $48\pm7$	& $62\pm16$	& $73\pm12$	& $81\pm24$    \\
	\midrule
	\multirow{5}{*}{1.5}	& $\left<\nch\right>$	& $11.8\pm0.4$	& $14.2\pm0.7$	& $17.5\pm0.6$	& $17.8\pm1.1$ \\
	{} 			& $C_2$			& $1.7\pm0.1$	& $1.8\pm0.1$	& $1.9\pm0.1$	& $1.9\pm0.1$  \\
	{} 			& $C_3$			& $3.9\pm0.3$	& $4.5\pm0.6$	& $5.0\pm0.4$	& $5.2\pm0.8$  \\
	{} 			& $C_4$			& $11\pm1$	& $14\pm3$	& $16\pm2$	& $18\pm4$     \\
	{} 			& $C_5$			& $36\pm6$	& $51\pm13$	& $64\pm11$	& $70\pm20$    \\
	\bottomrule
 \end{tabu}
  \caption{Mean charged-particle multiplicity and $q$-moments of the multiplicity distributions measured by ALICE at $\sqrt{s} =$ 0.9, 2.76, 7 and 8~TeV, for \nsd events in three different pseudorapidity intervals. The uncertainties are combined statistical and systematic uncertainties.\label{tab:pnch-results-moments}}
\end{table}

\section{Discussion of results and conclusion}
\label{section:summary-and-conclusion}
The ALICE Collaboration has carried out a detailed study of pseudorapidity densities and multiplicity distributions of primary charged particles produced in proton-proton collisions, at $\sqrt{s} =$ 0.9, 2.36, 2.76, 7 and 8~TeV, in the pseudorapidity range $|\eta| < 2$. A large increase of event sample size compared to previous ALICE publications, combined with improved measurement techniques, was used to study the evolution of charged-particle multiplicities over the whole centre-of-mass energy range covered so far by the LHC. The data at the highest energy appear as a smooth continuation of lower energy data, both in shape and in magnitude.

The pseudorapidity density of charged particles, $\dndetainline$, was measured as a function of pseudorapidity in the range $|\eta| \leq 2$. The relative precision achieved at $\eta = 0$ for $\sqrt{s} =$ 7~TeV is 5.5\%, 2.6\% and 1.3\% for \inel, \nsd and \inelgt event classes, respectively.

The power law parameterization of $\dndetainline$ at $\eta = 0$, $s^{\delta}$, provides a good description of the data from ISR to LHC energies: $\delta = 0.102 \pm 0.003$, $0.114 \pm 0.003$ and $0.114 \pm 0.001$, for the \inel, \nsd and \inelgt event classes, respectively, to be compared to $\delta \simeq 0.15$ for Pb--Pb collisions \cite{aamodt2010charged}. The ALICE Collaboration has shown clearly that the particle pseudorapidity density increases faster with energy in Pb--Pb collisions than in pp collisions. The extrapolation of $\dndetainline$ at $\eta = 0$ to the nominal LHC energy ($\sqrt{s} = 14$~TeV) is obtained with a precision of 4.6\%, 3.0\% and 1.3\% for \inel, \nsd and \inelgt event classes, respectively.

Multiplicity distributions of primary charged particles were measured in three pseudorapidity ranges: $|\eta| < 0.5$, $|\eta| < 1.0$ and $|\eta| < 1.5$. At $\sqrt{s} = 7$~TeV, $\nch$ reaches about 70, 120 and 150, in $|\eta| <$ 0.5, 1.0 and 1.5, respectively, with the present statistics. These correspond to multiplicity densities 8 to 10 times the average multiplicity density. Based on the J. D. Bjorken formula \cite{Bjorken:1982qr}, a characteristic collision energy density can be estimated, which increases by the same factor. For a qualitative estimate, assuming that the average energy density in pp collisions at $\sqrt{s} = 7$~TeV is of the order of 1 GeV/fm$^3$ (see for example \cite{Csanad:2013lba}), a density of 10 GeV/fm$^3$ should be reached with high multiplicity pp collisions, similar to the energy density of Au-Au central collisions at RHIC \cite{mclerran2003rhic}. When LHC runs at its nominal centre-of-mass energy of 14~TeV, high multiplicity proton-proton collisions will provide further direct comparisons of nuclear matter properties for interacting systems with similar energy densities but very different volumes.

At LHC energies, above 2~TeV, multiplicity distributions can no longer be represented by a single NBD, but a double NBD gives a good representation of the data. The deviation from single NBD is already visible in the tail of the distributions at $\sqrt{s} = 0.9$~TeV, but becomes increasingly large as the centre-of-mass energy increases.

A test of KNO scaling between $\sqrt{s} = 0.9$ and 8~TeV confirms that KNO scaling violation increases with increasing $\sqrt{s}$ and, at a given centre-of-mass energy, with increasing width of pseudorapidity intervals.

Comparisons with models in the pseudorapidity range $|\eta| \leq 1$ show that none of the event generators considered is able to describe all properties of charged particle production up to $\sqrt{s} = 8$~TeV. \pythia{6} Perugia 2011, \pythia{8} 4C and EPOS LHC describe the pseudorapidity densities fairly well at $\sqrt{s} =$~7~TeV, as well as the multiplicity distributions, but not above $\nch \sim 60$. The fact that \pythia{6} Perugia 2011, \pythia{8} 4C and EPOS LHC are in reasonable agreement with the data presented in this publication can probably be attributed to the fact that these generators were adjusted using the first LHC data.

Studies of charged-particle production are refining the understanding of global properties of proton-proton collisions at the LHC. It was already demonstrated that, as with lower energy data, there is a strong correlation between multiplicity and mean transverse momentum~$\left<\pt\right>$ \cite{Abelev:2013bla}. However, there is also evidence \cite{aamodt2011strange} that the high multiplicity events have a topology more spherical than expected from current event generators \pythia{6}, \pythia{8} and \phojet, suggesting that single hard-parton collisions may not be the only contributors to high multiplicity events. The general picture that emerges from this study is that, from $\sqrt{s} = 0.9$ to 8~TeV, multiplicity distributions and charged particle pseudorapidity densities follow a smooth evolution.
%
%

\newenvironment{acknowledgement}{\relax}{\relax}
\begin{acknowledgement}
\section*{Acknowledgements}

The ALICE Collaboration would like to thank all its engineers and technicians for their invaluable contributions to the construction of the experiment and the CERN accelerator teams for the outstanding performance of the LHC complex.
The ALICE Collaboration gratefully acknowledges the resources and support provided by all Grid centres and the Worldwide LHC Computing Grid (WLCG) collaboration.
The ALICE Collaboration acknowledges the following funding agencies for their support in building and
running the ALICE detector:
State Committee of Science,  World Federation of Scientists (WFS)
and Swiss Fonds Kidagan, Armenia;
Conselho Nacional de Desenvolvimento Cient\'{\i}fico e Tecnol\'{o}gico (CNPq), Financiadora de Estudos e Projetos (FINEP),
Funda\c{c}\~{a}o de Amparo \`{a} Pesquisa do Estado de S\~{a}o Paulo (FAPESP);
National Natural Science Foundation of China (NSFC), the Chinese Ministry of Education (CMOE)
and the Ministry of Science and Technology of China (MSTC);
Ministry of Education and Youth of the Czech Republic;
Danish Natural Science Research Council, the Carlsberg Foundation and the Danish National Research Foundation;
The European Research Council under the European Community's Seventh Framework Programme;
Helsinki Institute of Physics and the Academy of Finland;
French CNRS-IN2P3, the `Region Pays de Loire', `Region Alsace', `Region Auvergne' and CEA, France;
German Bundesministerium fur Bildung, Wissenschaft, Forschung und Technologie (BMBF) and the Helmholtz Association;
General Secretariat for Research and Technology, Ministry of Development, Greece;
Hungarian Orszagos Tudomanyos Kutatasi Alappgrammok (OTKA) and National Office for Research and Technology (NKTH);
Department of Atomic Energy and Department of Science and Technology of the Government of India;
Istituto Nazionale di Fisica Nucleare (INFN) and Centro Fermi -
Museo Storico della Fisica e Centro Studi e Ricerche ``Enrico Fermi'', Italy;
MEXT Grant-in-Aid for Specially Promoted Research, Ja\-pan;
Joint Institute for Nuclear Research, Dubna;
National Research Foundation of Korea (NRF);
Consejo Nacional de Cienca y Tecnologia (CONACYT), Direccion General de Asuntos del Personal Academico(DGAPA), M\'{e}xico, Amerique Latine Formation academique - 
European Commission~(ALFA-EC) and the EPLANET Program~(European Particle Physics Latin American Network);
Stichting voor Fundamenteel Onderzoek der Materie (FOM) and the Nederlandse Organisatie voor Wetenschappelijk Onderzoek (NWO), Netherlands;
Research Council of Norway (NFR);
National Science Centre, Poland;
Ministry of National Education/Institute for Atomic Physics and National Council of Scientific Research in Higher Education~(CNCSI-UEFISCDI), Romania;
Ministry of Education and Science of Russian Federation, Russian
Academy of Sciences, Russian Federal Agency of Atomic Energy,
Russian Federal Agency for Science and Innovations and The Russian
Foundation for Basic Research;
Ministry of Education of Slovakia;
Department of Science and Technology, South Africa;
Centro de Investigaciones Energeticas, Medioambientales y Tecnologicas (CIEMAT), E-Infrastructure shared between Europe and Latin America (EELA), 
Ministerio de Econom\'{i}a y Competitividad (MINECO) of Spain, Xunta de Galicia (Conseller\'{\i}a de Educaci\'{o}n),
Centro de Aplicaciones Tecnológicas y Desarrollo Nuclear (CEA\-DEN), Cubaenerg\'{\i}a, Cuba, and IAEA (International Atomic Energy Agency);
Swedish Research Council (VR) and Knut $\&$ Alice Wallenberg
Foundation (KAW);
Ukraine Ministry of Education and Science;
United Kingdom Science and Technology Facilities Council (STFC);
The United States Department of Energy, the United States National
Science Foundation, the State of Texas, and the State of Ohio;
Ministry of Science, Education and Sports of Croatia and  Unity through Knowledge Fund, Croatia;
Council of Scientific and Industrial Research (CSIR), New Delhi, India;
Pontificia Universidad Cat\'{o}lica del Per\'{u}.
\end{acknowledgement}

\mciteSetMidEndSepPunct{;\xspace}{\ifmciteBstWouldAddEndPunct.\else\fi}{\relax}
\bibliographystyle{utphys}   
\bibliography{paper_biblio}

\newpage
\appendix
%
%
\section{The ALICE Collaboration}
\label{app:collab}



\begingroup
\small
\begin{flushleft}
J.~Adam\Irefn{org40}\And
D.~Adamov\'{a}\Irefn{org83}\And
M.M.~Aggarwal\Irefn{org87}\And
G.~Aglieri Rinella\Irefn{org36}\And
M.~Agnello\Irefn{org110}\And
N.~Agrawal\Irefn{org48}\And
Z.~Ahammed\Irefn{org132}\And
I.~Ahmed\Irefn{org16}\And
S.U.~Ahn\Irefn{org68}\And
S.~Aiola\Irefn{org136}
% ...
\renewcommand\labelenumi{\textsuperscript{\theenumi}~}

\section*{Affiliation notes}
\renewcommand\theenumi{\roman{enumi}}
\begin{Authlist}
\item \Adef{0}Deceased
\item \Adef{idp1754576}{Also at: Georgia State University, Atlanta, Georgia, United States}
\item \Adef{idp3803328}{Also at: M.V. Lomonosov Moscow State University, D.V. Skobeltsyn Institute of Nuclear, Physics, Moscow, Russia}
\end{Authlist}

\section*{Collaboration Institutes}
\renewcommand\theenumi{\arabic{enumi}~}
\begin{Authlist}

\item \Idef{org1}A.I. Alikhanyan National Science Laboratory (Yerevan Physics Institute) Foundation, Yerevan, Armenia
\item \Idef{org2}Benem\'{e}rita Universidad Aut\'{o}noma de Puebla, Puebla, Mexico
\item \Idef{org3}Bogolyubov Institute for Theoretical Physics, Kiev, Ukraine
\item \Idef{org4}Bose Institute, Department of Physics and Centre for Astroparticle Physics and Space Science (CAPSS), Kolkata, India
\item \Idef{org5}Budker Institute for Nuclear Physics, Novosibirsk, Russia
\item \Idef{org6}California Polytechnic State University, San Luis Obispo, California, United States
\item \Idef{org7}Central China Normal University, Wuhan, China
\item \Idef{org8}Centre de Calcul de l'IN2P3, Villeurbanne, France
\item \Idef{org9}Centro de Aplicaciones Tecnol\'{o}gicas y Desarrollo Nuclear (CEADEN), Havana, Cuba
\item \Idef{org10}Centro de Investigaciones Energ\'{e}ticas Medioambientales y Tecnol\'{o}gicas (CIEMAT), Madrid, Spain
\item \Idef{org11}Centro de Investigaci\'{o}n y de Estudios Avanzados (CINVESTAV), Mexico City and M\'{e}rida, Mexico
\item \Idef{org12}Centro Fermi - Museo Storico della Fisica e Centro Studi e Ricerche ``Enrico Fermi'', Rome, Italy
\item \Idef{org13}Chicago State University, Chicago, Illinois, USA
\item \Idef{org14}China Institute of Atomic Energy, Beijing, China
\item \Idef{org15}Commissariat \`{a} l'Energie Atomique, IRFU, Saclay, France
\item \Idef{org16}COMSATS Institute of Information Technology (CIIT), Islamabad, Pakistan
\item \Idef{org17}Departamento de F\'{\i}sica de Part\'{\i}culas and IGFAE, Universidad de Santiago de Compostela, Santiago de Compostela, Spain
\item \Idef{org18}Department of Physics and Technology, University of Bergen, Bergen, Norway
\item \Idef{org19}Department of Physics, Aligarh Muslim University, Aligarh, India
\item \Idef{org20}Department of Physics, Ohio State University, Columbus, Ohio, United States
\item \Idef{org21}Department of Physics, Sejong University, Seoul, South Korea
\item \Idef{org22}Department of Physics, University of Oslo, Oslo, Norway
\item \Idef{org23}Dipartimento di Elettrotecnica ed Elettronica del Politecnico, Bari, Italy
\item \Idef{org24}Dipartimento di Fisica dell'Universit\`{a} 'La Sapienza' and Sezione INFN Rome, Italy
\item \Idef{org25}Dipartimento di Fisica dell'Universit\`{a} and Sezione INFN, Cagliari, Italy
\item \Idef{org26}Dipartimento di Fisica dell'Universit\`{a} and Sezione INFN, Trieste, Italy
\item \Idef{org27}Dipartimento di Fisica dell'Universit\`{a} and Sezione INFN, Turin, Italy
\item \Idef{org28}Dipartimento di Fisica e Astronomia dell'Universit\`{a} and Sezione INFN, Bologna, Italy
\item \Idef{org29}Dipartimento di Fisica e Astronomia dell'Universit\`{a} and Sezione INFN, Catania, Italy
\item \Idef{org30}Dipartimento di Fisica e Astronomia dell'Universit\`{a} and Sezione INFN, Padova, Italy
\item \Idef{org31}Dipartimento di Fisica `E.R.~Caianiello' dell'Universit\`{a} and Gruppo Collegato INFN, Salerno, Italy
\item \Idef{org32}Dipartimento di Scienze e Innovazione Tecnologica dell'Universit\`{a} del  Piemonte Orientale and Gruppo Collegato INFN, Alessandria, Italy
\item \Idef{org33}Dipartimento Interateneo di Fisica `M.~Merlin' and Sezione INFN, Bari, Italy
\item \Idef{org34}Division of Experimental High Energy Physics, University of Lund, Lund, Sweden
\item \Idef{org35}Eberhard Karls Universit\"{a}t T\"{u}bingen, T\"{u}bingen, Germany
\item \Idef{org36}European Organization for Nuclear Research (CERN), Geneva, Switzerland
\item \Idef{org37}Excellence Cluster Universe, Technische Universit\"{a}t M\"{u}nchen, Munich, Germany
\item \Idef{org38}Faculty of Engineering, Bergen University College, Bergen, Norway
\item \Idef{org39}Faculty of Mathematics, Physics and Informatics, Comenius University, Bratislava, Slovakia
\item \Idef{org40}Faculty of Nuclear Sciences and Physical Engineering, Czech Technical University in Prague, Prague, Czech Republic
\item \Idef{org41}Faculty of Science, P.J.~\v{S}af\'{a}rik University, Ko\v{s}ice, Slovakia
\item \Idef{org42}Faculty of Technology, Buskerud and Vestfold University College, Vestfold, Norway
\item \Idef{org43}Frankfurt Institute for Advanced Studies, Johann Wolfgang Goethe-Universit\"{a}t Frankfurt, Frankfurt, Germany
\item \Idef{org44}Gangneung-Wonju National University, Gangneung, South Korea
\item \Idef{org45}Gauhati University, Department of Physics, Guwahati, India
\item \Idef{org46}Helsinki Institute of Physics (HIP), Helsinki, Finland
\item \Idef{org47}Hiroshima University, Hiroshima, Japan
\item \Idef{org48}Indian Institute of Technology Bombay (IIT), Mumbai, India
\item \Idef{org49}Indian Institute of Technology Indore, Indore (IITI), India
\item \Idef{org50}Inha University, Incheon, South Korea
\item \Idef{org51}Institut de Physique Nucl\'eaire d'Orsay (IPNO), Universit\'e Paris-Sud, CNRS-IN2P3, Orsay, France
\item \Idef{org52}Institut f\"{u}r Informatik, Johann Wolfgang Goethe-Universit\"{a}t Frankfurt, Frankfurt, Germany
\item \Idef{org53}Institut f\"{u}r Kernphysik, Johann Wolfgang Goethe-Universit\"{a}t Frankfurt, Frankfurt, Germany
\item \Idef{org54}Institut f\"{u}r Kernphysik, Westf\"{a}lische Wilhelms-Universit\"{a}t M\"{u}nster, M\"{u}nster, Germany
\item \Idef{org55}Institut Pluridisciplinaire Hubert Curien (IPHC), Universit\'{e} de Strasbourg, CNRS-IN2P3, Strasbourg, France
\item \Idef{org56}Institute for Nuclear Research, Academy of Sciences, Moscow, Russia
\item \Idef{org57}Institute for Subatomic Physics of Utrecht University, Utrecht, Netherlands
\item \Idef{org58}Institute for Theoretical and Experimental Physics, Moscow, Russia
\item \Idef{org59}Institute of Experimental Physics, Slovak Academy of Sciences, Ko\v{s}ice, Slovakia
\item \Idef{org60}Institute of Physics, Academy of Sciences of the Czech Republic, Prague, Czech Republic
\item \Idef{org61}Institute of Physics, Bhubaneswar, India
\item \Idef{org62}Institute of Space Science (ISS), Bucharest, Romania
\item \Idef{org63}Instituto de Ciencias Nucleares, Universidad Nacional Aut\'{o}noma de M\'{e}xico, Mexico City, Mexico
\item \Idef{org64}Instituto de F\'{\i}sica, Universidad Nacional Aut\'{o}noma de M\'{e}xico, Mexico City, Mexico
\item \Idef{org65}iThemba LABS, National Research Foundation, Somerset West, South Africa
\item \Idef{org66}Joint Institute for Nuclear Research (JINR), Dubna, Russia
\item \Idef{org67}Konkuk University, Seoul, South Korea
\item \Idef{org68}Korea Institute of Science and Technology Information, Daejeon, South Korea
\item \Idef{org69}KTO Karatay University, Konya, Turkey
\item \Idef{org70}Laboratoire de Physique Corpusculaire (LPC), Clermont Universit\'{e}, Universit\'{e} Blaise Pascal, CNRS--IN2P3, Clermont-Ferrand, France
\item \Idef{org71}Laboratoire de Physique Subatomique et de Cosmologie, Universit\'{e} Grenoble-Alpes, CNRS-IN2P3, Grenoble, France
\item \Idef{org72}Laboratori Nazionali di Frascati, INFN, Frascati, Italy
\item \Idef{org73}Laboratori Nazionali di Legnaro, INFN, Legnaro, Italy
\item \Idef{org74}Lawrence Berkeley National Laboratory, Berkeley, California, United States
\item \Idef{org75}Moscow Engineering Physics Institute, Moscow, Russia
\item \Idef{org76}Nagasaki Institute of Applied Science, Nagasaki, Japan
\item \Idef{org77}National Centre for Nuclear Studies, Warsaw, Poland
\item \Idef{org78}National Institute for Physics and Nuclear Engineering, Bucharest, Romania
\item \Idef{org79}National Institute of Science Education and Research, Bhubaneswar, India
\item \Idef{org80}Niels Bohr Institute, University of Copenhagen, Copenhagen, Denmark
\item \Idef{org81}Nikhef, Nationaal instituut voor subatomaire fysica, Amsterdam, Netherlands
\item \Idef{org82}Nuclear Physics Group, STFC Daresbury Laboratory, Daresbury, United Kingdom
\item \Idef{org83}Nuclear Physics Institute, Academy of Sciences of the Czech Republic, \v{R}e\v{z} u Prahy, Czech Republic
\item \Idef{org84}Oak Ridge National Laboratory, Oak Ridge, Tennessee, United States
\item \Idef{org85}Petersburg Nuclear Physics Institute, Gatchina, Russia
\item \Idef{org86}Physics Department, Creighton University, Omaha, Nebraska, United States
\item \Idef{org87}Physics Department, Panjab University, Chandigarh, India
\item \Idef{org88}Physics Department, University of Athens, Athens, Greece
\item \Idef{org89}Physics Department, University of Cape Town, Cape Town, South Africa
\item \Idef{org90}Physics Department, University of Jammu, Jammu, India
\item \Idef{org91}Physics Department, University of Rajasthan, Jaipur, India
\item \Idef{org92}Physik Department, Technische Universit\"{a}t M\"{u}nchen, Munich, Germany
\item \Idef{org93}Physikalisches Institut, Ruprecht-Karls-Universit\"{a}t Heidelberg, Heidelberg, Germany
\item \Idef{org94}Purdue University, West Lafayette, Indiana, United States
\item \Idef{org95}Pusan National University, Pusan, South Korea
\item \Idef{org96}Research Division and ExtreMe Matter Institute EMMI, GSI Helmholtzzentrum f\"ur Schwerionenforschung, Darmstadt, Germany
\item \Idef{org97}Rudjer Bo\v{s}kovi\'{c} Institute, Zagreb, Croatia
\item \Idef{org98}Russian Federal Nuclear Center (VNIIEF), Sarov, Russia
\item \Idef{org99}Russian Research Centre Kurchatov Institute, Moscow, Russia
\item \Idef{org100}Saha Institute of Nuclear Physics, Kolkata, India
\item \Idef{org101}School of Physics and Astronomy, University of Birmingham, Birmingham, United Kingdom
\item \Idef{org102}Secci\'{o}n F\'{\i}sica, Departamento de Ciencias, Pontificia Universidad Cat\'{o}lica del Per\'{u}, Lima, Peru
\item \Idef{org103}Sezione INFN, Bari, Italy
\item \Idef{org104}Sezione INFN, Bologna, Italy
\item \Idef{org105}Sezione INFN, Cagliari, Italy
\item \Idef{org106}Sezione INFN, Catania, Italy
\item \Idef{org107}Sezione INFN, Padova, Italy
\item \Idef{org108}Sezione INFN, Rome, Italy
\item \Idef{org109}Sezione INFN, Trieste, Italy
\item \Idef{org110}Sezione INFN, Turin, Italy
\item \Idef{org111}SSC IHEP of NRC Kurchatov institute, Protvino, Russia
\item \Idef{org112}Stefan Meyer Institut f\"{u}r Subatomare Physik (SMI), Vienna, Austria
\item \Idef{org113}SUBATECH, Ecole des Mines de Nantes, Universit\'{e} de Nantes, CNRS-IN2P3, Nantes, France
\item \Idef{org114}Suranaree University of Technology, Nakhon Ratchasima, Thailand
\item \Idef{org115}Technical University of Ko\v{s}ice, Ko\v{s}ice, Slovakia
\item \Idef{org116}Technical University of Split FESB, Split, Croatia
\item \Idef{org117}The Henryk Niewodniczanski Institute of Nuclear Physics, Polish Academy of Sciences, Cracow, Poland
\item \Idef{org118}The University of Texas at Austin, Physics Department, Austin, Texas, USA
\item \Idef{org119}Universidad Aut\'{o}noma de Sinaloa, Culiac\'{a}n, Mexico
\item \Idef{org120}Universidade de S\~{a}o Paulo (USP), S\~{a}o Paulo, Brazil
\item \Idef{org121}Universidade Estadual de Campinas (UNICAMP), Campinas, Brazil
\item \Idef{org122}University of Houston, Houston, Texas, United States
\item \Idef{org123}University of Jyv\"{a}skyl\"{a}, Jyv\"{a}skyl\"{a}, Finland
\item \Idef{org124}University of Liverpool, Liverpool, United Kingdom
\item \Idef{org125}University of Tennessee, Knoxville, Tennessee, United States
\item \Idef{org126}University of the Witwatersrand, Johannesburg, South Africa
\item \Idef{org127}University of Tokyo, Tokyo, Japan
\item \Idef{org128}University of Tsukuba, Tsukuba, Japan
\item \Idef{org129}University of Zagreb, Zagreb, Croatia
\item \Idef{org130}Universit\'{e} de Lyon, Universit\'{e} Lyon 1, CNRS/IN2P3, IPN-Lyon, Villeurbanne, France
\item \Idef{org131}V.~Fock Institute for Physics, St. Petersburg State University, St. Petersburg, Russia
\item \Idef{org132}Variable Energy Cyclotron Centre, Kolkata, India
\item \Idef{org133}Warsaw University of Technology, Warsaw, Poland
\item \Idef{org134}Wayne State University, Detroit, Michigan, United States
\item \Idef{org135}Wigner Research Centre for Physics, Hungarian Academy of Sciences, Budapest, Hungary
\item \Idef{org136}Yale University, New Haven, Connecticut, United States
\item \Idef{org137}Yonsei University, Seoul, South Korea
\item \Idef{org138}Zentrum f\"{u}r Technologietransfer und Telekommunikation (ZTT), Fachhochschule Worms, Worms, Germany
\end{Authlist}
\endgroup

\end{document}